\documentclass[%
reprint,
nofootinbib,
mathtools,
amsmath,
amssymb,
prd,
longbibliography,
superscriptaddress]{revtex4-2}

\usepackage{booktabs}
\usepackage{mathrsfs}
\usepackage{graphicx}
\usepackage{dcolumn}
\usepackage{bm}
\usepackage{lipsum}
\usepackage{graphics}
\usepackage{bbm}
\usepackage[dvipsnames]{xcolor}
\usepackage{color}
\usepackage{microtype}
\usepackage{IEEEtrantools}
\usepackage{mathrsfs}
\usepackage{amsthm}
\usepackage{enumitem}
\usepackage[normalem]{ulem}
\usepackage[pdftex,breaklinks,colorlinks,
linkcolor=Blue,
citecolor=teal,
anchorcolor=red,
urlcolor=cyan]{hyperref}
\usepackage[capitalise]{cleveref}
\usepackage{url}
\usepackage{orcidlink}
\hypersetup{breaklinks=true}
\usepackage[T1]{fontenc}
\usepackage{lmodern}
\usepackage{dsfont}
\usepackage{empheq} 
\usepackage{tensor}
\usepackage[latin9]{inputenc}
\usepackage{multirow}
\usepackage{tabularray}
\usepackage{esint}
\usepackage{soul}
\usepackage{amsfonts}
\usepackage{babel}
\usepackage{verbatim} 

\graphicspath{{Plots/}}

\newcommand{\pth}{\text{\normalfont\sffamily\itshape\th\/}}

\newcommand{\bracket}[1]{\left( #1 \right)}

\def\GHPwt{\ensuremath{\circeq}}
\font\ec=ecrm0800 at 12pt
\def\thorn{\hbox{\ec\char'336}}
\def\edth{\hbox{\ec\char'360}}

\def\mb{{\bar{m}}}

\def\NPs{\mathtt{s}}
\def\lp{s}
\def\l{\ell}
\def\m{m}
\def\n{n}
\def\q{q}
\def\Q{Q}
\def\lOne{\ell_1}
\def\mOne{m_1}
\def\qOne{q_1}
\def\QOne{Q_1}
\def\lTwo{\ell_2}
\def\mTwo{m_2}
\def\qTwo{q_2}
\def\QTwo{Q_2}

\def\pertorder{i}
\def\L{L}
\def\M{M}

\newcommand{\T}{{\mathcal T}}

\renewcommand{\O}{\mathcal{O}}

\def\scri{\mathscr{I}}

\begin{document}
\title{Quadratic quasinormal modes at null infinity on a Schwarzschild spacetime}

\author{Patrick Bourg\orcidlink{0000-0003-0015-0861}}
\affiliation{Institute for Mathematics, Astrophysics and Particle Physics, Radboud University, Heyendaalseweg 135, 6525 AJ Nijmegen, The Netherlands}
\author{Rodrigo Panosso Macedo\orcidlink{0000-0003-2942-5080}}
\affiliation{%
Center of Gravity, Niels Bohr Institute, Blegdamsvej 17, 2100 Copenhagen, Denmark
}%
\author{Andrew Spiers\orcidlink{0000-0003-0222-7578}}
\affiliation{%
 School of Mathematical Sciences \& School of Physics and Astronomy,
University of Nottingham, University Park, Nottingham, NG7 2RD, UK
}%
\affiliation{Nottingham Centre of Gravity, University of Nottingham, University Park, Nottingham, NG7 2RD, UK}
\author{Benjamin Leather\orcidlink{0000-0001-6186-7271}}
\affiliation{%
Max Planck Institute for Gravitational Physics (Albert Einstein Institute), Am M\"uhlenberg 1, Potsdam 14476, Germany
}%

\author{B\'eatrice Bonga\orcidlink{0000-0002-5808-9517}}
\affiliation{Institute for Mathematics, Astrophysics and Particle Physics, Radboud University, Heyendaalseweg 135, 6525 AJ Nijmegen, The Netherlands}

\author{Adam Pound\orcidlink{0000-0001-9446-0638}}
\affiliation{%
 School of Mathematical Sciences and STAG Research Centre, University of Southampton, Southampton, SO17 1BJ, United Kingdom
}

\date{\today}

\begin{abstract}
The ringdown of perturbed black holes has been studied since the 1970s, but until recently, studies have focused on linear perturbations. There is now burgeoning interest in \emph{nonlinear} perturbative effects during ringdown. Here, using a hyperboloidal framework, we provide a complete treatment of linear and quadratic quasinormal modes (QNMs and QQNMs) in second-order perturbation theory, in Schwarzschild spacetime. We include novel methods for extracting QNMs and QQNMs amplitudes using a Laplace transform treatment, allowing for the inclusion of arbitrary initial data. We produce both time- and frequency-domain codes. From these codes, we present new results further exploring the unforeseen dependence of QQNMs amplitudes on the parity of the progenitor system, as demonstrated in our letter~[Phys. Rev. Lett. 134, 061401 (2025)]. Our numerical results are restricted to perturbations of a Schwarzschild black hole, but our methods extend straightforwardly to the astrophysically realistic case of a Kerr black hole.
\end{abstract}

\maketitle

\tableofcontents

\section{Introduction}

In the era of gravitational wave (GW) astronomy, black hole (BH) spectroscopy plays a pivotal role in testing the nature of BHs and general relativity (GR) ~\cite{Nollert99_qnmReview,Kokkotas:1999bd,Berti:2009kk,Konoplya:2011qq,Barausse:2014tra,Dreyer:2003bv,Berti:2005ys}. 
The goal of this research program is to probe the BH geometry by measuring the characteristic frequencies of the BH spacetime in the so-called ringdown regime. The ringdown phase starts after the merger of two compact objects when the distorted final remnant settles down into a stationary BH with GWs dissipating energy and linear and angular momentum. The resulting waveform is (at least for some time) well described by a superposition of exponentially damped sinusoids, whose typical oscillatory and decaying time scales are encoded into the complex quasi-normal mode (QNM) frequencies. Most importantly, these QNM frequencies are uniquely determined by the mass and spin of the final BH in vacuum GR~\cite{Nollert99_qnmReview,Kokkotas:1999bd,Berti:2009kk,Konoplya:2011qq}.

While the BH spectroscopy program is mature and dates back to 1971 \cite{press1971long,Detweiler:1980gk,Dreyer:2003bv}, recently, there have been exciting developments on all fronts: observations, new techniques for data analysis, and new theoretical insights. Observationally, the measurement of the dominant QNM is confirmed in various gravitational wave events~\cite{LIGOScientific:2016aoc,LIGOScientific:2020tif,LIGOScientific:2021sio}. More remarkably, there is one event, GW190512, for which two QNM frequencies have been identified (although which two modes are present in the data is still contested) ~\cite{Isi:2019aib,Capano:2020dix,Capano:2021etf,Cotesta:2022pci,Capano:2022zqm,Forteza:2022tgq,Finch:2022ynt,Abedi:2023kot,Carullo:2023gtf,Baibhav:2023clw,Nee:2023osy,Zhu:2023mzv,Siegel:2023lxl, Gennari:2023gmx}. 
Looking into the near future, data from the next generation of GW detectors, such as the Einstein Telescope, Cosmic Explorer, and LISA, will contribute significantly to the BH spectroscopy program: not only will there be more events observed, the signal-to-noise ratio (SNR) will also be significantly higher and thereby allow for more stringent tests. The latest estimates suggest that ET shall observe at least two detectable QNMs from stellar-mass binary mergers in 20-50 events per year ~\cite{Maggiore:2019uih,Cabero:2019zyt} and LISA will observe even $\sim 5-8$ QNMs from multiple massive BH binary mergers~\cite{Berti:2016lat,Toubiana:2023cwr}.

Developments on the observational side are complemented by new algorithms for detecting QNMs on the data analysis side, such as the QNM (rational) filters~\cite{Ma:2022wpv,Ma:2023cwe}. Other improvements in inferring the parameters using Bayesian analyses include marginalizing over the starting time of the ringdown phase and a deeper understanding of data conditioning operations~\cite{Finch:2022ynt,Correia:2023bfn,Correia:2023ipz,Siegel:2024jqd}.

In addition to all these advancements, future detection will allow us to test the nonlinear nature of gravity using the ringdown regime. Since GR is a nonlinear theory of gravity, it has been recently emphasised that second-order effects in BH perturbation theory (BHPT) also contribute to the GW signal in the ringdown phase and may even dominate over some linear contributions during parts of the ringdown ~\cite{London:2014cma,Cheung:2022rbm, Mitman:2022qdl, Zlochower:2003yh,Baibhav:2023clw,Redondo-Yuste:2023seq,Cheung:2023vki,ma2024excitation}. In particular, by analyzing numerical simulations of BH mergers, it was indeed found that this is the case \cite{London:2014cma,Mitman:2022qdl,Cheung:2022rbm,Cheung:2023vki}. 

More specifically, the QNM frequencies arising from BHPT at first order are usually labelled by three integers, as in $\omega_{\l \m \n}$. The pair $(\l, \m)$ corresponds to polar and azimuthal indices resulting from projecting the underlying field variables onto spherical harmonics on the celestial sphere, while the overtone index $\n=0,1,\ldots$ lists the characteristic frequencies within a fixed angular mode. The dynamics at second order in BHPT yield the same set of frequencies as the linear equation (from the homogeneous solution), but also an additional set of new frequencies. These new frequencies arise from the coupling of two linear QNMs $\omega_{\l_1 \m_1 \n_1}$ and $\omega_{\l_2 \m_2 \n_2}$, yielding the so-called quadratic QNMs (QQNMs): $\omega_{\l_1 \m_1 \n_1\times\l_2 \m_2 \n_2}=\omega_{\l_1 \m_1 \n_1}+\omega_{\l_2 \m_2 \n_2}$~\cite{London:2014cma}. Since the QNM spectrum in the frequency space is sparse, the detection of these QQNMs is a clear signature of a nonlinear effect. In other words, the QQNMs are essentially fingerprinted with the order of perturbation theory that they arose from. This gives a clean split of the GW signal into linear and non-linear pieces and thus would allow for new tests of BHs and GR. 

While QQNMs have so far only been identified in numerical simulations, forecasts indicate that QQNMs will be detected in up to a few tens of events per year~\cite{Yi:2024elj, Khera:2024yrk} by the Einstein Telescope and Cosmic Explorer. Moreover, the most optimistic predictions for LISA indicate that up to ${\cal O}(1000)$ events may offer accurate data for BH spectroscopy at second order in BHPT~\cite{Yi:2024elj}. A recent study~\cite{Lagos:2024ekd} also found that including the GR predictions for QQNM amplitudes dramatically improved fitting techniques and the likelihood of detecting QQNMs.

This article focuses on QQNMs. It was argued in the literature that the ratio of the amplitudes of the QQNM and its linear parent modes would be initial-data independent and would only depend on the mass and spin of the final BH \cite{Mitman:2022qdl,Zhu:2024rej,Kehagias:2023ctr,Redondo-Yuste:2023seq,Perrone:2023jzq}. However, we have recently shown~\cite{Bourg:2024jme} that this ratio is not initial-data independent but, in fact, depends on the ratio between even- and odd-parity linear QNMs. This ratio of even- to odd-parity modes depends on the degree to which the progenitor system possessed equatorial (up-down) symmetry. In other words, contrary to previous expectations that the QQNM/QNM ratio is independent of the past history of the binary, we demonstrated that this ratio depends on what created the ringing BH~\cite{Bourg:2024jme}. 
Our new understanding of the QQNM/QNM ratio allowed us to reconcile some contradictory results in the literature. Using second-order BHPT in the simple case of a Schwarzschild background, Ref.~\cite{ma2024excitation} found a QQNM/QNM ratio of magnitude $\approx 0.137$ for a particular mode combination, while other works observed the value $\approx 0.15$ for the same mode combination~\cite{Redondo-Yuste:2023seq,Bucciotti:2024zyp}. This difference is entirely due to the different choices for the ratio of the linear even- and odd-parity modes sourcing the quadratic mode. Our new understanding has recently been used to study QQNMs in Kerr~\cite{Khera:2024yrk}.

The insight that the QQNM/QNM ratio depends on the progenitor system through the even- to odd-parity mode ratio impacts GW data analysis and astronomy directly. In particular, analyses of numerical simulations with up-down symmetry should use this knowledge in their fitting procedures. Moreover, in real data, the measurements of QQNMs can be used to extract the individual excitation of first-order even- and odd-sectors, offering a route to explore the breaking of isospectrality in beyond-GR theories or due to environmental effects.

This work expands the relevant results from~\cite{Bourg:2024jme}, provides a detailed description of the underlying geometrical and numerical infrastructure, and is accompanied by publicly available codes. Specifically, to obtain the semi-analytical results in~\cite{Bourg:2024jme}, we have developed a novel code that fully controls the core geometrical aspects of the problem. For this purpose, we have employed the covariant second-order BHPT formalism from~\cite{Spiers:2023mor,Spiers:2023cip,Green:2019nam}, initially developed to address challenges within the gravitational self-force program~\cite{Pound:2021qin}.

Black hole spacetimes are commonly studied in terms of Schwarzschild (or Boyer-Lindquist) coordinates $(t,r,\theta, \varphi)$. Even though the time evolution of the ringdown dynamics shows a field decaying in time with the characteristic QNM frequencies, it has long been known that the corresponding QNM radial eigenfunctions diverge in the limit $r\rightarrow \infty$, and at the BH horizon $r=r_h$. Because the source term dictating the dynamics at second order depends on quadratic products of these QNM eigenfunctions, it inherits such divergence, making second-order BHPT particularly challenging. Apart from being counter-intuitive on physical grounds, these divergences make calculations at second order extremely challenging at a practical level (see, e.g., \cite{Bucciotti:2024jrv}).

Such divergences naturally appear in the standard coordinate chart $(t,r,\theta, \varphi)$, in which $r=r_h$ and $r\to\infty$ correspond to the bifurcation two-sphere and spatial infinity, where the time-domain QNM solutions diverge. 
The hyperboloidal framework~\cite{Zenginoglu:2011jz,PanossoMacedo:2023qzp} solves this issue altogether as hyperboloidal slices do not reach the surfaces where the QNM solutions are singular: the past horizon, past null infinity and spatial infinity.\footnote{The blow-up there is a consequence of assuming that the QNM solutions extend to the infinite past, whereas only their ``late time'' behaviour is physically relevant following whatever event has perturbed the black hole (or created the perturbed black hole).} Thus, this framework naturally restricts to regions where the QNM solutions are physically meaningful.
Indeed, hyperboloidal slicings are currently employed in several studies of QQNMs~\cite{Redondo-Yuste:2023seq,Zhu:2024rej,Khera:2024yrk}.

Starting from the covariant second-order BHPT formalism of Refs.~\cite{Spiers:2023mor,Spiers:2023cip,Green:2019nam}, we have specialised the equations to a time- and 
frequency-domain hyperboloidal framework that allows us to directly and accurately compute the physical waveform without requiring regularization~\cite{PanossoMacedo:2014dnr,Ansorg:2016ztf,Ammon:2016fru,PanossoMacedo:2018hab,PanossoMacedo:2019npm,Jaramillo:2020tuu,PanossoMacedo:2023qzp,PanossoMacedo:2022fdi}.
Such a combination of time- and frequency-domain techniques is a unique feature of our framework. Indeed, the majority of studies in nonlinear GW spectroscopy are performed in the time domain, where the QQNM amplitudes are extracted by fitting the expected model to the time series resulting either from NR~\cite{London:2014cma,Mitman:2022qdl,Cheung:2022rbm,Cheung:2023vki} or BHPT~\cite{Redondo-Yuste:2023seq,Zhu:2024rej}. In BHPT, however, the QQNM amplitude is not a free parameter. Instead, it is determined by the second-order source, whose content is fully fixed by the even- and odd-parity amplitudes of the linear QNMs. Thus, frequency-domain calculations, such as those performed by~\cite{ma2024excitation,Bucciotti:2024jrv,Khera:2024yrk}, have the potential to optimize the analyses of the QQNMs by {\em predicting} the amplitude coefficients directly for the source term, and avoiding the time evolution and the systematic errors associated with a numerical fitting altogether.

Our frequency-domain analysis complements other, similar recent QQNM studies~\cite{Bucciotti:2024zyp,Khera:2024yrk}, particularly Ref.~\cite{Khera:2024yrk}, which also used hyperboloidal methods. We build on the techniques originally introduced by Refs.~\cite{Ansorg:2016ztf,Ammon:2016fru}, in which QNM excitation amplitudes are calculated directly from the hyperboloidal source. We detail how to use this approach to cleanly extract the linear and quadratic QNM content from the full second-order solution, without resorting to bi-linear orthogonality operators that require the analytical extension of the radial coordinate into the complex plane~\cite{Zimmerman:2014aha,Green:2022htq,London:2023aeo}. Additionally, our Laplace transformation approach allows for the inclusion of initial data in our analysis.

Of course, all the above considerations rely on the frequency-domain QQNM calculations being equivalent to the QQNM excitation in gravitational wave signals. Thus, a consistency check between time- and frequency-domain approaches is necessary. To validate our frequency-domain results, we use a fully spectral, hyperboloidal time-domain code.

Even though the current work focuses on the Schwarzschild spacetime, our approach offers an independent infrastructure to consistently cross-check the results in the field. By using well-established results from self-force theory and the hyperboloidal framework, we make different choices for the calculations (e.g. gauge and coordinates employed, variables under consideration, and the resulting source). Our independent analysis makes us confident in the final results, and it allows us to develop a self-consistent code to directly study QQNMs resulting from the coupling of {\em any} parent QNM pair.

This paper is organized as follows. Section~\ref{sec:BHPT_Review} reviews the covariant formalism for BHPT at first and second order. Section~\ref{sec:HypFramework} introduces the specific hyperboloidal coordinate system we use. Then, Sections~\ref{sec:QNMs} and \ref{sec:QQNMs} detail the techniques to calculate the relevant quantities at first and second order: QNM frequencies and eigenfunctions, as well as excitation factors associated with linear and quadratic QNMs. Finally, Section~\ref{sec:codes} outlines the numerical methods used in the time- and frequency-domain codes before we present our results in Section~\ref{sec:Results}. We conclude the work and discuss future directions in Section~\ref{sec:Conclusion}. Some review material and technical details are relegated to appendices.

\section{Perturbation theory at first and second order}\label{sec:BHPT_Review}

\subsection{Black hole perturbation theory}

To study linear and quadratic QNMs, we work with BHPT up to second order. In BHPT, approximate solutions to the Einstein field equation are found perturbatively around a background BH spacetime.
We expand the metric ($g_{ab}$) around the background metric ($g^{(0)}_{ab}$) in orders of a small parameter $\varepsilon$,
\begin{align}\label{eq:metricexpansion}
g_{ab} &= g^{(0)}_{ab}+\varepsilon h^{(1)}_{ab} +\varepsilon^2 h^{(2)}_{ab} +\ldots+ \varepsilon^n h^{(n)}_{ab}+\ldots,
\end{align}
where $h^{(i)}_{ab}$ is the $i$th-order metric perturbation. In this work, Schwarzschild spacetime is our background; however, we build our calculation using methods that naturally extend to Kerr.

We can use~\cref{eq:metricexpansion} to perturbatively expand the vacuum Einstein field equations, $G_{ab}[g_{ab}]= 0$; 
up to second order in $\varepsilon$, this gives~\cite{Pound:2021qin,Spiers:2023cip},
\begin{align} 
\delta G_{ab}[h^{(1)}_{ab}]&=0,\label{eq:linearEFE}\\
\delta G_{ab}[h^{(2)}_{ab}]&=-\delta^2G_{ab}[h^{(1)}_{ab},h^{(1)}_{ab}],\label{eq:quadraticEFE}
\end{align}
where $\delta G_{ab}$ is the linearised Einstein operator and $\delta^2 G_{ab}$ is the quadratic Einstein operator~\cite{Spiers:2023cip}. We use geometrized units with $G=c=1$.
The left-hand side of~\cref{eq:linearEFE,eq:quadraticEFE} shows the same operator acting on $h^{(1)}_{ab}$ and $h^{(2)}_{ab}$. Thus, the homogeneous second-order solutions are identical to linear QNM solutions; we assume that the second-order homogeneous solutions are absorbed into the first-order solution without loss of generality. At second order, the particular solutions remain; these solutions have distinct frequencies from the homogeneous solutions of ~\cref{eq:quadraticEFE}. The particular solution frequencies are the so-called quadratic QNMs or QQNMs for short (the labelling ``quadratic'' is due to the source in ~\cref{eq:quadraticEFE} being quadratic in $h_{ab}^{(1)}$). 

\cref{eq:linearEFE,eq:quadraticEFE} are two sets of ten coupled differential equations (with six independent and four constraint equations each); they are \textit{generally} non-separable on a Kerr background. Hence, as we build our formalism to naturally extend to Kerr, we opt to solve the Teukolsky equations, rather than \cref{eq:linearEFE,eq:quadraticEFE} directly. The Teukolsky equations offer multiple advantages over the linearised Einstein field equations~\eqref{eq:linearEFE} and \eqref{eq:quadraticEFE}: Instead of ten coupled equations, one solves a complex second-order differential equation, called the Teukolsky (master) equation, for the perturbed Weyl curvature scalar. The Teukolsky equation is separable into radial ordinary differential equations by decomposing the angular dependence into spin-weighted spheroidal harmonic modes. The perturbed Weyl curvature scalars encode most of the information of the metric perturbation~\cite{Wald:1973wwa}. In particular, the gravitational wave strain at future null infinity is directly related to one of the Weyl scalars~\cite{Teukolsky:1972my,Campanelli:1998jv,Spiers:2023cip}.

In the next section, we provide a synopsis of the covariant formulation of the Teukolsky equation.

\subsection{The Teukolsky equation}
\label{sec:teukolsky-equation}

The Teukolsky equation is expressed within the Newman--Penrose (NP) or Geroch--Held--Penrose (GHP) formalisms~\cite{Geroch:1973am}, reviewed in~\cref{app:np-ghp}. The basic ingredient in these formalisms is an orthonormal null tetrad $\{l^a,n^a,m^a,\bar m^a\}$.

Due to the symmetry of Petrov type-D spacetimes, Schwarzschild (and Kerr) admits decoupled spin $+2$ and $-2$ equations constructed from the perturbative Einstein equations $\delta G_{ab}[h^{(i)}_{ab}]=S_{ab}^{(i)}$. We write these decoupled equations as 
\begin{align}\label{eq:teuk}
    \hat\O\left[ \Psi_{0L}^{(i)} \right] = {\rm S}_0^{(\pertorder)}, \ \ \ \ \hat\O^\prime \left[ \Psi^{(\pertorder)}_{4L} \right] = {\rm S}_4^{(\pertorder)},
\end{align}
respectively, at each perturbative order $(\pertorder)$, where $\hat\O$ and $\hat\O^\prime$ are second-order linear partial differential wave-like operators. The sources are related to those in the Einstein equation by
\begin{equation}
{\rm S}_0^{(i)}={\cal S}_0\left[S^{(i)}_{ab}\right], \qquad {\rm S}_4^{(i)}={\cal S}_4\left[S^{(i)}_{ab}\right],
\end{equation}
where ${\cal S}_0$ and ${\cal S}_4$ are second-order linear differential operators. The curvature scalars $\Psi^{(i)}_{0L}$ and $\Psi^{(i)}_{4L}$ are similarly related to the metric perturbations by
\begin{equation}\label{eq:PsiL def}
\Psi_{0L}^{(\pertorder)}=\T_0\left[ h^{(\pertorder)}_{ab} \right], \quad \Psi_{4L}^{(\pertorder)}=\T_4 \left[ h^{(\pertorder)}_{ab} \right],
\end{equation}
where $\T_0$, $\T_4$ are also second-order linear differential operators. Explicit (tetrad- and coordinate-invariant) definitions of the operators $\T_0$, $\T_4$, ${\cal S}_0$, ${\cal S}_4$, and $\hat\O$ and $\hat\O'$ are available in Ref.~\cite{Pound:2021qin}.

In Eq.~\eqref{eq:PsiL def}, we have isolated the \emph{linear} parts of the $i$th-order Weyl scalars. At first order, these constitute the entire Weyl scalars: $\Psi_{0}^{(1)} = \Psi_{0L}^{(1)}$ and $\Psi_{4}^{(1)}=\Psi_{4L}^{(1)}$. At second order,
\begin{equation}
\Psi_{0}^{(2)} = \Psi_{0L}^{(2)} + \Psi_{0Q}^{(2)}, \quad \Psi_{4}^{(2)} = \Psi_{4L}^{(2)} + \Psi_{4Q}^{(2)};
\end{equation}
the label $L$ indicates the contribution that is linear in the second-order metric perturbation, whereas the label $Q$ indicates quadratic contributions from first-order perturbations.
We discuss the relationship between the different variables $\Psi^{(2)}_{0}$, $\Psi^{(2)}_{4}$ and $\Psi^{(2)}_{0L}$, $\Psi^{(2)}_{4L}$ in Appendix~\ref{sec:psi4 vs psi4L}.

In Kerr spacetime, \cref{eq:teuk} is separable when expressed in terms of a master Teukolsky equation form~\cite{Teukolsky:1972my, Teukolsky:1973ha, Pound:2021qin}.
The master Teukolsky equations are tetrad independent, although the relationship between the Weyl scalars and the master variable is tetrad dependent. From now on, we will work with the Kinnersley tetrad; see Eq.~\eqref{eq:tetrad,Kinn,t}.
In the Kinnersley tetrad, the master Teukolsky equation is
\begin{align}\label{eq:masterTeuk}
{}_{\NPs} \hat{\mathcal{O}}[{}_{\NPs} \Psi^{(\pertorder)}]= {}_{\NPs} {\rm S}^{(\pertorder)},
\end{align}
where ${}_{\NPs} \hat{\mathcal{O}}$ is the spin-$\NPs$ master Teukolsky wave operator~\cite{Pound:2021qin}. The master functions ${}_{\NPs} \Psi^{(\pertorder)}$ and source terms ${}_{\NPs} {\rm S}^{(\pertorder)}$ relate to their counterparts in \cref{eq:teuk} via
\begin{eqnarray}
{}_{+2} \Psi^{(\pertorder)} = \Psi_{0L}^{(\pertorder)}, \quad {}_{+2} {\rm S}^{(\pertorder)} = -2r^2 {\rm S}_0^{(\pertorder)}, \\
{}_{-2} \Psi^{(\pertorder)} = r^4 \Psi_{4L}^{(\pertorder)}, \quad {}_{-2} {\rm S}^{(\pertorder)} = -2r^6 {\rm S}_4^{(\pertorder)}. \label{eqn:MasterFunctions}
\end{eqnarray}
with $r$ a radial coordinate in a spherical polar coordinate system such as the Schwarzschild coordinates $(t,r,\theta, \varphi)$.

At first order, $(\pertorder)=(1)$, the source term vanishes: ${}_{\NPs} {\rm S}^{(1)} = 0$. At second order, $(\pertorder)=(2)$, ${}_{\NPs} {\rm S}^{(2)}$ depends quadratically on the first-order metric perturbation:
\begin{align}\label{eq:masterreducedTeuk}
{}_{+2}{\rm S}^{(2)}&=2r^2 \mathcal{S}_0\Big[\delta^2 G_{ab}[h^{(1)}_{ab},h^{(1)}_{ab}]\Big],\\ 
{}_{-2}{\rm S}^{(2)}&=2r^{6}\mathcal{S}_4\Big[\delta^2 G_{ab}[h^{(1)}_{ab},h^{(1)}_{ab}]\Big].\label{eq:masterreducedTeuk s=-2}
\end{align}
These quadratic sources are given explicitly, in mode-decomposed form, in the \texttt{PerturbationEquations} package~\cite{warburton2023perturbationequations} in the Black Hole Perturbation Toolkit~\cite{BHPToolkit}. Note that they require the complete first-order metric perturbation as input, not simply the first-order Weyl scalar ${}_{\pm2}\Psi^{(1)}$. We review the procedure of reconstructing $h^{(1)}_{ab}$ from ${}_{\pm2}\Psi^{(1)}$, and further discuss the calculation of the quadratic source, in~\cref{app:MetricReconstruction}.

We are primarily interested in the perturbations of $\Psi_4$, rather than $\Psi_0$, as they directly relate to the gravitational wave strain at future null infinity.
In the Kinnersley tetrad, the limit towards future null infinity ($r\rightarrow\infty$ taken along slices of constant retarded time $u$) yields
\begin{align}\label{eq:strain-psi4L}
    \Psi_{4L}^{(\pertorder)}=-\frac{1}{2}\partial_u^2 h^{(\pertorder)}_{\mb\mb}=-\frac{1}{2}\partial_u^2 \left(h^{(\pertorder)}_+ -ih^{(\pertorder)}_\times \right),
\end{align}
where $h^{(i)}_+$ and $h^{(i)}_\times$ are the two independent polarizations of the gravitational wave strain.\footnote{This result is derived by examining the leading-order behaviour in $\T_4$; see Refs.~\cite{Pound:2021qin,Newman:1961qr, Newman:1962cia,Szekeres:1965ux}.}
The variables $\Psi_4^{(\pertorder)}$ and $\Psi_{4L}^{(\pertorder)}$ are identical at leading order in $\frac{1}{r}$ (see Appendix~\ref{sec:psi4 vs psi4L}), so that 
\begin{align}
    \Psi_{4}^{(\pertorder)}=-\frac{1}{2}\partial_u^2 h^{(\pertorder)}_{\mb\mb}=-\frac{1}{2}\partial_u^2 \left(h^{(\pertorder)}_+-ih^{(\pertorder)}_\times \right)
\end{align}
also holds for $r\rightarrow\infty$ along constant $u$.

In the rest of this paper, we will suppress the spin index $\NPs$ for most variables, and will restore it when a potential confusion could arise. In particular, we have chosen to keep the spin index $\NPs$ for the spherical harmonics for clarity.

\section{
The hyperboloidal framework}\label{sec:HypFramework}

\subsection{Hyperboloidal equations in the time domain}
\begin{figure}[t] \centering
\includegraphics[width=\columnwidth]{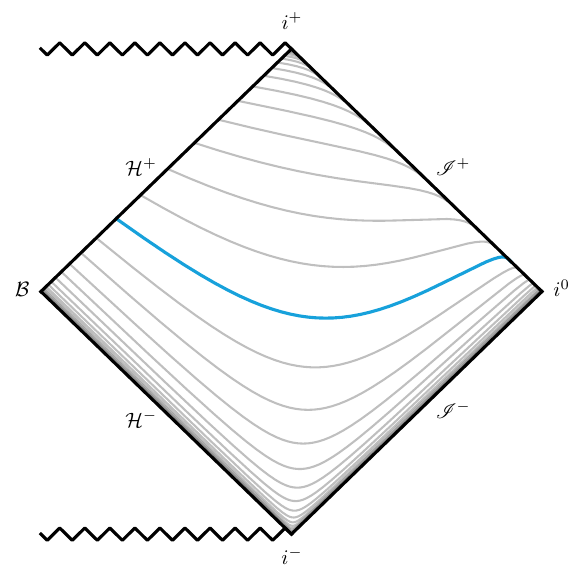}
    \caption{Penrose diagram of the Schwarzschild exterior
    demonstrating the minimal-gauge hyperboloidal slicing we employ in 
    this work. The solid curves each depict surfaces of constant hyperboloidal time $\tau$ that extend smoothly from the black-hole horizon, 
    ${\cal H}^{+}$, to future null infinity, $\scri^{+}$.
    In this choice of hyperboloidal coordinate system, with 
    $\lambda = 4M$, the future horizon and future null infinity are 
    located at $\sigma = 1$ and $\sigma = 0$, respectively.
    The ``initial time surface'' $\tau=0$ is highlighted in blue.
    }
    \label{fig:minimal_gauge}
\end{figure}

We now specialise the formalism in the previous section to a coordinate system adapted to the geometry at the black-hole horizon and the wave zone. For this purpose, we introduce the hyperboloidal coordinates $(\tau, \sigma, \theta, \varphi)$ via the scri-fixing technique~\cite{Zenginoglu:2007jw} with notation from Ref.~\cite{PanossoMacedo:2023qzp}:
\begin{equation}
t = \lambda\bigg( \tau - H(\sigma) \bigg), \quad r = \dfrac{r_h}{\sigma},
\end{equation}
with $\lambda$ a convenient length scale for the spacetime and $r_h = 2 M$. The so-called height function $H(\sigma)$ ensures that, along $\tau=\text{constant}$, the surfaces $r\rightarrow \infty$ ($\sigma=0$) and $r=r_h$ ($\sigma=1$) correspond, respectively, to future null infinity $\scri^+$ and the black-hole horizon ${\cal H}^+$; see Fig.~\ref{fig:minimal_gauge}. 
Note that in a Laplace framework, one is interested in the region $\tau \geq 0$, where $\tau=0$ (the blue curve) represents the ``initial time-slice''; see Eq.~\eqref{eq:def_Laplace}. This is in contrast to a Fourier decomposition, where the entire past and future history is taken into account; that is, $-\infty < \tau < \infty$ instead.

In the following, we will work in the minimal gauge, where the height function reads~\cite{PanossoMacedo:2023qzp}
\begin{equation}
H(\sigma) = \dfrac{r_h}{\lambda}\bigg( -\dfrac{1}{\sigma} + \ln(\sigma) + \ln(1-\sigma) \bigg).
\end{equation}

The hyperboloidal coordinates $(\tau, \sigma, \theta, \varphi)$ allow for a straightforward conformal decomposition of the physical spacetime $g_{ab}$ into $g_{ab} = \Omega^{-2} \tilde g_{ab}$, with $\Omega = \sigma/\lambda$ and $\tilde g_{ab}$ a conformal metric with regular components in the entire region $\sigma\in[0,1]$. 

In the hyperboloidal framework, one aims to work with only regular quantities. In particular, the Teukolsky master function undergoes the conformal transformation~\cite{Zenginoglu:2011jz,PanossoMacedo:2019npm,Gasperin2025}
\begin{equation}
\label{eq:ConfMasterFunc}
\Psi^{(\pertorder)} = {\cal Z} \, \tilde \Psi^{(\pertorder)}, \quad {\cal Z} = \lambda^{-\NPs} \sigma^{1+2\NPs} (1- \sigma)^{-\NPs},
\end{equation}
The function ${\cal Z}$ ensures that the conformal master function $ \tilde \Psi^{(\pertorder)}$ is dimensionless and regular in the entire domain $\sigma\in[0,1]$, assuming non-vanishing values at the boundaries.

The conformal master function satisfies the equation
\begin{equation}
\label{eq:ConfTeukMast_Gen}
{\tilde \O}[  \tilde \Psi^{(\pertorder)}]  = \tilde {\rm S}^{(\pertorder)}, \quad \tilde {\rm S}^{(\pertorder)} = {\cal Z}^{-1} \, {\rm S}^{(\pertorder)},
\end{equation}
where the conformal operator ${\tilde \O}$ is defined by
\begin{equation}
    {\tilde \O} := {\cal Z}^{-1} {}_{\NPs} {\hat \O} {\cal Z}.
\end{equation}

Taking advantage of the angular separability of the operator $ {\tilde \O}$ in the Schwarzschild background, we decompose the conformal master function and the conformal source into spin-weighted spherical harmonic modes via
\begin{align}
\label{eq:Psi_angularDecomposition}
    \tilde \Psi^{(\pertorder)} &= \sum_{\l, \m} \tilde \Psi^{(\pertorder)}_{\l \m}(\tau, \sigma) {}_{\NPs} Y_{\l \m}(\theta, \varphi), \\
   \tilde S^{(\pertorder)} &= \sum_{\l, \m} \tilde S^{(\pertorder)}_{\l \m}(\tau, \sigma) {}_{\NPs} Y_{\l \m}(\theta, \varphi).
\end{align}
By construction, the spherical harmonics satisfy the angular equation, and we are left with two-dimensional partial differential equations for each mode coefficient:
\begin{equation}
\label{eq:ConfTeukMast_Gen_lm}
  {\tilde \O}_{\l \m}[ \tilde \Psi^{(\pertorder)}_{\l \m}] = \tilde {\rm S}^{(\pertorder)}_{\l \m}.
\end{equation}

In the hyperboloidal coordinates $(\tau, \sigma)$, the operator $ {\tilde \O}_{\l \m}$ assumes the explicit form 
\begin{equation}
    \label{eq:O_operator}
     {\tilde \O}_{\l \m} = -w(\sigma)\partial^2_{\tau} + \hat L_2 \partial_\tau + \hat L_1,
\end{equation}
with 
\begin{align}
    \label{eq:L1}
    \hat L_1 &= \sigma^2(1-\sigma)\partial^2_{\sigma} + \sigma\Big[ 2 (1+\NPs) - (3+\NPs) \sigma \Big]\partial_\sigma \nonumber \\
    &\quad- \Big[\l(\l+1) - \NPs(\NPs+1) + (1+\NPs) \sigma \Big], \\
    \label{eq:L2}
    \hat L_2 &= \dfrac{2r_h}{\lambda}\left(1- 2\sigma^2 \right)\partial_\sigma - \dfrac{2r_h}{\lambda} \Big[ \NPs - (2+\NPs) \sigma \Big], \\
    \label{eq:w}
    w(\sigma) &= \left(\dfrac{2 r_h}{\lambda}\right)^{\!2}(1+ \sigma).
\end{align}

In the time domain, Eq.~\eqref{eq:ConfTeukMast_Gen_lm} is solved after a prescription of regular data on the initial time slice $\tau = 0$:
\begin{align}
\label{eq:InitiaData}
\begin{split}
 \tilde \Psi^{(\pertorder)}_{\l \m}(0,\sigma) &= \tilde \Psi^{(\pertorder)}_o{}_{\l \m}(\sigma), \\ \dfrac{\partial\, \tilde \Psi^{(\pertorder)}_{\l \m}}{\partial \tau} (0,\sigma) &= \dot{\tilde\Psi}^{(\pertorder)}_o{}_{\l \m}(\sigma),
\end{split}
\end{align}
which suffices to uniquely determine the dynamics of the regular fields $ \tilde \Psi^{(\pertorder)}_{\l \m}$ for $\tau >0$ and $\sigma\in[0,1]$.

\subsection{Frequency-domain approach: the Laplace transform}

The Laplace framework allows us to incorporate the initial data \eqref{eq:InitiaData} into a frequency-domain formulation.
In what follows, for an arbitrary function $F(\tau)$, its Laplace transform $\mathcal{L}[F(\tau)](\lp)$ will be defined by
\begin{equation}
\label{eq:def_Laplace}
f(\lp) := \mathcal{L}[F(\tau)](\lp) := \int_0^\infty e^{-\lp \tau} F(\tau) d\tau.
\end{equation}
The Laplace parameter $\lp$ is related to the commonly used Fourier frequency $\omega$~\cite{Berti:2009kk} by $\lp=-i\lambda\omega$. We will informally refer to $\lp$ as a frequency.

We introduce the frequency-domain field $ \tilde \psi^{(i)}_{\l \m}$ via
\begin{equation}
    \tilde \psi^{(\pertorder)}_{\l \m}(\sigma; \lp) = {\cal L}\left[ \tilde \Psi^{(\pertorder)}_{\l \m}(\tau,\sigma)\right](\lp).
\end{equation}
By applying the Laplace transform to Eq.~\eqref{eq:ConfTeukMast_Gen_lm} and integrating by parts, we find $ \tilde \psi^{(i)}_{\l \m}$
satisfies the ordinary differential equation
\begin{equation}
    \label{eq:freq_domain_hyp_eq}
     \tilde {\cal D}_{\l\m}\left[ \tilde \psi^{(\pertorder)}_{\l \m}\right] = \tilde {\rm R}^{(\pertorder)}_{\l \m}, \quad \tilde {\cal D}_{\l\m} = -\lp^2 w(\sigma) + \lp \hat L_2 + \hat L_1.
\end{equation}
The conformal source term in the frequency domain is composed of two pieces, 
\begin{equation}
\label{eq:FrequencyDomainSource}
     \tilde {\rm R}^{(\pertorder)}_{\l \m} = \tilde {\rm I}^{(\pertorder)}_{\l \m} + \tilde {\mathfrak s}^{(\pertorder)}_{\l \m}.
\end{equation}
The first term,
\begin{align}
\label{eq:Source_ID}
 \tilde {\rm I}^{(\pertorder)}_{\l \m}(\sigma;\lp) & = -w(\sigma) \bigg( \tilde \Psi^{(\pertorder)}_o{}_{\l \m}(\sigma) \, \lp + \dot{\tilde\Psi}^{(\pertorder)}_o{}_{\l \m}(\sigma) \bigg) \notag \\
&\quad + \hat L_2 \left[ \tilde \Psi^{(\pertorder)}_o{}_{\l \m}(\sigma) \right],
\end{align}
results from incorporating the initial data \eqref{eq:InitiaData} into the frequency domain.
The second term follows directly from the Laplace transform of the time-domain source,
\begin{equation}
    \label{eq:LaplaceTransfSource}
     \tilde {\mathfrak s}^{(\pertorder)}_{\l \m}(\sigma; \lp) := {\cal L}\left[ \tilde {\rm S}^{(\pertorder)}_{\l \m}(\tau,\sigma)\right](\lp).
\end{equation}

Once a solution to the inhomogeneous equation is available for ${\rm Re}(\lp) > 0$, the time-domain evolution
follows from the inverse Laplace transform
\begin{equation}
    \label{eq:InverseLaplace}
     \tilde \Psi^{(\pertorder)}_{\l \m}(\tau, \sigma) = \dfrac{1}{2 \pi i} \int_{\Gamma} \tilde \psi^{(\pertorder)}_{\l \m}(\sigma; \lp) \, e^{\lp \tau} d\lp,
\end{equation}
where the integration path $\Gamma$ is parametrised by $\lp = \lp^{\rm R} + i \lp^{\rm I}$, with an arbitrary constant real part $\lp^{\rm R}>0$ and $\lp^{\rm I}\in(-\infty, \infty)$.

To pin down the specific elements of the ringdown dynamics, we analytically extend $\tilde \psi^{(\pertorder)}_{\l \m}$ to the region ${\rm Re}(\lp) < 0$. QNMs and QQNMs are associated with poles of the function $\tilde \psi^{(\pertorder)}_{\l \m}$: 
(i) QNMs correspond to poles of the Green's functions associated with the Teukolsky wave operator $ \tilde{\cal O}$~\cite{Leaver:1986gd,Kokkotas:1999bd}; (ii) QQNMs correspond to poles of the Laplace transformed source $ \tilde{\sf R}^{(2)}_{\l \m}(\sigma; \lp)$ \cite{Lagos:2022otp}.

In the next sections, we discuss how to control these contributions at first and second order in perturbation theory.

\section{First-order quasinormal modes}\label{sec:QNMs}

\begin{figure*}[tb] \centering
\includegraphics[width=\textwidth]{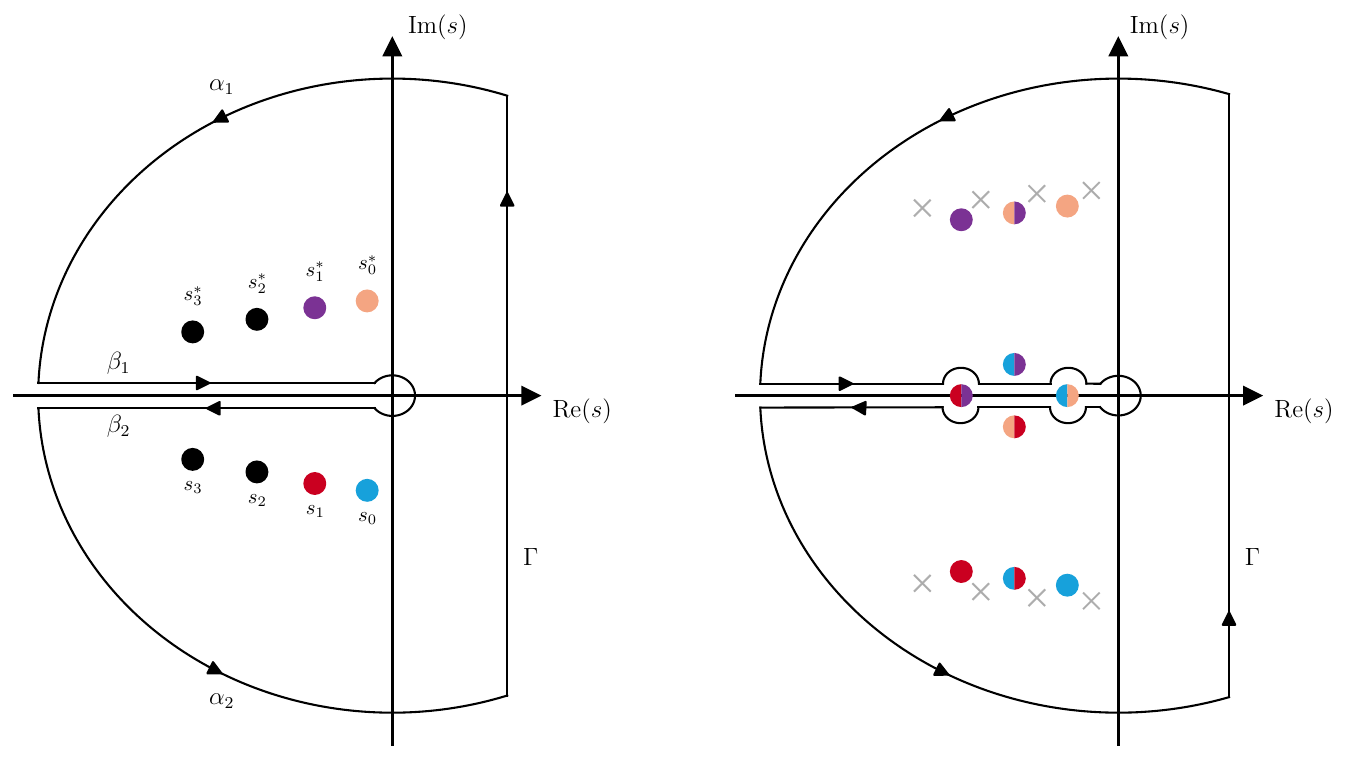}
  \caption{{\em Left Panel:} The first-order solution in the time domain is obtained via the inverse Laplace transform~\eqref{eq:InverseLaplace}, consisting of an integration over the frequency parameter $\lp$ along the path $\Gamma$. In the complex $\lp$ plane, the factor $\tilde \psi^{(1)}_{\l \m}(\sigma; \lp)$ in the integrand has poles at each QNM frequency $\lp_{\l\m\n}$ and a branch cut along the negative real axis, inherited from the Green's function for the wave operator ${\tilde{\cal O}}$.
  {\em Right Panel:} The second-order Laplace-transformed solution, $\tilde{\psi}^{(2)}_{\l \m}(\sigma; \lp)$, shows poles from two sources. Grey crosses mark the Green's function poles of the frequency-domain operator ($\tilde{\mathcal{D}}_{\l \m}$ in Eq.~\eqref{eq:freq_domain_hyp_eq}), corresponding to the linear QNMs, while circles denote the quadratic QNMs.
  Colour coding distinguishes the fundamental mode (dark blue) and its mirror mode (orange) as well as the first overtone (red) and its mirror (purple). Quadratic QNMs arise from the product of linear QNMs -- for example, $(2, 2, 0) \times (2, 2, 0)$, is a solid blue circle, while $(2, 2, 0) \times (2, 2, 1)$, is a half-filled blue and orange.
  Here, QQNMs arise from linear QNMs with $\l_{1} = \m_{1} = 2$ and $\l_{2} = \m_{2} = 2$. Consequently, the poles of the QQNMs lie in the $(L, M) = (4, 4)$ plane and the associated linear QNMs likewise are characterized by $(L, M) = (4, 4)$.
  Since some QQNMs lie on the branch cut, we have deformed $\beta_1$ and $\beta_2$ to indicate that the integral over $\beta_1\cup\beta_2$ is no longer simply a branch cut contribution.
  }
    \label{fig:QNM_LaplaceContour}
\end{figure*}

At first order in perturbation theory, the wave equation~\eqref{eq:ConfTeukMast_Gen}, has a vanishing right-hand side, and therefore, the source term in the Laplace transformed equation~\eqref{eq:freq_domain_hyp_eq}, has contributions only from the field's initial data. The time evolution is then obtained via the inverse Laplace transform through Eq.~\eqref{eq:InverseLaplace}, where the 
Green's function associated with the operator $ \tilde{\cal D}_{\l\m}$ determines the solution $\tilde \psi^{(\pertorder)}_{\l \m}(\sigma; \lp)$. 
Using the residue theorem as illustrated on the left panel of Fig.~\ref{fig:QNM_LaplaceContour}, we can express the integral along the path $\Gamma$ in Eq.~\eqref{eq:InverseLaplace} as a sum of contributions from the Green's function in the region ${\rm Re}(\lp) < 0$~\cite{Leaver:1986gd,Berti:2009kk}: (i) residues at the discrete poles $\lp^{(1)}_{\l\m\n}$ in the half-plane ${\rm Re}(\lp) < 0$, which give rise to QNM dynamics; (ii) a branch cut contribution along the negative real axis (in the $\lp$-plane); and (iii) a contribution from the high-frequency arc that closes the contour. 
More explicitly, we express the integral along $\Gamma$ as 
\begin{equation} \label{QNM_Contour_Integral}
    \int_{\Gamma} f d\lp = 2\pi i\sum_{\n}{\rm Res}(f,\lp_{\l\m\n}) - \int_{\rm b.c.}f d\lp - \int_{\rm arc}f d\lp, 
\end{equation}
where $f=\frac{1}{2\pi i}\,\tilde \psi^{(1)}_{\l \m}e^{\lp\tau}$, ``b.c.'' stands for the path $\beta_1\cup\beta_2$ around the branch cut, and ``arc'' denotes the path $\alpha_1\cup\alpha_2$ in Fig.~\ref{fig:QNM_LaplaceContour} along the high-frequency arc.
If we neglect the branch cut and high-frequency arc contributions, the time-domain solution then reads
\begin{align}
\label{eq:DynSol_Hom+Part_PsiLinear}
    \tilde \Psi^{(1)}_{\l\m}(\tau, \sigma) &\approx \sum_{\n}  A^{(1)}_{\l\m\n} \, \tilde\psi_{\l\m\n}^{(1)}(\sigma) \, e^{\lp^{(1)}_{\l\m\n} \tau}
\end{align}
for some constant amplitudes $ A^{(1)}_{\l\m\n}$. 
As we will see below, the function $ \tilde\psi_{\l\m\n}^{(1)}(\sigma)$
satisfies the homogeneous version of Eq.~\eqref{eq:freq_domain_hyp_eq}.

In the next section, we discuss how to make Eq.~\eqref{eq:DynSol_Hom+Part_PsiLinear} {\em exact}; i.e., how to specialize the dynamics at first order to uniquely single out the contributions of individual QNM frequencies.

\subsection{First-order QNMs: initial-data-driven dynamics with pure quasinormal modes}
\label{sec:first_order_qnms}
A key advantage of the hyperboloidal framework is that the QNM eigenfunctions $ \tilde\psi_{\l\m\n}^{(1)}(\sigma)$ are regular in the entire BH exterior region $\sigma\in[0,1]$. We exploit this property to fine-tune first-order initial data that excites QNMs individually, therefore yielding a single QNM dynamic. In other words, our goal is to impose a first-order dynamics exactly of the form
\begin{align} \label{QNM_TD_ansatz}
     \tilde{\Psi}_{\l\m}^{(1)}(\tau,\sigma) = A_{\l\m\n}^{(1)} \tilde{\psi}_{\l\m\n}^{(1)}(\sigma) e^{\lp_{\l\m\n}^{(1)} \tau},
\end{align}
where $\lp_{\l\m\n}^{(1)}$ is a (for now arbitrary) frequency, $ A_{\l\m\n}^{(1)}$ is an arbitrary complex constant, and $ \tilde{\psi}_{\l\m\n}^{(1)}$ is, without loss of generality, normalised to unity at future null infinity:
\begin{equation} \label{eq:QNM_func_norm}
     \tilde\psi_{\l \m \n}^{(1)}(0) = 1.
\end{equation}
With the above ansatz, the initial data~\eqref{eq:InitiaData} are given by
\begin{align} \label{QNM_ID1}
    \tilde \Psi^{(1)}_o{}_{\l \m}(\sigma) &=  A_{\l\m\n}^{(1)} \tilde{\psi}_{\l\m\n}^{(1)}(\sigma), \\
    \dot{\tilde{\Psi}}^{(1)}_o{}_{\l \m}(\sigma) &= A_{\l\m\n}^{(1)} \lp_{\l\m\n}^{(1)} \tilde{\psi}_{\l\m\n}^{(1)}(\sigma). \label{QNM_ID2}
\end{align}

In the Laplace domain, the time-domain ansatz~\eqref{QNM_TD_ansatz} transforms into
\begin{equation} \label{QNM_FD_ansatz}
    \tilde{\psi}_{\l\m}^{(1)}(\sigma;\lp) = A_{\l\m\n}^{(1)} \frac{ \tilde{\psi}_{\l\m\n}^{(1)}(\sigma)}{\lp - \lp_{\l\m\n}^{(1)}}.
\end{equation}
Plugging this ansatz into Eq.~\eqref{eq:freq_domain_hyp_eq}, we obtain
\begin{equation} \label{QNM_equation1}
    A_{\l\m\n}^{(1)} \tilde {\cal D}_{\l\m}\left[ \frac{ \tilde{\psi}_{\l\m\n}^{(1)}(\sigma)}{\lp - \lp_{\l\m\n}^{(1)}} \right] = \tilde {\rm I}^{(1)}_{\l \m},
\end{equation}
with the right-hand side containing information only from the initial data~\eqref{QNM_ID1} via $ \tilde {\rm I}^{(1)}_{\l \m}$, cf. Eq.~\eqref{eq:Source_ID}, because ${\mathfrak s}^{(1)}_{\l \m} = 0$.
Without loss of generality, it is convenient to re-express the frequency-domain operator $ \tilde{\cal D}_{\l\m}$, defined in Eq.~\eqref{eq:freq_domain_hyp_eq}, as
\begin{equation}
\label{eq:FreqOperator_QNMDecomp}
    \tilde{\cal D}_{\l\m} = \tilde {\cal D}_{\l\m\n} + \left(\lp - \lp^{(1)}_{\l\m\n}\right) \tilde{\partial {\cal D}}_{\l\m},
\end{equation}
with 
\begin{eqnarray}
    \tilde {\cal D}_{\l\m\n} &=& \left. \tilde {\cal D}_{\l\m} \right|_{\lp = \lp^{(1)}_{\l\m\n}} \notag \\ 
    &=& -\left( \lp^{(1)}_{\l\m\n} \right)^2 w(\sigma) + \hat L_1 + \lp^{(1)}_{\l\m\n} \, \hat L_2, \\
    \tilde{\partial {\cal D}}_{\l\m} &=& \hat L_2 - \left(\lp + \lp^{(1)}_{\l\m\n}\right) w(\sigma). \label{dD_formula}
\end{eqnarray}
Equation~\eqref{QNM_equation1} then reads
\begin{equation} \label{QNM_equation2}
     A_{\l\m\n}^{(1)} \frac{ \tilde {\cal D}_{\l\m\n}\left[ \tilde{\psi}_{\l\m\n}^{(1)}(\sigma) \right]}{\lp - \lp_{\l\m\n}^{(1)}} +  \tilde{\partial {\cal D}}_{\l\m} \left[ \tilde{\psi}_{\l\m\n}^{(1)}(\sigma) \right] =  \tilde {\rm I}^{(1)}_{\l \m}.
\end{equation}
Assuming that the solution is meromorphic, its Laurent series around $\lp = \lp_{\l\m\n}^{(1)}$ must satisfy the governing equation at every order. In particular, this implies
\begin{equation}
    \label{eq:hom_eq_QNM}
      \tilde{\cal D}_{\l\m\n}[  \tilde \psi_{\l \m\n}^{(1)}(\sigma)]  = 0.
\end{equation}
Assuming Eq.~\eqref{eq:hom_eq_QNM} holds, and using Eqs.~\eqref{eq:Source_ID}, \eqref{QNM_ID1}, \eqref{QNM_ID2} and \eqref{dD_formula}, it is straightforward to check that \eqref{QNM_equation2} is then satisfied in the limit $\lp \to \lp_{\l\m\n}^{(1)}$. 

We are therefore left with finding all the regular solutions $ \tilde \psi_{\l \m\n}^{(1)}(\sigma)$, and their associated frequencies $\lp_{\l\m\n}^{(1)}$ to Eq.~\eqref{eq:hom_eq_QNM}.
Due to regularity conditions at the boundaries, only a countable number of such solutions exist, which can be conveniently labelled by the index $\n$ and referred to as the (first-order) QNMs~\cite{Ansorg:2016ztf,Gajic:2019qdd,Gajic:2019oem,galkowski2021outgoing,PanossoMacedo:2023qzp,PanossoMacedo:2024nkw}. Note that if $\lp_{\l\m\n}$ is \emph{not} a QNM frequency, then Eq.~\eqref{eq:hom_eq_QNM} has no solution that is regular at the boundaries $\sigma=0$ and $1$.

By introducing the auxiliary variable $ \tilde\Xi_{\l \m\n}^{(1)}(\sigma) = \lp^{(1)}_{\l \m\n}\, \tilde\psi_{\l \m\n}^{(1)}(\sigma)$, we recast Eq.~\eqref{eq:hom_eq_QNM} as an eigenvalue problem~\cite{PanossoMacedo:2023qzp},
\begin{equation} \label{QNMEvalueProblem}
     \boldsymbol{L}_{\l \m \n} \left[ \vec{u}_{\l \m\n}^{(1)} \right] = \lp^{(1)}_{\l \m\n} \vec{u}_{\l \m\n}^{(1)}
\end{equation}
with 
\begin{equation}
    \label{eq:QNM_EigenvalueSystem}
     \boldsymbol{L}_{\l \m\n} = 
    \left( 
        \begin{array}{cc}
            0 & 1 \\
            w^{-1}\hat{L}_1 & w^{-1}\hat{L}_2
        \end{array}    
    \right), \quad 
     \vec{u}_{\l \m\n}^{(1)} = 
    \left( 
        \begin{array}{c}
             \tilde\psi_{\l \m\n}^{(1)}  \\
             \tilde\Xi_{\l \m\n}^{(1)}
        \end{array}    
    \right).
\end{equation}
The functions $w(\sigma)$, $\hat{L}_1$ and $\hat{L}_2$ are defined in Eqs.~\eqref{eq:L1}--\eqref{eq:w}.

In Sec.~\ref{sec:codes}, we discuss the numerical techniques to discretize the operator $ \boldsymbol{L}_{\l \m \n}$, from which the first-order QNMs are computed as the eigensolutions of the resulting discrete matrix~\cite{Jaramillo:2020tuu}. 

In later sections, we will be interested in the individual mode contributions of the QNM to the strain. For a given $(\l,\m,\n)$, we define the first-order QNM amplitude to be 
\begin{equation} \label{eq:FirstOrder_ampltidue_at_infinity}
    \tilde {\cal A}^{(1)}_{\l\m\n}(\sigma) = A^{(1)}_{\l\m\n}\,  \tilde\psi_{\l\m\n}^{(1)}(\sigma).
\end{equation}

\subsection{Regular and mirror modes}
\label{section:Regular_and_mirror_modes}

As mentioned in the previous section, the index $\n$ labels the countable set of QNM frequencies of the independent operator $ \tilde {
\cal D}_{\l \m}$ at a fixed $(\l, \m)$ mode. What is the most natural choice for labelling these frequencies?

\begin{figure}
\includegraphics[width=0.95\columnwidth]{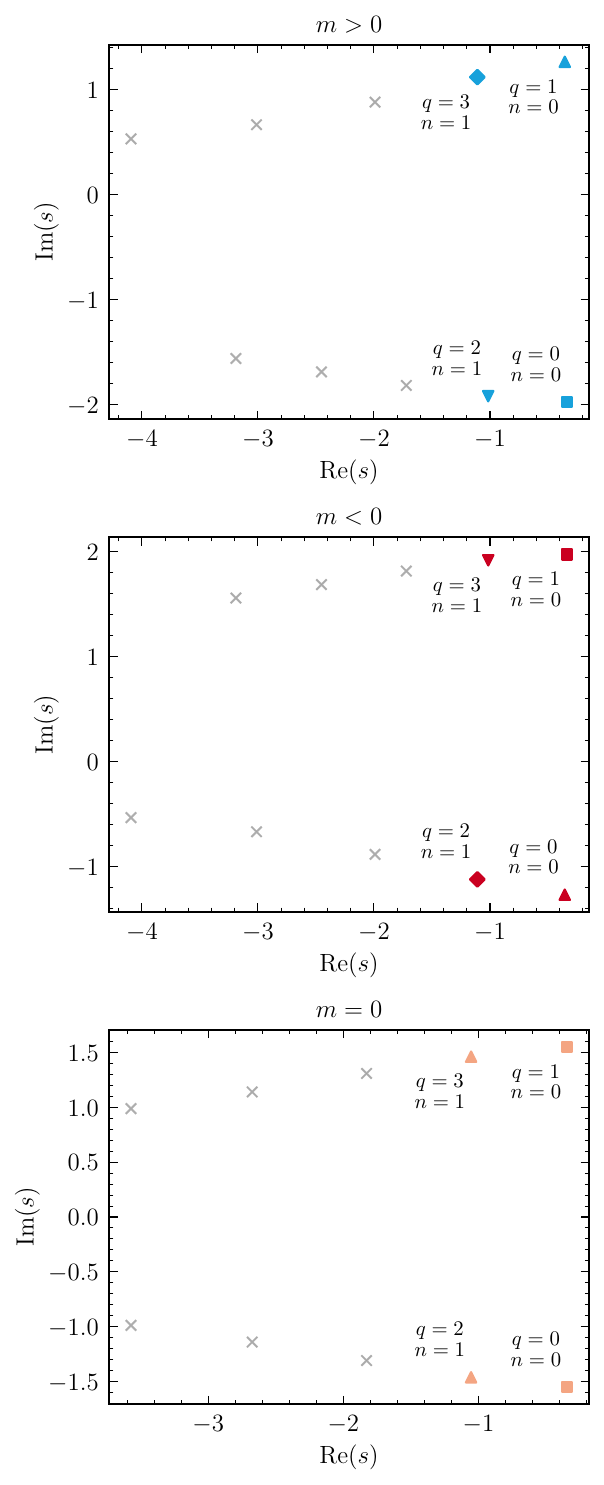}
    \caption{The first few QNM frequencies of a Kerr BH in the complex $\lp$-plane for spin $a/M = 0.6$ with $l=2$, $\m=\pm 2$ and $\m=0$. The frequencies for $\m<0$ and $\m>0$ are related by complex conjugation, indicated in the figure by matching shapes. The mode numbers for $\n$ and $\q$ have been labelled next to relevant shapes to clarify our notation.}
    \label{fig:QNMFrequencies}
\end{figure}

In Fig.~\ref{fig:QNMFrequencies}, we show a plot in the complex $\lp$-plane of the first few QNM frequencies in Kerr with the smallest decay rate (given by $\left| \left| {\rm Re}(\lp^{(1)}_{\l\m\n}) \right| \right|$), for $\l=2$ and $\m=\pm 2$, and $\m=0$.

Focusing first on the top plot, corresponding to $\m>0$, it is natural to label the slowest decaying mode with $\n=0$, the second slowest by $\n=1$, and so on. As a result, increasing overtone number $\n$ directly corresponds to an increase in the decay rate of the associated mode.
It is then typical to proceed in the same manner for the $\m<0$ frequencies (middle plot).
As is shown in the figure, an important point to realise is that a given $\n$ actually refers to \emph{two} complex frequencies, which have the same real part (and thus the same decay rate), but different imaginary parts.

An important symmetry between the set of QNM frequencies in Kerr corresponding to $\m>0$ and $\m<0$ (at a fixed $\l$) is that both sets are related by complex conjugation. 
More precisely, we have drawn the first four frequencies with the smallest decay rate as square, triangle, diamond and inverted triangle shapes. The frequencies between $\m>0$ and $\m<0$ with the matching shapes are related by complex conjugation. As a result of the above choice of labelling the overtones $\n$, this symmetry can be simply expressed by
\begin{equation} \label{QNM_frequency_traditional_relation}
    \lp^{(1)}_{\l-\m \n}= \bracket{\lp^{(1)}_{\l \m \n}}^\star.
\end{equation}
This motivates the definition of \textit{regular} and \textit{mirror} QNMs, depending, respectively, on whether ${\rm Im}(\lp^{(1)}_{\l\m \n}) < 0$ or ${\rm Im}(\lp^{(1)}_{\l\m \n}) > 0$.
Some care needs to be taken when working with an expression like Eq.~\eqref{QNM_frequency_traditional_relation}, since $\n$ refers to two frequencies. This is particularly evident when setting $\m=0$ in Eq.~\eqref{QNM_frequency_traditional_relation}, which seemingly implies that the QNMs there are real. One way to clarify this ambiguity is to denote by $\lp^{(1)}_{\l\m \n,+}$ and $\lp^{(1)}_{\l \m \n,-}$ the QNMs with ${\rm Im}(\lp^{(1)}_{\l \m \n}) > 0$ and ${\rm Im}(\lp^{(1)}_{\l \m \n}) < 0$ respectively. In particular, instead of the relation~\eqref{QNM_frequency_traditional_relation}, one would have the more accurate statement
\begin{equation} \label{QNM_frequency_traditional_relation-Clarified}
    \lp^{(1)}_{\l-\m \n,+}= \bracket{\lp^{(1)}_{\l \m \n,-}}^\star
\end{equation}
The case $\m=0$ now simply reveals that the two frequencies $\lp^{(1)}_{\l 0 \n,\pm}$ are related by complex conjugation.

As a result of this notational ambiguity, certain summations over $\n$ can be rather confusing. Consider, for example, the sum over the full QNM spectrum for spin $\NPs = \pm 2$:
\begin{align}
     \sum_{\l=0}^\infty \sum_{\m=-\l}^\l \sum_{\n=0}^\infty A^{(1)}_{\l\m\n} \, \tilde \psi_{\l\m\n}^{(1)}(\sigma) \, e^{\lp^{(1)}_{\l\m\n} \tau} {}_{\NPs} Y_{\l\m}(\theta, \varphi).
\end{align}
The sum over $\n$ might be misleading and cause confusion since the contribution from a single index $\n$ is implicitly understood as the sum over all the eigenvalues with the same prescribed decay rate, given by ${\rm Re}(\lp)$. In Kerr, for a given $(\l,\m)$ mode, this would correspond to \emph{two} frequencies.
The above sum, therefore is understood to mean more explicitly
\begin{align} \label{eq:ringdown_waveform_Kerr}
    \sum_{\l=0}^\infty \sum_{\m=-\l}^{\l} \sum_{\n=0}^\infty & \Bigl[ A^{(1)}_{\l\m\n,+} \, \tilde \psi_{\l\m\n,+}^{(1)}(\sigma) \, e^{\lp^{(1)}_{\l\m\n,+} \tau} \nonumber \\ 
    &+ A^{(1)}_{\l\m\n,-} \, \tilde \psi_{\l\m\n,-}^{(1)}(\sigma) \, e^{\lp^{(1)}_{\l\m\n,-} \tau} \Bigr] {}_{\NPs} Y_{\l\m}(\theta, \varphi) ,
\end{align}
The amplitudes $ A^{(1)}_{\l\m\n, \pm}$ are, in general, independent and associated with the excitation of the regular and mirror modes $\lp_{\l \m \n, \pm}^{(1)}$. In the above, the label $\n$ can now be interpreted as referring to a unique QNM frequency.
In Schwarzschild, the degeneracy of the $\m$-mode occurs at all values of $\m$.

In the interest of clarity, we will, from now on, refrain from using the index $\n$ and the associated relation~\eqref{QNM_frequency_traditional_relation} (or \eqref{QNM_frequency_traditional_relation-Clarified}).
Instead, we will use a new overtone label $\q$, which will {\em always} uniquely be associated to one single eigenvalue of $ \tilde {\cal D}_{\l \m}$. In order to distinguish this new notation, we resort to a semicolon between each mode indices $(\l$, $\m)$ and $\q$ to distinguish it from the $(\l \m \n)$ notation. For instance $\lp^{(1)}_{220}$ is understood as $(\l, \m, \n) = (2,2,0)$, whereas $\lp^{(1)}_{2;2;0}$ has labels $(\l, \m, \q) = (2,2,0)$.

In order for $\q$ to always correspond to a unique QNM frequency, we have chosen the following labelling rule: (i) for $\m>0$, the $\q$-labelling follows the convention: $\lp^{(1)}_{\l;\m;2 \n} = \lp^{(1)}_{\l \m \n,-}$, $\lp^{(1)}_{\l;\m;2 \n+1} = \lp^{(1)}_{\l \m \n,+}$; (ii) for $\m<0$, we instead label the QNM frequencies according to the defining relation
\begin{equation} \label{QNM_frequency_new_relation_tmp}
    \lp^{(1)}_{\l;\m;2\n} = \bracket{\lp^{(1)}_{\l;-\m; 2\n+1}}^\star \,.
\end{equation}
Note that this choice of labelling the $\m<0$ mode can be consistently extended to the case $\m=0$; see Fig.~\ref{fig:QNMFrequencies}.

In this $\q$ notation a sum such as Eq.~\eqref{eq:ringdown_waveform_Kerr} is then unambiguously expressed as
\begin{equation}
\label{eq:QNM_sum_q}
    \sum_{\l=0}^\infty \sum_{\m=-\l}^\l \sum_{\q=0}^\infty A^{(1)}_{\l;\m;\q} \, \tilde \psi_{\l;\m;\q}^{(1)}(\sigma) \, e^{ \lp^{(1)}_{\l;\m;\q} \tau} {}_{\NPs} Y_{\l\m}(\theta, \varphi).
\end{equation}

In the following, the above relation~\eqref{QNM_frequency_new_relation_tmp} can be written more succinctly as
\begin{equation} \label{QNM_frequency_new_relation}
    \lp_{\l;\m;\q} = \bracket{\lp^{(1)}_{\l;-\m; \Q}}^\star,
\end{equation}
where $\Q \equiv \Q(\q) = \q + 1$ if $\q$ is even, and $\Q \equiv \Q(\q) = \q - 1$ if $\q$ is odd.
In particular, note the useful identity
\begin{equation} \label{Q_relation}
    \Q(\Q(\q)) = \q.
\end{equation}
Note that we also get an additional relation in Schwarzschild, $\lp^{(1)}_{\l;\m;\q} = \bracket{\lp^{(1)}_{\l;\m;\Q}}^\star$; that is $\lp^{(1)}_{\l \m \n,\pm}$ are related by complex conjugation.

As a result of our choice of normalisation~\eqref{eq:QNM_func_norm}, the associated eigenfunctions also follow a similar relation as in Eq.~\eqref{QNM_frequency_new_relation},
\begin{equation}
     \tilde{\psi}^{(1)}_{\l;\m;\q} = \bracket{ \tilde{\psi}^{(1)}_{\l; -\m;\Q}}^\star \, .
\end{equation}

The sum over $\q$ in Eq.~\eqref{eq:QNM_sum_q} can be re-expressed as a sum over $\n$ as
\begin{align}
    \sum_{\l=0}^\infty &\sum_{\m=-\l}^\l \sum_{\n=0}^\infty \Bigl[ A^{(1)}_{\l;\m;2\n} \, \tilde \psi_{\l;\m;2\n}^{(1)}(\sigma) \, e^{\lp^{(1)}_{\l;\m;2\n} \tau} \nonumber \\
    & + A^{(1)}_{\l;\m;2\n+1} \, \tilde \psi_{\l;\m;2\n+1}^{(1)}(\sigma) \, e^{\lp^{(1)}_{\l;\m;2\n+1} \tau} \Bigr] {}_{\NPs} Y_{\l\m}(\theta, \varphi).
\end{align}

\section{Quasinormal modes at second order}\label{sec:QQNMs}

\subsection{Quadratic QNMs: source-driven dynamics with pure quadratic quasinormal modes}

The Teukolsky equation at second order in the Laplace domain is given by Eq.~\eqref{eq:freq_domain_hyp_eq}. As we wish to later study the impact of second-order effects on the gravitational strain at null infinity, we will consider the case $s=-2$. We recall the source term depends on two contributions: the initial data at that order and the Laplace transform of the time-domain second-order source, which is driven by a quadratic combination of first-order dynamics. 
To ensure that the second-order dynamics is driven purely by the source term, we require the second-order initial data to vanish; that is,
\begin{equation}
    \label{eq:2ndOrder_InitiaData}
 \tilde \Psi^{(2)}_o{}_{\l \m}(\sigma) = 0, \quad  
 \dot{\tilde\Psi}^{(2)}_o{}_{\l \m}(\sigma) = 0,
\end{equation}
for all $(\l,\m)$.
This choice implies that $\tilde {\rm I}^{(2)}_{\l \m} = 0$; see Eq.~\eqref{eq:Source_ID}.

In Eq.~\eqref{eq:DynSol_Hom+Part_PsiLinear}, we have taken our first-order solution to be a sum of QNM solutions. We now consider any two of those modes, $I=(\l;\m;\q)$ and $I'=(\l';\m';\q')$. The construction of the second-order source term will involve the quadratic combinations $I \times I$, $I' \times I'$, and $I \times I'$, subject to selection rules when these combinations are projected back into a specific second-order mode, which we label with mode numbers $(\L, \M)$. 

Utilizing the compact notation $(I \times I')_{\L\M}$ to encapsulate these quadratic combinations, subject to the underlying $(\L, \M)$ selection rules, we can write the second-order source~\eqref{eq:masterreducedTeuk} with an explicit time dependence of the following form:
\begin{align}
    \label{eq:SecOrdSource_TimeDep}
     \tilde{\rm S}^{(2)}_{\L \M}(\tau,\sigma) &= \sum_{ (I \times I')_{\L\M} }  \tilde {\cal S}^{(2)}_{(I \times I')_{\L\M}}(\sigma) \, e^{\tau \lp^{(2)}_{(I \times I')_{\L\M}}}, \\
    \lp^{(2)}_{(I \times I')_{\L\M}}  &= \lp^{(1)}_{I} + \lp^{(1)}_{I'}. \label{eq:SecOrdFreq}
\end{align}
Applying the Laplace transform~\eqref{eq:LaplaceTransfSource} to Eq.~\eqref{eq:SecOrdSource_TimeDep} yields
\begin{equation}
    \label{eq:SecOrdSource_LaplTras}
     \tilde{\mathfrak s}^{(2)}_{\L\M}(\sigma;\lp) = \sum_{ (I \times I')_{\L\M} }\dfrac{ \tilde {\cal S}^{(2)}_{(I \times I')_{\L\M}}(\sigma)}{\lp - \lp^{(2)}_{(I \times I')_{\L\M}}},
\end{equation}
where one identifies the QQNMs as poles in the source term, $\lp = \lp^{(2)}_{(I \times I')_{\L\M}}$, in contrast to the first-order QNMs that arise as poles of the Green's function associated with the operator in Eq.~\eqref{eq:freq_domain_hyp_eq}.

\subsection{Spectral representation of the second-order solution}\label{sec:second order dynamics}

As we did at first order, we can use the residue theorem to write the inverse Laplace transform~\eqref{eq:InverseLaplace} as a sum of contributions from the region ${\rm Re}(\lp) < 0$.

In particular, the integral along the path $\Gamma$ in Eq.~\eqref{eq:InverseLaplace} can also be written as a sum of contributions from (i) residues at the discrete poles in the half-plane ${\rm Re}(\lp) < 0$. This will include the same poles at first order that gives rise to QNM dynamics, the new QQNM poles that appear only at second order, and finally, the special poles that lie on the branch cut ${\rm Im}(\lp) = 0$, which contribute to gravitational memory~\cite{Khera:2024yrk}; (ii) a branch cut contribution along the negative real axis (in the $\lp$-plane); and (iii) a contribution from the high-frequency arc that closes the contour. 

In the following, we will only focus on the QNM poles, as well as the QQNM poles, excluding those lying on the branch cut.
With this in mind, the dynamical evolution can be written as
\begin{align}
    \label{eq:DynSol_Hom+Part}
     \tilde \Psi^{(2)}_{\L\M}(\tau, \sigma) &=  \tilde \chi^{(2)}_{\L\M}(\tau, \sigma) +  \tilde \Upsilon^{(2)}_{\L\M}(\tau, \sigma), \\
    \label{eq:HomSol}
     \tilde \chi^{(2)}_{\L\M}(\tau, \sigma) &\approx \sum_{\q}  A^{(2)}_{\L;\M; \q} \,  \tilde\psi_{\L;\M ;\q}^{(1)}(\sigma) \, e^{\tau \lp^{(1)}_{\L;\M;\q}}, \\
     \tilde\Upsilon^{(2)}_{\L\M}(\tau, \sigma) &= \sum_{ (I \times I' )_{\L\M} }  \tilde {\cal A}^{(2)}_{(I \times I')_{\L\M}}(\sigma) \, e^{\tau \lp^{(2)}_{(I \times I' )_{\L\M}}}. \label{eqn:Upsilon_TD}
\end{align}
The function $ \tilde \chi^{(2)}_{\L\M}(\tau, \sigma)$ shares the same form as the first-order solution~\eqref{eq:DynSol_Hom+Part_PsiLinear}; it is a solution to the homogeneous equation
\begin{equation}
     {\tilde \O}_{\L\M}[  \tilde \chi^{(2)}_{\L\M}]  = 0,
\end{equation}
with $ \tilde\psi^{(1)}_{\L;\M;\q}(\sigma)$ and $\lp^{(1)}_{\L;\M;\q}$ associated with the first-order QNMs resulting from the QNM eigenvalue problem~\eqref{eq:QNM_EigenvalueSystem}. Since the functions $ \tilde\psi_{\L;\M;\q}^{(1)}(\sigma)$ are solutions to the frequency domain homogeneous equations~\eqref{eq:hom_eq_QNM}, they do not incorporate any property from the frequency domain source term. Instead, the effects of the initial data and source dynamics are captured by the coefficients $ A^{(2)}_{\L;\M;\q}$,
which measure the excitation of linear QNMs.
In the following, these amplitudes will be referred to as second-order QNM amplitudes.

$\tilde\Upsilon^{(2)}_{\L\M}(\tau, \sigma)$ represents a particular solution to the inhomogeneous equation~\eqref{eq:ConfTeukMast_Gen_lm}, with $ \tilde {\cal A}^{(2)}_{(I \times I' )_{\L\M}}(\sigma)$ the corresponding particular solution to the frequency-domain equation~\eqref{eq:freq_domain_hyp_eq}.
This contribution to the second-order solution
arises from the poles in the source~\eqref{eq:SecOrdSource_LaplTras} and can therefore be seen as a purely second-order effect. The associated amplitude will be referred to as the QQNM amplitude.

In the next sections, we lay out the methods to calculate the second-order QNM amplitudes $A^{(2)}_{\L;\M; \q}$ and the QQNM amplitude $ \tilde {\cal A}^{(2)}_{(I \times I' )_{\L\M}}(\sigma)$.

\subsubsection{Second-order QNM amplitudes} 
To calculate the second-order QNM amplitudes $A^{(2)}_{\L;\M; \q}$, we re-express some of the methods introduced by Ref.~\cite{Ansorg:2016ztf} in terms of the differential operator $ \tilde {\cal D}_{\L\M}$ (see also \cite{Ammon:2016fru}). We first concentrate on the behaviour of $\tilde \psi^{(2)}_{\L\M}(\sigma; \lp)$ around a fixed pole $\lp^{(1)}_{\L;\M;\q}$; i.e., around a given first-order QNM that is excited by the second-order source $ \tilde{\mathfrak s}^{(2)}_{\L\M}(\sigma;\lp)$. For that purpose, we assume the ansatz
\begin{equation}
    \label{eq:Ansatz_AmpQNM_1stOrder}
    \tilde \psi^{(2)}_{\L\M}(\sigma; \lp) = \dfrac{ A^{(2)}_{\L;\M;\q} \,  \tilde\psi_{\L;\M;\q}^{(1)}(\sigma)}{\lp - \lp^{(1)}_{\L;\M;\q}} + \tilde W^{(2)}_{\L\M}(\sigma; \lp).
\end{equation}
Akin to Eqs.~\eqref{QNM_TD_ansatz} and \eqref{QNM_FD_ansatz}, the first term will contribute to the specific QNM dynamics with frequency $\lp^{(1)}_{\L;\M;\q}$, whereas the auxiliary function $\tilde W^{(2)}_{\L\M}(\sigma; \lp)$, which is regular at the pole $\lp^{(1)}_{\L;\M;\q}$, accounts for the rest of the solution $\tilde \psi^{(2)}_{\L\M}(\sigma; \lp)$ in the neighborhood of the pole. Indeed, via the inverse Laplace transform~\eqref{eq:InverseLaplace} and use of the residue theorem as in Fig.~\ref{fig:QNM_LaplaceContour}, it is straightforward to see that the first term in Eq.~\eqref{eq:Ansatz_AmpQNM_1stOrder} recovers the discrete contribution in Eq.~\eqref{eq:HomSol}.

The ansatz~\eqref{eq:Ansatz_AmpQNM_1stOrder} introduces two unknowns into the problem: the excitation amplitude $ A^{(2)}_{\L;\M; \q}$ and the auxiliary function $\tilde W^{(2)}_{\L\M}(\sigma; \lp)$. In principle, both of these quantities are unique. To see this, observe that for all $\lp$ with ${\rm Re}(\lp)>0$, there is a unique solution to Eq.~\eqref{eq:freq_domain_hyp_eq} satisfying regular boundary conditions. By analytical continuation, the solution $\tilde\psi^{(2)}_{LM}(\sigma,\lp)$ is uniquely determined for ${\rm Re}(\lp)<0$, except for the ambiguity in choosing the branch cut emanating from the branch point at $\lp = 0$; we consistently place this cut along the negative real axis.
This suffices to make $ A^{(2)}_{\L;\M; \q}$ and $\tilde W^{(2)}_{\L\M}(\sigma; \lp)$ unique.

To obtain an equation relating both of them to the source term $\tilde {\rm R}^{(2)}_{\L\M}$, it is convenient to re-express the frequency operator as in Eqs.~\eqref{eq:FreqOperator_QNMDecomp}--\eqref{dD_formula}.
When Eq.~\eqref{eq:Ansatz_AmpQNM_1stOrder} is combined with ~\eqref{eq:FreqOperator_QNMDecomp}, the inhomogeneous Eq.~\eqref{eq:freq_domain_hyp_eq} assumes the form
\begin{align}
    &\dfrac{ A^{(2)}_{\L;\M;\q} \,  \tilde {\cal D}_{\L;\M;\q} \left[ \tilde\psi^{(1)}_{\L;\M;\q}(\sigma)\right]}{\lp - \lp^{(1)}_{\L;\M;\q}} +  \tilde {\cal D}_{\L;\M;\q}\left[\tilde W^{(2)}_{\L\M}(\sigma; \lp)\right] \nonumber \\
    &+  A^{(2)}_{\L;\M;\q}\,   \tilde{\partial {\cal D}}_{\L\M} \left[ \tilde\psi^{(1)}_{\L;\M;\q}(\sigma)\right] \nonumber \\
    &+ (\lp- \lp^{(1)}_{\L;\M;\q})\,  \tilde{\partial {\cal D}}_{\L\M}\left[\tilde W^{(2)}_{\L\M}(\sigma; \lp)\right] = \tilde {\rm R}^{(2)}_{\L\M}(\sigma; \lp).
\end{align}
Since the QNM eigenfunction $ \tilde\psi^{(1)}_{\L; \M; \q}(\sigma)$ satisfies the homogeneous Eq.~\eqref{eq:hom_eq_QNM}, the term proportional to $\left(\lp - \lp^{(1)}_{\L;\M;\q} \right)^{-1}$ vanishes, and the limit $\lp \rightarrow \lp^{(1)}_{\L;\M;\q}$ yields
\begin{eqnarray}
    \label{eq:Eq_QNM_Amp}
    &&  \tilde {\cal D}_{\L;\M;\q}\left[\tilde W^{(2)}_{\L;\M;\q}(\sigma)\right] \nonumber \\
    && +  A^{(2)}_{\L;\M;\q} \,  \tilde{\partial {\cal D}}_{\L;\M;\q}\left[  \tilde\psi^{(1)}_{\L;\M;\q}(\sigma)\right] = \tilde {\rm R}^{(2)}_{\L;\M;\q}(\sigma),
\end{eqnarray}
with 
\begin{eqnarray}
\tilde W^{(2)}_{\L;\M;\q}(\sigma) &=& \tilde W^{(2)}_{\L\M}(\sigma; \lp^{(1)}_{\L;\M;\q}), \\
  \tilde{\partial {\cal D}}_{\L;\M;\q} &=& \left.   \tilde{\partial {\cal D}}_{\L\M}\right|_{\lp = \lp^{(1)}_{\L;\M;\q}}, \\
\tilde {\rm R}^{(2)}_{\L;\M;\q}(\sigma) &=& \tilde {\rm R}^{(2)}_{\L\M}(\sigma; \lp^{(1)}_{\L;\M;\q}).
\end{eqnarray}

We are ultimately interested in determining $ A^{(2)}_{\L;\M;\q}$. However, it will be useful to first think of Eq.~\eqref{eq:Eq_QNM_Amp} as an equation for $\tilde W^{(2)}_{\L;\M;\q}(\sigma)$. Even if we take $ A^{(2)}_{\L;\M;\q}$ as known, Eq.~\eqref{eq:Eq_QNM_Amp} does not uniquely determine $\tilde W^{(2)}_{\L;\M;\q}(\sigma)$. Indeed, the operator $ \tilde {\cal D}_{\L;\M;\q}$ has a non-vanishing kernel spanned by the QNM eigenfunction $ \tilde\psi^{(1)}_{\L;\M;\q}$ satisfying the homogeneous equation~\eqref{eq:hom_eq_QNM}. Thus, if $\tilde W^{(2)}_{\L;\M;\q}$ is a solution to Eq.~\eqref{eq:Eq_QNM_Amp}, then $\tilde W^{(2)}_{\L;\M;\q} \rightarrow \tilde W^{(2)}_{\L;\M;\q} + \alpha \,\, \tilde\psi^{(1)}_{\L;\M;\q}$ also satisfies Eq.~\eqref{eq:Eq_QNM_Amp} for an arbitrary complex constant $\alpha$. This freedom can be removed by setting $\tilde W^{(2)}_{\L;\M;\q}$ at some point $\sigma_o\in[0,1]$ to a fixed value
\begin{equation}
    \label{eq:Norm_W}
     \tilde W^{(2)}_{\L;\M;\q}(\sigma_o) =   W_{o}{}^{(2)}_{\L;\M;\q}.
\end{equation}
As discussed above, $ \tilde W^{(2)}_{\L;\M;\q}$ is unique, implying $ W_{o}{}^{(2)}_{\L;\M;\q}$ is as well. However, we will demonstrate below that $ W_{o}{}^{(2)}_{\L;\M;\q}$ can be freely specified for the purposes of finding $ A^{(2)}_{\L;\M;\q}$.

Condition~\eqref{eq:Norm_W}, together with Eq.~\eqref{eq:Eq_QNM_Amp} completely fixes a unique solution for the pair $\left( \tilde W^{(2)}_{\L;\M;\q}(\sigma), A^{(2)}_{\L;\M;\q} \right)$. Indeed, Eqs.~\eqref{eq:Norm_W} and \eqref{eq:Eq_QNM_Amp} form the linear system
\begin{eqnarray} \label{eq:2ndOrderQNM}
&&    \left( 
        \begin{array}{cc}
             \tilde {\cal D}_{\L;\M;\q} &  \tilde{\partial {\cal D}}_{\L;\M;\q}\left[  \tilde\psi^{(1)}_{\L;\M;\q}(\sigma)\right] \\
            \delta(\sigma - \sigma_o) & 0
        \end{array}    
    \right)
    \left( 
        \begin{array}{c}
            \tilde W^{(2)}_{\L;\M;\q} \\
             A^{(2)}_{\L;\M;\q}
        \end{array}    
    \right) \nonumber \\
  &&    =
    \left( 
        \begin{array}{c}
            \tilde {\rm R}^{(2)}_{\L;\M;\q} \\
            \delta(\sigma-\sigma_o) W_{o}{}^{(2)}_{\L;\M;\q}
        \end{array}    
    \right).
\end{eqnarray}
This matrix is invertible, yielding a unique solution. To make this more transparent, we can expand $\tilde W^{(2)}_{\L;\M;\q}$ and $\tilde \psi^{(1)}_{\L;\M;\q}$ in a spectral basis and convert the operators to spectral ones, making the linear system algebraic. In Sec.~\ref{sec:Results}, we empirically show that the resulting value of $ A^{(2)}_{\L;\M;\q}$ is independent of both the normalisation $ W_{o}{}^{(2)}_{\L;\M;\q}$ and $\sigma_o$, although we have not yet established a rigorous proof that this holds universally.

The only possible remaining freedom on the value $ A^{(2)}_{\L;\M;\q}$ comes from re-scaling the QNM eigenfunction $ \tilde\psi^{(1)}_{\L;\M;\q}$. If we change the normalisation~\eqref{eq:QNM_func_norm} into $ \tilde\psi_{\L;\M;\q} \rightarrow \eta \,\,  \tilde\psi_{\L;\M;\q}$, then the amplitude $ A^{(2)}_{\L;\M;\q} \rightarrow \eta^{-1}\,\,  A^{(2)}_{\L;\M;\q}$ still yields a solution to Eq.~\eqref{eq:Eq_QNM_Amp}. Therefore, similarly to the first-order amplitude~\eqref{eq:FirstOrder_ampltidue_at_infinity}, we introduce the uniquely defined second-order QNM excitation factor
\begin{equation} \label{eq:2ndOrder_ampltidue_at_infinity}
    \tilde {\cal A}^{(2)}_{\L;\M;\q}(\sigma) =  A^{(2)}_{\L;\M;\q}\,  \tilde\psi^{(1)}_{\L;\M;\q}(\sigma).
\end{equation}


\subsubsection{Quadratic QNM amplitudes} 

We now turn our attention to the calculation of the QQNM excitation factor $ \tilde {\cal A}^{(2)}_{(I \times I' )_{\L\M}}(\sigma)$. By applying the Laplace transform~\eqref{eq:def_Laplace} to the time-domain solution~\eqref{eqn:Upsilon_TD}, we find the frequency-domain solution to the inhomogeneous equation must have a pole at $\lp = \lp^{(2)}_{(I \times I' )_{\L\M}}$. Consequently, we consider the ansatz
\begin{equation}
    \label{eq:Ansatz_AmpQNM_2stOrder}
    \tilde \psi^{(2)}_{\L\M}(\sigma; \lp) = \dfrac{  \tilde {\cal A}^{(2)}_{(I \times I' )_{\L\M}}(\sigma)}{\lp - \lp^{(2)}_{(I \times I' )_{\L\M}}} + \tilde \Theta^{(2)}_{\L\M}(\sigma; \lp).
\end{equation}
As in Eq.~\eqref{eq:Ansatz_AmpQNM_1stOrder}, application of the inverse Laplace transform~\eqref{eq:InverseLaplace} and use of the residue theorem shows the first term in Eq.~\eqref{eq:Ansatz_AmpQNM_2stOrder} contributes to the specific QQNM dynamics with frequency $\lp^{(2)}_{(I \times I' )_{\L\M}}$. The function $\tilde \Theta^{(2)}_{\L\M}(\sigma; \lp)$ accounts for all other contributions to the solution $\tilde \psi^{(2)}_{\L\M}(\sigma; \lp)$, which are necessarily regular at $\lp^{(2)}_{(I \times I' )_{\L\M}}$.

However, contrary to Eq.~\eqref{eq:Ansatz_AmpQNM_1stOrder}, the unknowns $ \tilde {\cal A}^{(2)}_{(I \times I' )_{\L\M}}(\sigma)$ and $\tilde \Theta^{(2)}_{\L\M}(\sigma; \lp)$ in Eq.~\eqref{eq:Ansatz_AmpQNM_2stOrder} do not couple at $\lp = \lp^{(2)}_{(I \times I' )_{\L\M}}$: we can find $ \tilde {\cal A}^{(2)}_{(I \times I' )_{\L\M}}(\sigma)$ without requiring any information from $\tilde \Theta^{(2)}_{\L\M}(\sigma; \lp)$. 
Indeed, if we identify the pole in the source term $\tilde {\rm R}^{(2)}_{\L\M}$ by combining Eqs.~\eqref{eq:FrequencyDomainSource} and \eqref{eq:SecOrdSource_LaplTras} into 
\begin{equation}
\label{eq:PolesSecondOrder_Source_Amp}
    \tilde {\rm R}^{(2)}_{\L\M} = \tilde {\rm I}^{(2)}_{\L\M}(\sigma;\lp) + \sum_{ (J \times J' )_{\L\M} }\dfrac{  \tilde {\cal S}^{(2)}_{(J \times J' )_{\L\M}}(\sigma)}{\lp - \lp^{(2)}_{(J \times J' )_{\L\M}}},
\end{equation}
a short manipulation of Eq.~\eqref{eq:freq_domain_hyp_eq} considering Eqs.~\eqref{eq:Ansatz_AmpQNM_2stOrder} and \eqref{eq:PolesSecondOrder_Source_Amp} yields
\begin{align}
    & \tilde {\cal D}_{\L\M}\left[ \tilde {\cal A}^{(2)}_{(I \times I' )_{\L\M}}(\sigma) \right] \nonumber\\
    &+  \left( \lp - \lp^{(2)}_{(I \times I' )_{\L\M}}\right)  \tilde {\cal D}_{\L\M}\left[ \tilde \Theta^{(2)}_{\L\M}(\sigma;\lp) \right]  \nonumber\\[.5em]
    &\qquad =  
    \left( \lp - \lp^{(2)}_{(I \times I' )_{\L\M}}\right) \tilde {\rm I}^{(2)}_{\L\M}(\sigma;\lp) \nonumber\\
   &\qquad\quad+ \sum_{(J \times J' )_{\L\M} }  \tilde {\cal S}^{(2)}_{(J \times J' )_{\L\M}}(\sigma) \dfrac{\lp - \lp^{(2)}_{(I \times I' )_{\L\M}}}{\lp - \lp^{(2)}_{(J \times J' )_{\L\M}}}.
\end{align}
In the limit $\lp\rightarrow \lp^{(2)}_{(I \times I' )_{\L\M}}$, the only surviving term on the right-hand side of the above expression is that with $(J \times J' )_{\L\M}=(I \times I' )_{\L\M}$. Thus, the equation for the QQNM excitation factor is
\begin{equation} \label{eqn:QQNM_equation}
     \tilde {\cal D}_{\L\M}\left[  \tilde {\cal A}^{(2)}_{(I \times I' )_{\L\M}} (\sigma) \right] = \tilde {\cal S}^{(2)}_{(I \times I' )_{\L\M}}(\sigma).
\end{equation}

Note that while we set the initial-data source $\tilde {\rm I}^{(2)}_{\L\M}$ to zero in our general discussion of the second-order solution, we leave it unspecified in the equations above to make clear that the QQNM amplitudes are independent of it.

\subsubsection{Summary}

\begin{figure}[t] \centering
\includegraphics[width=\columnwidth]{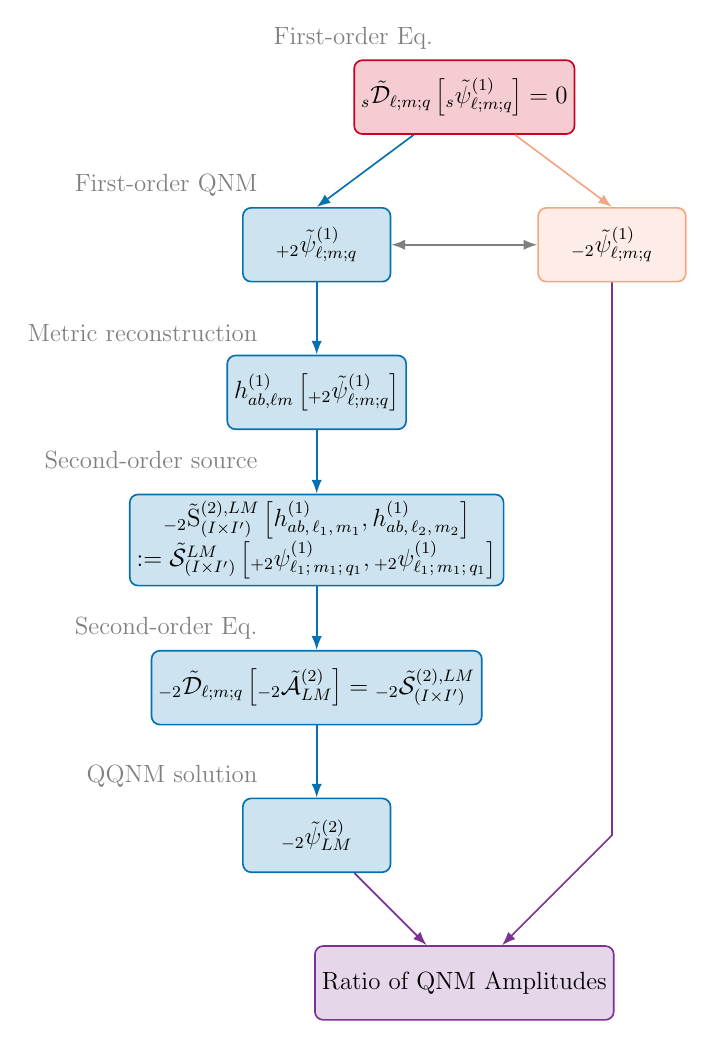}
  \caption{Flowchart illustrating our workflow for calculating the QQNM/QNM amplitude ratio. 
  Our method for computing first-order QNMs -- specifically, exciting a single QNM using finely 
  tuned initial data -- is described in Sec.~\ref{sec:first_order_qnms}. For the purpose of calculating the waveform strain, we calculate the Weyl scalar of spin-weight $-2$. However, for the purpose of calculating the second-order quadratic source, we calculate the first-order Weyl scalar of spin-weight $+2$, ${}_{+2}\tilde\psi^{(1)}_{\l;\m;\q}$. Of course, both first-order Weyl scalars are compatible with each other and satisfy the Teukolsky-Starobinsky identities (as indicated by the horizontal arrow). The first-order metric perturbation is calculated from ${}_{+2}\tilde\psi^{(1)}_{\l;\m;\q}$ following the standard metric reconstruction procedure, as reviewed in \cref{app:MetricReconstruction}. The second-order source~\eqref{eq:SecOrdSource_LaplTras} is then calculated from the reconstructed 
  metric perturbation, as also outlined in \cref{app:MetricReconstruction}. 
  Finally, the QQNM amplitudes are calculated from the second-order 
  Teukolsky equation following the procedure explained in Sec.~\ref{sec:QQNMs}.}
    \label{fig:flowchart}
\end{figure}

In Fig.~\ref{fig:flowchart}, we present a flow chart summarizing the calculation of the 
QQNM/QNM amplitude ratio. 

Our approach begins by selectively exciting a single first-order QNM (or a finite set of such QNMs) through carefully tuned initial
conditions, as detailed in Sec.~\ref{sec:first_order_qnms}. We solve Eq.~\eqref{eq:hom_eq_QNM} for both ${}_{+2}\tilde\psi^{(1)}_{\l;\m;\q}$ and ${}_{-2}\tilde\psi^{(1)}_{\l;\m;\q}$, ensuring that they represent the same physical metric perturbation by enforcing the relationship~\eqref{eq:QNM_ampltidue_relation_between_opposite_spin} between their amplitudes. We use the amplitude of ${}_{-2}\tilde\psi^{(1)}_{\l;\m;\q}$ to calculate the QQNM/QNM ratio, while we use ${}_{+2}\tilde\psi^{(1)}_{\l;\m;\q}$ as a helper function to calculate the second-order Weyl scalar.

The second-order source requires the complete first-order metric perturbation $h^{(1)}_{ab}$. We reconstruct $h^{(1)}_{ab}$ from ${}_{+2}\tilde\psi^{(1)}_{\l;\m;\q}$ in the outgoing radiation gauge following the standard Chrzanowski-Cohen-Kegeles procedure; interested readers may consult Appendix~\ref{app:MetricReconstruction} for a more comprehensive discussion. This choice of gauge ensures that $h^{(1)}_{ab}$, and therefore the second-order source constructed from it, is well-behaved at the boundaries.

Next, from the reconstructed metric perturbation, we construct the second-order source term in Eq.~\eqref{eqn:QQNM_equation}. This calculation is outlined in Appendix~\ref{app:source calculation}. Finally, we solve Eq.~\eqref{eqn:QQNM_equation} to extract the QQNM amplitude and the QQNM/QNM ratio.

In the next section, we introduce the numerical methods to solve all equations needed in this process: the QNM 
eigenvalue problem~\eqref{QNMEvalueProblem}, and the QQNM amplitude equation~\eqref{eqn:QQNM_equation}. We also present our method of solving the time-domain equation~\eqref{eq:ConfTeukMast_Gen_lm} at first and second order. The comparison between time- and frequency-domain results, in Sec.~\ref{sec:Results}, also requires solving Eq.~\eqref{eq:2ndOrderQNM} for the second-order linear QNM amplitudes.

\section{Teukolsky code in frequency and time domain}\label{sec:codes}

\subsection{Laplace domain code}

In the preceding sections, the equations governing the QNMs at first and second order were formulated in a frequency-domain approach through three different ordinary differential equations. First, we have the first-order QNMs, which were formulated as an eigenvalue problem; see Eq.~\eqref{QNMEvalueProblem}. At second order, we used two similar but distinct ansatze to construct the second order~\eqref{eq:2ndOrderQNM}, and quadratic~\eqref{eqn:QQNM_equation}, QNMs.
These three equations are solved via a spectral collocation method.
Specifically, all three equations can be cast into the general linear system,
\begin{align} \label{eq:AUb}
    {\bf A} U(x) = b(x),
\end{align}
where $U(x)$ are the unknowns and $\bf A$ is a linear differential operator.
In the above, we defined the coordinate 
\begin{equation}
\label{eq:x_of_sigma}
x = 1-2 \sigma
\end{equation}  
with domain $x \in [-1,1]$, which will be useful when decomposing the unknown variable $U$ into a sum of Chebyshev polynomials.
Equations~\eqref{eq:2ndOrderQNM} and \eqref{eqn:QQNM_equation} are already in the above form, where, respectively, the operator $\bf A$ is simply the operator on the left-hand side of Eqs.~\eqref{eq:2ndOrderQNM} and~\eqref{eqn:QQNM_equation}, 
$U$ denotes the unknowns $( \tilde{W}^{(2)}_{\L; \M; \q}, A^{(2)}_{\L; \M; \q})$ and $ \tilde {\cal A}^{(2)}_{(I \times I' )_{\L\M}}$, and $b$ is the right-hand side of these equations. For the first-order QNM eigenvalue problem~\eqref{QNMEvalueProblem}, the unknown $U(x)$ is the QNM amplitude $\vec{u}_{\l \m}$, while the operator ${\bf A} = {\bf L}_{\l\m} - \lp_{\l\m}^{(1)} \mathds{1}$, with $\mathds{1}$ the identity operator, and $b=0$.

For a fixed numerical truncation $N$, the numerical scheme approximates the function $U(x)$ via the expansion
\begin{equation} \label{eqn:Spectral}
    U(x) \approx U_N(x) := \sum_{k=0}^N c_k T_k(x),
\end{equation}
where $T_k(x) := \cos (k \arccos(x))$ are the Chebyshev polynomials of the first kind. The constant coefficients $c_k$ are fixed via collocation points. Specifically, let us consider the Chebyshev-Lobatto grid $\left\{ x_i \right\}_{i=0}^N$, along the $x$ direction
\begin{equation}
\label{eq:LobattoGrid}
    x_i = \cos \left( \frac{\pi i}{N} \right), \quad i=0,\ldots,N.
\end{equation}
An important property of the Chebyshev-Lobatto grid is that it includes the boundaries $x = \pm 1$ as grid points with labels $i = 0$ and $i = N$. Therefore, it is ideal to solve boundary-value problems.

The Chebyshev coefficients $c_k$ are fixed by imposing that the above approximate representation of $U(x)$, $U_N(x)$, is exact at the collocation points,
\begin{equation}
    U(x_i) = U_N(x_i).
\end{equation}
By differentiating Eq.~\eqref{eqn:Spectral} with respect to $x$ and evaluating at the collocation points, the derivative of $U(x)$ at $x_i$ is computed as a simple matrix multiplication applied to the discrete values $U_N(x_i)$:
\begin{equation}
(\partial_x U)(x_i) = \sum_{j=0}^N D_{ij} U_N(x_j).
\end{equation}
The matrix $D_{ij}$ is explicitly given by
\begin{equation}
\label{eq:D_Lobatto}
    {D}_{ij} = 
\begin{cases}
\displaystyle
        \dfrac{k_i (-1)^{i-j}}{k_j (x_i-x_j)} & i \neq j\\[1.25em]
\displaystyle
        -\dfrac{x_j}{2 (1-x_j^2)} & 0 < i=j < N\\[1.25em]
\displaystyle
        \dfrac{2N^2+1}{6} & i=j=0 \\[1.25em]
\displaystyle
       -\dfrac{2N^2+1}{6}& i=j=N
    \end{cases},
\end{equation}
where
\begin{equation}
    k_i = 
    \begin{cases}
        2 & i=0,N \\
        1 & i \neq 0,N
    \end{cases}.
\end{equation}

Therefore, the spectral collocation method allows us to recast the differential equation~\eqref{eq:AUb} into a system of linear algebraic equations for the unknowns $U_N(x_i) = U(x_i)$,
\begin{equation} \label{eq:MasterSpectralEq}
    \sum_{j=0}^N A_{ij} U_N(x_j) = b(x_i),
\end{equation}
where $A_{ij}$ denotes the matrix corresponding to the discretization of the operator ${\bf A}$.

A point worth emphasizing here is that thanks to the hyperboloidal formulation, the system~\eqref{eq:AUb} does not need to be supplemented with external boundary conditions, such as, for example, the usual ingoing/outgoing boundary conditions for the first-order QNM problem~\eqref{QNMEvalueProblem}. Indeed, such a prescription occurs when using a timelike coordinate $t$, for which the limits $r\to r_h$ and $r \to \infty$ connect the bifurcation sphere to spacelike infinity, where the QNM solutions (and any other object constructed from them, such as the second-order source) become singular.
The usual outgoing boundary conditions at large $r$ and ingoing boundary conditions near $r_h$ on constant-$t$ slices are equivalent to the condition that the perturbed metric is regular at the future horizon and future null infinity.
Thanks to the hyperboloidal formulation, a constant-$\tau$ slice instead connects the (future) black-hole horizon to (future) null infinity, avoiding the points where QNM solutions become singular. Together with the fact that we are working with regularized quantities everywhere [see Eq.~\eqref{eq:ConfMasterFunc}], this means that we can simply seek globally regular solutions in our (compact) domain~\cite{PanossoMacedo:2019npm}.

\subsection{Time domain code}
We also employ a time-domain approach to benchmark the predictions and results from the theoretical and numerical infrastructure in the frequency domain. For that purpose, a first-order reduction in time, with $ \tilde \Phi^{(\pertorder)}_{\l \m} = \tilde \Psi^{(\pertorder)}_{\l \m}{}_{,\tau}$, allows the representation of Eqs.~\eqref{eq:ConfTeukMast_Gen_lm} and \eqref{eq:O_operator} as
\begin{equation}
\label{eq:time evolution}
\left( \partial_{\tau} - \boldsymbol{L}_{\l \m \n} \right) \vec U^{(\pertorder)}_{\l\m\n} = {} \vec B^{(\pertorder)}_{\l\m\n},
\end{equation}
with $ \boldsymbol{L}_{\l \m \n} $ as in Eq.~\eqref{eq:QNM_EigenvalueSystem}, $ \vec U^{(i)}_{\l\m} = \left( \tilde \Psi^{(\pertorder)}_{\l \m} \, \, \, \tilde \Phi^{(\pertorder)}_{\l \m} \right)^T$, and $ {} \vec B^{(\pertorder)}_{\l\m} = \left( 0 \,\,\, \tilde {\rm S}^{(\pertorder)}_{\l \m} \right)^T$. Equation~\eqref{eq:time evolution} is to be solved with prescribed initial conditions $\vec U^{(i)}_{0}{}_{\l\m} = \left.  \vec U^{(i)}_{\l\m} \right|_{\tau = \tau_0}$ at an initial hyperboloidal slice $\tau = \tau_0$, and it has the formal solution
\begin{multline}
  \vec U^{(\pertorder)}_{\l\m}(\tau, \sigma) = e^{L(\sigma) \, \tau} \biggl(  \vec U^{(i)}_{0}{}_{\l\m} (\sigma)  \\ 
 + \int_{0}^{\tau}  e^{-L(\sigma) \, \tau'} {} \vec B^{(\pertorder)}_{\l\m}(\tau',\sigma) d\tau'  \biggr).
\end{multline}

We employ a fully spectral code~\cite{PanossoMacedo:2014dnr} to solve Eq.~\eqref{eq:time evolution}. Within this strategy, we divide the time interval $\tau \in [\tau_0, \tau_{\rm final}]$ into $n^{\rm max}_{\tau}$ strips with size $\delta \tau = \dfrac{\tau_{\rm final} -\tau_0}{n^{\rm max}_{\tau}}$. Within each time strip $n_{\tau} = 0, \dots, n^{\rm max}_{\tau}-1$, we introduce an auxiliary field $  \vec V^{(i)}_{n_{\tau}}{}_{\l\m} $ via
\begin{equation}
\label{eq:Ansatz}
 \vec U^{(i)}_{\l\m} (\tau, \sigma) =   \vec U^{(i)}_{n_{\tau}}{}_{\l\m}(\sigma) + \left( \tau - \tau_n\right)  \vec V^{(i)}_{n_{\tau}}{}_{\l\m} (\tau, \sigma).
\end{equation}
In the above expression $  \vec U^{(i)}_{n_{\tau}}{}_{\l\m}(\sigma)$ is the field's value at the time step $\tau_n = \tau_0 + n_{\tau} \delta \tau $. In particular, for $n_{\tau} = 0$ it corresponds to the prescribed initial data  $ \vec U^{(i)}_{0}{}_{\l\m}(\sigma)$. Once the solution is known in the strip $n_{\tau} = 0$, the information at the surface $\tau_1 = \tau_0 +  \delta \tau$ serves as the initial data for the next time strip with $n_{\tau} = 1$. This process is then repeated recursively for all time strips.

With the ansatz~\eqref{eq:Ansatz}, the wave equation~\eqref{eq:time evolution} becomes
\begin{multline}
\label{eq:wave equation aux}
(\tau - \tau_{n_\tau})\left( \partial_{\tau}   -   \boldsymbol{L}_{\l \m \n} \right)  \vec V^{(\pertorder)}_{n_{\tau}}{}_{\l\m} \\ +  \vec V^{(\pertorder)}_{n_{\tau}}{}_{\l\m} =  
{} \vec B^{(\pertorder)}_{\l\m} +  \boldsymbol{L}_{\l \m \n}   \vec U^{(\pertorder)}_{n_{\tau}}{}_{\l\m};
\end{multline}
i.e., information about initial data enters as a source of the singular wave equation for the auxiliary field $  \vec V^{(\pertorder)}_{n_{\tau}}{}_{\l\m}$. In particular, $\left.  \vec V^{(\pertorder)}_{n_{\tau}}{}_{\l\m}\right|_{\tau = \tau_{n_\tau}}$ is fully fixed by Eq.~\eqref{eq:wave equation aux}.

Equation~\eqref{eq:wave equation aux} is in a form akin to Eq.~\eqref{eq:AUb}, with the operator identified as ${\bf A} \leftrightarrow (\tau - \tau_{n_\tau})\left( \partial_{\tau} - \boldsymbol{L}_{\l \m \n} \right) + \mathds{1} $, the unknown variable as $U \leftrightarrow  \vec V^{(\pertorder)}_{n_{\tau}}{}_{\l\m}$, and the source term given by $b \leftrightarrow {} \vec B^{(\pertorder)}_{\l\m} +  \boldsymbol{L}_{\l \m \n}   \vec U^{(\pertorder)}_{n_{\tau}}{}_{\l\m} $. A fundamental difference, though, is that the functions' domain now is $(x^0,x^1)\in [-1,1]^2$, with $x^1$ as in Eq.~\eqref{eq:x_of_sigma} and $x^0$ mapping the time coordinate $\tau\in[\tau_{n_{\tau}}, \tau_{n_{\tau}+1}]$ in a given strip $n_{\tau}$ into $[-1,1]$ via
\begin{equation}
x^0 = \dfrac{2 \tau - \tau_{n_\tau +1} - \tau_{n_\tau}   }{\delta \tau}.
\end{equation}

Expanding the spectral decomposition~\eqref{eqn:Spectral} to the domain $(x^0,x^1)$, the numerical discretisation is now given by
\begin{equation}
U_{N_0, N_1}(x^0, x^1) = \sum_{k_0 = 0}^{N_0}\sum_{k_1=0}^{N_1} c_{k_0 k_1} T_{k_0}(x^0) T_{k_1}(x^1).
\end{equation}
As before, the coefficients $c_{k_0 k_1}$ are fixed by imposing that the approximated solution $U_{N_0, N_1}(x^0, x^1)$ coincides with the exact field $U(x^0, x^1)$ at a given set of grid points $\left\{ x_{i_0} \right\}_{i_0=0}^{N_0}$ and $\left\{ x_{i_1} \right\}_{i_1=0}^{N_1}$. 

The spatial grid $x^1_{i_1}$ is still given by the Chebyshev-Lobatto points~\eqref{eq:LobattoGrid}, implying the same differentiation matrix~\eqref{eq:D_Lobatto} for derivatives along the $\sigma$-direction. For the time grid $ x_{i_0}$ we employ the Chebyshev-Radau nodes
\begin{equation}
\label{chebyshev_radau_nodes}
    x^0_{i_0} = \cos\left(  2\pi \frac{i_0}{2N_0+1} \right), \ i_0 = 0, \ldots, N_0.
\end{equation}
At a given time strip $n_{\tau}$, this choice makes the grid point with label $i_0 = N_0$ to be slightly above the initial data surface $x^0 = -1 \leftrightarrow \tau = \tau_{n_\tau}$. In other words, the initial data surface is not within the grid as this information is contained in the right-hand side of Eq.~\eqref{eq:wave equation aux}

However, the grid includes the final surface $x^0 = 1 \leftrightarrow \tau = \tau_{n_\tau+1}$ precisely at the grid with $i_0 = 0$. The solution thereon determines the initial data for the next time strip $n_\tau +1$. For the Chebyshev-Radau grid, the differentiation matrix reads
\begin{equation}
    {D}_{ij} = 
\begin{cases}
\displaystyle
        \dfrac{N(N + 1)}{3} & i = j = 0\\[1.25em]
\displaystyle
        (-1)^j \dfrac{\sqrt{2(1+x_j)}}{1-x_j} & i=0,\, j \neq 0\\[1.25em]
\displaystyle
        \dfrac{(-1)^{i+1}}{\sqrt{2(1+x_i)}(1-x_i)} & i\neq0, \,  j = 0\\[1.25em]
\displaystyle
       -\dfrac{1}{2\left(1-\left(x_i\right)^2\right)}& i = j \neq 0
\\[1.75em]
\displaystyle
	\dfrac{(-1)^{i-j}}{x_i-x_j} \sqrt{\dfrac{1+x_j}{1+x_i}} &  0 \neq i \neq j \neq 0  
    \end{cases}.
    \label{eq:D_radau} 
\end{equation}
We then obtain a linear algebraic system of equations similar to Eq.~\eqref{eq:MasterSpectralEq}, but with a discrete square matrix $A_{IJ}$ and vectors $U_{I}$ and $B_J$ with size $n_{\rm total} = (N_0+1)(N_1+1)$ corresponding to the tensorial product of the discretisation along $\tau$ and $\sigma$.

\section{Results}\label{sec:Results}

We now present results from both our frequency- and time-domain codes. Firstly, we benchmark in Sec.~\ref{sec:first_order_dynamics} the first-order dynamics triggering a pure QNM evolution. In Secs.~\ref{sec:Single_mode_excitation} and \ref{sec:two modes} we focus on frequency-domain calculations, where we are able to cleanly isolate QQNMs sourced by a single or multiple first-order QNMs. In Sec.~\ref{sec:comparison} we compare the frequency-domain results to a full time-domain solution, exploring how the QQNMs and linear second-order QNMs contribute to the complete time-domain signal.

Throughout this section, we set the spin $\NPs=-2$, as we are interested in the associated strain evaluated at (future) null infinity.

We recall that the Laplace parameter $\lp$ relates do the commonly used frequency $\omega$ via
\begin{equation}
    \lp = - i \lambda \omega.
\end{equation}
For all explicit results and plots, we have set $\lambda = 4M$ and $M=1$.

\subsection{First-order dynamics} \label{sec:first_order_dynamics}
\begin{figure}[t!] \centering
\includegraphics[width= \columnwidth]{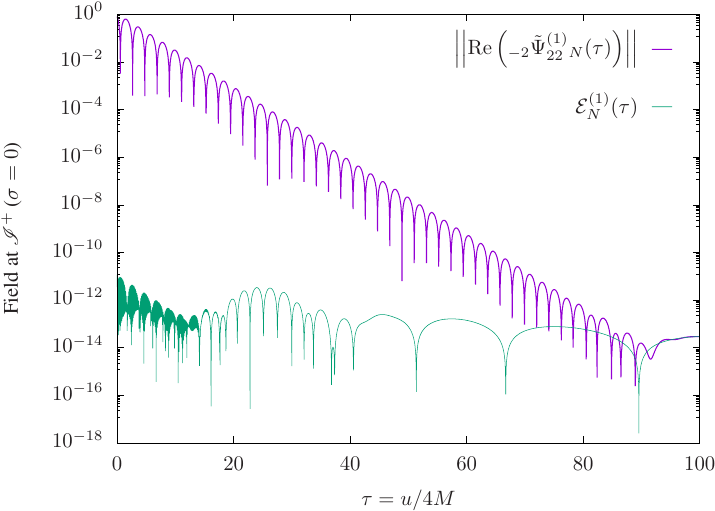}
  \caption{Time-domain evolution of fields at future null infinity. For our choice of initial data, we obtain pure QNM dynamics at first order in perturbation theory (purple). The error ${\cal E}_N^{(1)}(\tau)$ around machine precision (green line) confirms that the signal consists of a single QNM frequency, here $\lp^{(1)}_{2;2;0} \approx -0.356 - 1.495\,i$.} \label{fig:timeXfrequency_benchmark}
\end{figure}

To verify that the procedure outlined in Sec.~\ref{sec:first_order_qnms} does yield a pure QNM evolution at first perturbative order, we study the time evolution beginning from the initial data~\eqref{QNM_ID1} and~\eqref{QNM_ID2}. Figure~\ref{fig:timeXfrequency_benchmark} compares the numerical solution $ \tilde{\Psi}_{\l\m}^{(1)}{}_{N}(\tau,\sigma)$ against the expected exact solution~\eqref{QNM_TD_ansatz}. In particular, the purple line is the time-domain solution, as generated from our time-domain code, evaluated at future null infinity $\sigma=0$ for the fundamental quadrupole mode $(\l, \m, \q) = (2,2,0)$ of the spin $\NPs=-2$ field. One observes a consistent ringdown with the expected QNM frequencies $\lp^{(1)}_{2;2;0} \approx -0.356 - 1.495\,i$ up to amplitudes $\sim 10^{-12}$. The green line shows the relative error 
\begin{equation}
{\cal E}_N^{(1)}(\tau) = \left| 1- \dfrac{ \tilde{\Psi}_{\l\m}^{(1)}{}_{N}(\tau,0)}{ \tilde{\Psi}_{\l\m}^{(1)}(\tau,0)}\right|,
\end{equation} 
which remains at round-off level $\sim 10^{-12}$ along the entire evolution. Moreover, we also observed that the QNM amplitude predicted by Eq.~\eqref{eq:2ndOrderQNM} within the frequency domain recovers $A_{\l;\m;\q}^{(1)}=1$ if $(\l, \m, \q) = (2,2,0)$, but it vanishes otherwise. These results illustrate and confirm that (i) the theoretical framework indeed allows for clean QNM dynamics at first order, and (ii) the numerics are robust and stable to recover the theoretical predictions close to machine precision level for the time and frequency domain approaches.

With the framework benchmarked to allow the exact excitation of individual QNMs at first order, we now proceed to investigate the effects on the wave signal at second order. 
\subsection{Single first-order mode excitation} \label{sec:Single_mode_excitation}

To calculate the QQNM solution~\eqref{eqn:Upsilon_TD}, we solve the second-order Teukolsky equation Eq.~\eqref{eqn:QQNM_equation}, which yields the amplitude of the QQNM at the frequencies $\lp_{(I \times I')_{\L \M}}^{(2)}$. 

Since the QQNM amplitude $\tilde{\cal{A}}_{(I \times I')_{\L \M}}^{(2)}$ directly depends on the second-order source $\tilde{\cal S}_{(I \times I')_{\L \M}}^{(2)}$, we need to first specify which linear QNM modes are excited. In this section, we consider the source term generated by a single regular mode excitation $(\lOne,\mOne,\qOne)$ at first order, together with its associated mirror mode $(\lOne,-\mOne,\QOne)$,
\begin{align} \label{eqn:Psi4_Ansatz1}
    & \tilde{\Psi}^{(1)}(\tau,\sigma,\theta,\varphi)\nonumber\\ 
    &\quad= A_{\lOne; \mOne; \qOne}^{(1)} \tilde{\psi}_{\lOne ;\mOne; \qOne}^{(1)}(\sigma) e^{\lp^{(1)}_{\lOne ;\mOne; \qOne} \tau} {}_{-2} Y_{\lOne \mOne}(\theta, \varphi) \nonumber\\
     &\quad + A^{(1)}_{\lOne; -\mOne; \QOne} \tilde{\psi}_{\lOne; -\mOne;\QOne}^{(1)}(\sigma)e^{\lp^{(1)}_{\lOne; -\mOne; \QOne} \tau} {}_{-2} Y_{\lOne -\mOne} (\theta,\varphi).
\end{align}

According to the decomposition~\eqref{eq:SecOrdSource_TimeDep}, the full, four-dimensional second-order source takes the form
\begin{align}
    \label{eq:SecOrdSource_FullDecomposition}
     \tilde{\cal S}^{(2)}(\tau,\sigma,\theta, \varphi) &= \sum_{\L,\M} \sum_{ (I \times I')_{\L\M} } \tilde {\cal S}^{(2)}_{(I \times I')_{\L\M}}(\sigma) \times \nonumber \\
    &\qquad e^{\tau \lp^{(2)}_{(I \times I')_{\L\M}}} {}_{-2} Y_{\L\M}(\theta,\varphi).
\end{align}
Since we are considering here a single mode excitation, the above can be written in a more explicit form. 
First, due to the coupling of the spherical harmonics, the $\L$ follows the usual ``triangle inequality'' rule, while $\M$ can only take values $-2\mOne,0$, or $2\mOne$.

Equation~\eqref{eq:SecOrdSource_FullDecomposition} can hence be written more explicitly as\footnote{We added a comma between the mode indices in some variables for clarity}
\begin{align} \label{eqn:2ndOrderSourceDecomposition_1}
    \tilde{\cal S}^{(2)}&(\tau,\sigma,\theta, \varphi) \nonumber\\
    &= \sum_{\L}\Bigl\{\tilde{S}_{\L, 2\mOne}^{(2)}(\sigma) e^{2 \lp_{\lOne; \mOne; \qOne}^{(1)} \tau}  {}_{-2} Y_{\L, 2\mOne}(\theta, \varphi) \nonumber\\
    &\quad+ 2 \times \tilde{S}_{\L 0}^{(2)}(\sigma) e^{2 {\rm Re}(\lp_{\lOne; \mOne; \qOne}^{(1)}) \tau} {}_{-2} Y_{\L 0}(\theta, \varphi) \nonumber \\
    &\quad+ \tilde{S}_{\L, -2\mOne}^{(2)}(\sigma) e^{2 \lp_{\lOne; -\mOne;\QOne}^{(1)} \tau} {}_{-2} Y_{\L, -2\mOne}(\theta, \varphi) \Bigr\},
\end{align}
where we used Eq.~\eqref{eq:SecOrdFreq} to write the QQNM frequencies $\lp^{(2)}_{(I \times I')_{\L\M}}$ in terms of the linear QNM frequencies, and $ \tilde {\cal S}^{(2)}_{\L\M}(\sigma)$ contains all the individual contributions $ \tilde {\cal S}^{(2)}_{(I \times I')_{\L\M}}(\sigma)$. We added a factor of two in the definition of $ \tilde{S}_{\L, 0}^{(2)}(\sigma)$ to reflect the fact that it is real (so that for each mode contribution, its complex conjugate also contributes).

The calculation of the source coefficients $\tilde S^{(2)}_{LM}(\sigma)$ is described in Appendix~\ref{app:MetricReconstruction}. In Fig.~\ref{fig:2ndOrderSource}, we plot the source term $ \tilde{S}^{(2)}_{44}(\sigma)$, generated by $(\lOne,\mOne,\qOne) = (2,2,0)$ and the mirror mode $(\lOne;-\mOne;\QOne) = (2,-2,1)$. Thanks to the hyperboloidal formulation and the rescaling of the source~\eqref{eq:ConfTeukMast_Gen}, $ \tilde{S}^{(2)}_{44}(\sigma)$ (and any other modes of the source) is well behaved everywhere, including at both endpoints. In particular, by inspecting the functional form of the source in terms of the first-order metric perturbation, one can show in general that for any $(\L, \M)$, 
\begin{equation}
\lim_{\sigma \to 0} \left( \sigma^{-1} \tilde{S}^{(2)}_{\L \M}(\sigma) \right)
\end{equation}
is a (non-zero) constant.

\begin{figure}
    \includegraphics[width= 0.48\textwidth]{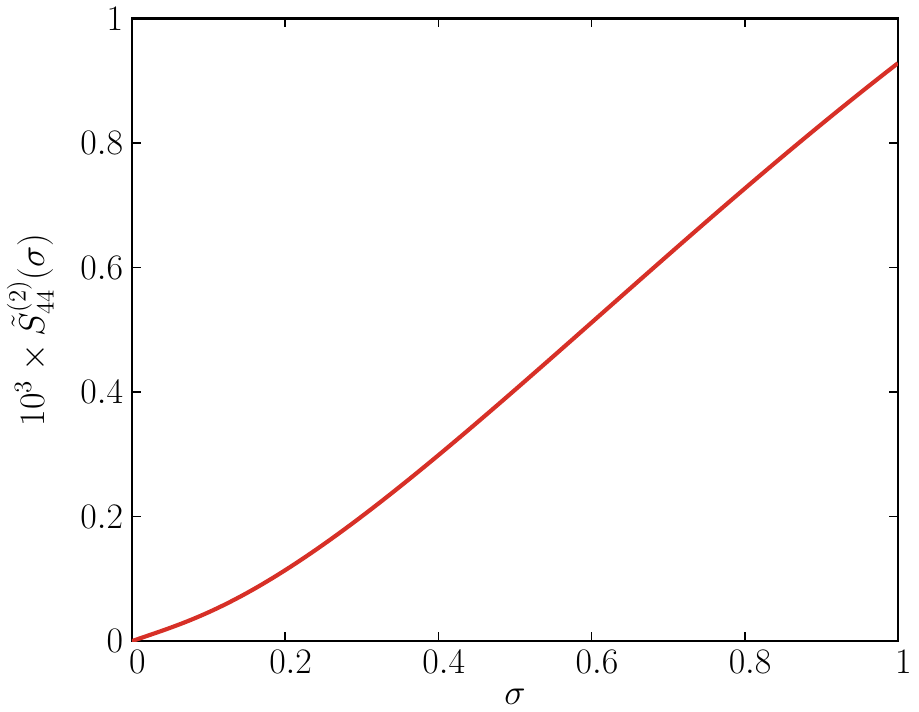}
\caption{Radial part of the second-order source $ \tilde{S}^{(2)}_{44}(\sigma)$. Note that the full expression of $ \tilde{S}^{(2)}_{44}(\sigma)$ depends on the two linear amplitudes $ A_{2;2;0}^{(1)}$ and $A_{2;-2;1}^{(1)}$ in the same way as displayed in Eq.~\eqref{eqn:Psi42ndOrder++}. For this plot, we chose $ A_{2;2;0}^{(1)} =  \bracket{A_{2;-2;1}^{(1)}}^\star=1$, although other choices give the same qualitative features.}
\label{fig:2ndOrderSource}
\end{figure}

With the expression for the source term~\eqref{eqn:2ndOrderSourceDecomposition_1} at hand, we can now solve Eq.~\eqref{eqn:QQNM_equation}. The QQNM solution inherits the same structure as the second-order source~\eqref{eqn:2ndOrderSourceDecomposition_1},
\begin{align} \label{eqn:2ndOrderPsiDecomposition_1}
    \tilde{\Psi}^{(2)}&(\tau,\sigma,\theta,\varphi) \nonumber\\*
    &= \sum_{\L}\Bigl\{\tilde{\cal A}_{\L ,2\mOne}^{(2)}(\sigma) e^{2 \lp^{(1)}_{\lOne; \mOne; \qOne} \tau} {}_{-2} Y_{\L ,2\mOne}(\theta, \varphi) \nonumber\\
    &\quad+ 2 \times \tilde{\cal A}_{\L 0}^{(2)}(\sigma) e^{2 {\rm Re}(\lp^{(1)}_{\lOne; \mOne; \qOne}) \tau} {}_{-2} Y_{\L 0}(\theta, \varphi) \nonumber \\
    &\quad+ \tilde{\cal A}_{\L, -2\mOne}^{(2)}(\sigma) e^{2 \lp^{(1)}_{\lOne; -\mOne;\QOne} \tau} {}_{-2} Y_{\L, -2\mOne}(\theta, \varphi) \Bigr\}.
\end{align}
By solving Eq.~\eqref{eqn:QQNM_equation}, one finds that the QQNM amplitudes $ \tilde{\cal A}_{\L, 2\mOne}^{(2)}(\sigma)$, $ \tilde{\cal A}_{\L, 0}^{(2)}(\sigma)$, and $ \tilde{\cal A}_{\L, -2\mOne}^{(2)}(\sigma)$ depend on the first-order excitation coefficients, $ A^{(1)}_{\lOne; \mOne; \qOne}$ and $ A^{(1)}_{\lOne; -\mOne; \QOne}$, in the following way:
\begingroup%
\allowdisplaybreaks%
\begin{align}
\label{eqn:Psi42ndOrder++}
     \tilde{\cal A}_{\L, 2\mOne}^{(2)}(\sigma) &=  a_{\L, 2\mOne}  (\sigma) \bracket{ A^{(1)}_{\lOne; \mOne; \qOne}}^2 \nonumber \\*
    &+  b_{\L, 2\mOne}  (\sigma)  A^{(1)}_{\lOne; \mOne; \qOne} \bracket{ A^{(1)}_{\lOne; -\mOne; \QOne}}^{\!\star}, \\
     \tilde{\cal A}_{\L 0}^{(2)}(\sigma) &= c_{\L 0} (\sigma)  A^{(1)}_{\lOne; \mOne; \qOne}  A^{(1)}_{\lOne; -\mOne; \QOne} \nonumber \\*
     &+d_{\L 0} (\sigma)  A^{(1)}_{\lOne; \mOne; \qOne} \bracket{ A^{(1)}_{\lOne; \mOne; \qOne}}^{\!\star} \nonumber \\*
    &+ \left( d_{\L 0}  (\sigma)\right)^\star  A^{(1)}_{\lOne; -\mOne; \QOne} \bracket{ A^{(1)}_{\lOne; -\mOne; \QOne}}^{\!\star}, \label{eqn:Psi42ndOrder+-} \\
     \tilde{\cal A}_{\L, -2\mOne}^{(2)}(\sigma) &= (-1)^\L \bigg[ \left( a_{\L, 2\mOne} (\sigma) \right)^\star \bracket{ A^{(1)}_{\lOne; -\mOne; \QOne}}^2 \nonumber \\*
    &+ \left( b_{\L, 2\mOne} (\sigma)\right)^\star  A^{(1)}_{\lOne; -\mOne; \QOne}\! \bracket{ A^{(1)}_{\lOne; \mOne; \qOne}}^{\!\star}\! \bigg]. \label{eqn:Psi42ndOrder--}
\end{align}
\endgroup
We note that the factor $(-1)^L$ appearing in the expression of $ \tilde{\cal A}^{(2)}_{\L, -2\mOne}(\sigma)$ follows from the symmetries of 3$j$ symbols, which appear when expanding products of spherical harmonics into a sum of single harmonics; see Eq.~(94) in~\cite{Spiers:2023mor}.

We provide the coefficients $a,b,c,d$, evaluated at $\scri^+$, in Table~\ref{table:abcd_1}.

\begin{table}[tb] \centering
\begin{tabular}{ c c | c c } 
\toprule
$(\lOne,\mOne,\qOne)$ & $\L$ & \multicolumn{2}{c}{at $\scri^+ \, (\sigma=0)$} \\
\midrule
\multirow{5}{4em}{$(2,0,0)$} & \multirow{5}{2.5em}{\ \ \ $2$}
& $ a_{20}$ & $ 0.790153 - 2.66674\,i $ \\ 
&& $ b_{20}$ & $ -0.127560 - 0.211491\,i $ \\ 
&& $ c_{20}$ & $ 0.139479 $ \\ 
&& $ d_{20}$ & $ -0.0275200 + 0.239190\,i $ \\
\midrule
\multirow{5}{4em}{$(2,2,0)$} & \multirow{5}{2.5em}{\ \ \ $2$}
& $ a_{22}$ & $ 0 $ \\ 
&& $ b_{22}$ & $ 0 $ \\ 
&& $ c_{20}$ & $ -0.139479 $ \\ 
&& $ d_{20}$ & $ 0.0275200 - 0.239190\,i $ \\
\midrule
\multirow{5}{4em}{$(2,0,0)$} & \multirow{5}{2.5em}{\ \ \ $4$}
& $ a_{40}$ & $ 9.88388 + 3.99254\,i $ \\ 
&& $ b_{40}$ & $ 1.11259 + 0.648263\,i $ \\ 
&& $ c_{40}$ & $ 0.00361875 $ \\ 
&& $ d_{40}$ & $ 0.000131353 - 0.00843725\,i $ \\
\midrule
\multirow{5}{4em}{$(2,2,0)$} & \multirow{5}{2.5em}{\ \ \ $4$}
& $ a_{44}$ & $ 13.7824 + 5.56733\,i $ \\ 
&& $ b_{44}$ & $ 1.55143 + 0.90396\,i $ \\ 
&& $ c_{40}$ & $ 0.000603125 $ \\ 
&& $ d_{40}$ & $ 0.0000218922 - 0.00140621\,i $ \\
\midrule
\multirow{5}{4em}{$(2,2,2)$} & \multirow{5}{2.5em}{\ \ \ $4$}
& $ a_{44}$ & $ 3.93713 +11.0992\,i $ \\ 
&& $ b_{44}$ & $ 0.112633 +1.32345\,i $ \\ 
&& $ c_{40}$ & $ -0.602210*10^6 $ \\ 
&& $ d_{40}$ & $ (-0.490123-2.36369\,i)*10^7 $ \\
\bottomrule
\end{tabular}
\caption{Values of the coefficients at $\scri^+$ appearing in Eqs.~\eqref{eqn:Psi42ndOrder++}--\eqref{eqn:Psi42ndOrder--}, for a selection of mode numbers $(\l,\m,\q)$.}
\label{table:abcd_1}
\end{table}

Typically, one is not directly interested in the quadratic amplitude $ \tilde{\cal A}_{\L \M}^{(2)}(\sigma)$. Instead, one is interested in $ \tilde{\cal A}_{\L \M}^{(2)}(\sigma)$ relative to the first-order amplitude that generated it. We define the following ratio,
\begin{equation} \label{eqn:QQNMRatio_Psi4Reg_OneMode}
     \mathcal{R}^{ \tilde{\Psi}}_{\L, 2\mOne}(\sigma) := \frac{\left( \tilde{\Psi}^{(2)}\right)_{\L, 2\mOne}}{\left( \tilde{\Psi}^{(1)}\right)^2_{\lOne \mOne}} = \frac{ \tilde{\cal A}_{\L, 2\mOne}^{(2)}(\sigma)}{ \left( \tilde{\cal A}_{\lOne; \mOne; \qOne}^{(1)}(\sigma) \right)^2},
\end{equation}
where for the second equality, we used Eqs.~\eqref{eqn:2ndOrderPsiDecomposition_1}, \eqref{eqn:Psi4_Ansatz1}, and \eqref{eq:FirstOrder_ampltidue_at_infinity}.

We highlighted in Sec.~\ref{sec:QQNMs} that the QQNM amplitudes are independent of the initial-data source term in the Laplace-domain field equation. However, the QQNM amplitudes \emph{do} depend on the progenitor system through a dependence on the ratio between mirror and regular mode amplitudes, $\bracket{ A^{(1)}_{\lOne; -\mOne;\QOne}}^\star/ A^{(1)}_{\lOne;\mOne;\qOne}$. As we discuss below, this ratio is associated with the degree to which the system is up-down asymmetric. The QQNM amplitudes' dependence on this ratio is their \emph{only} dependence on what created the ringing BH. We write this dependence as
\begin{equation} \label{eqn:RPsi4_1}
     \mathcal{R}^{ \tilde{\Psi}}_{\L, 2\mOne}(\sigma) =  a_{\L, 2\mOne} (\sigma) + b_{\L, 2\mOne} (\sigma) z_{\lOne; \mOne; \qOne},
\end{equation}
where we define
\begin{equation} \label{eqn:z1}
     z_{\lOne; \mOne; \qOne} := \frac{\bracket{ A^{(1)}_{\lOne; -\mOne; \QOne}}^\star}{ A^{(1)}_{\lOne; \mOne; \qOne}}.
\end{equation}
The ratio between regular and mirror modes is equivalent to the ratio between even- and odd-parity components of the GW strain; see Appendix~\ref{app:even and odd}. Therefore, using Eqs.~\eqref{eqn:AtoC1} and \eqref{eqn:AtoC2}, we can alternatively express $ \mathcal{R}^{ \tilde{\Psi}}_{\L, 2\mOne}(\sigma)$ in terms of the ratio of odd- to even-parity amplitudes, $C^-_{\lOne; \mOne;\qOne} / C^+_{\lOne; \mOne;\qOne}$. For a system that is up-down symmetric (i.e., symmetric under reflection through the equatorial plane), the ratio $C^-_{\lOne; \mOne;\qOne} / C^+_{\lOne; \mOne;\qOne}$ vanishes for even values of $\lOne+\mOne$, while it becomes infinite for odd values of $\lOne+\mOne$. For up-down \emph{anti}symmetric systems, the opposite is true: $C^-_{\lOne; \mOne;\qOne}/C^+_{\lOne; \mOne;\qOne}$ vanishes for odd values of $\lOne+\mOne$, while it becomes infinite for even values of $\lOne+\mOne$. We review these properties in Appendix~\ref{app:even and odd}.

We remark that this construction also follows through in Kerr spacetime, where $ \mathcal{R}^{ \tilde{\Psi}}_{\L, 2\mOne}(\sigma)$ is also a function of the BH parameters and $C^-_{\lOne; \mOne;\qOne} / C^+_{\lOne; \mOne;\qOne}$. This can be derived, for example, by expressing the source in Kerr, $ \tilde{\mathcal{S}}$, available in Ref.~\cite{2nd-order-notebook} in terms of the Weyl scalar (see Appendix~\ref{app:MetricReconstruction}). It still consists of a sum of terms $\propto \Psi^2$ and $\propto \Psi\bar\Psi$. Hence, using Eqs.~\eqref{eqn:Psi4_Ansatz1} and~\eqref{QNM_frequency_new_relation}, similar relations as 
Eqs.~\eqref{eqn:Psi42ndOrder++}--\eqref{eqn:Psi42ndOrder--} hold in Kerr (up to angular mode mixing). As a result, the ratio
$ \mathcal{R}^{ \tilde{\Psi}}_{\L, 2\mOne}(\sigma)$ will still depend on the progenitor's up-down (a)symmetry via the ratio $C^-_{\lOne ;\mOne;\qOne}/C^+_{\lOne; \mOne;\qOne}$.

Note that the ratio $ \mathcal{R}^{ \tilde{\Psi}}_{\L, 2\mOne}(\sigma)$ is defined in terms of the normalized field $ \tilde{\Psi}^{(i)}$.
It is useful to relate this ratio to the analogous ratio from the gravitational strain $h$ evaluated at $\scri^+$.
As detailed in Appendix~\ref{Appendix:SecondOrderQNMAmplitude}, the analogous expression for the gravitational strain evaluated at null infinity is given by
\begin{equation} \label{eqn:Rh_1}
         M \mathcal{R}^h_{\L, 2\mOne} = -\frac{M^4}{\lambda^4} \bracket{\lp^{(1)}_{\lOne; \mOne; \qOne}}^2 \mathcal{R}^{ \tilde{\Psi}}_{\L, 2\mOne}\Big|_{\sigma=0}.
\end{equation}

\begin{figure}
    \includegraphics[width=\columnwidth]{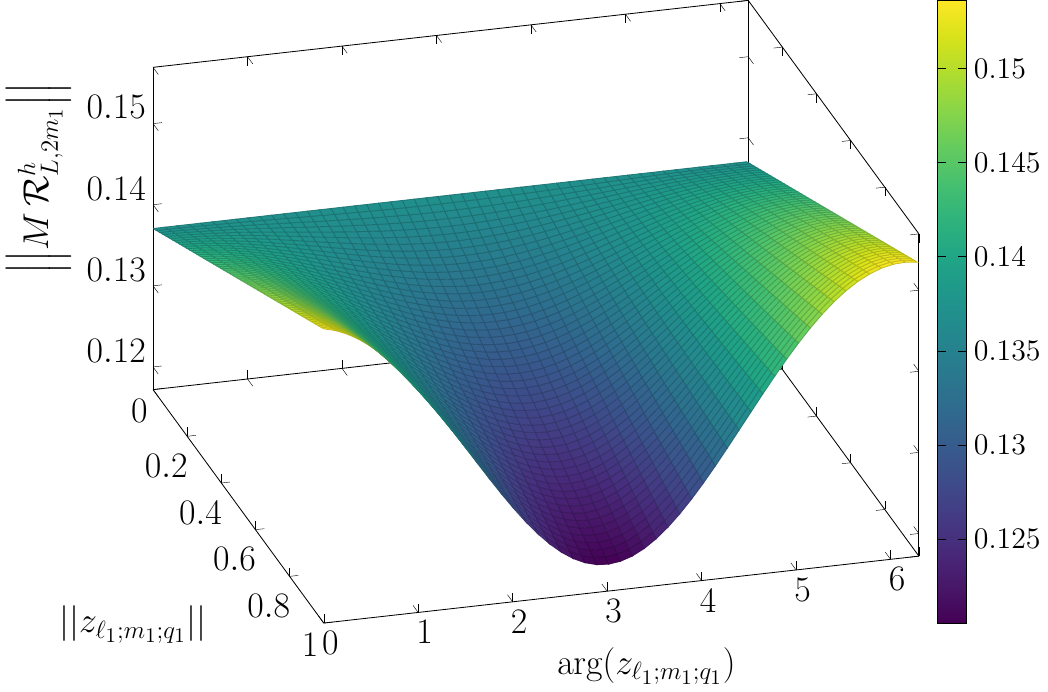}
    \caption{ Surface plot of $|| M\,\mathcal{R}^h_{\L, 2\mOne}||$, given in Eqs.~\eqref{eqn:RPsi4_1} and \eqref{eqn:Rh_1}, in terms of the amplitude and phase of the complex number, $ z_{\lOne; \mOne; \qOne}$; see Eq.~\eqref{eqn:z1}. We used $(\lOne,\mOne, \qOne) = (2,2,0)$, and $L=4$.}
    \label{fig:Rh_3DPlot}
\end{figure}

In Fig.~\ref{fig:Rh_3DPlot}, we show a surface plot of the absolute value of $ \mathcal{R}^h_{\L, 2\mOne}$, as a function of the amplitude and phase of the complex number $ z_{\lOne; \mOne; \qOne}$ for the case $(\lOne, \mOne,\qOne) = (2,2,0)$ and $\L=4$. At $ z_{\lOne; \mOne; \qOne} = 0$, corresponding to $ A_{\lOne; -\mOne; \QOne} = 0$, $ \mathcal{R}^h_{\L, 2\mOne}$ is a constant, proportional to $ a_{\L, 2\m_1}(0)$ in Eq.~\eqref{eqn:RPsi4_1}, that depends only on the background parameters. From Eq.~\eqref{eqn:Rh_1} and the value of $ a_{\L, 2\m_1}(0)$ in Table~\ref{table:abcd_1}, we find $ \mathcal{R}^h_{44} \approx 0.137 \exp(- 0.0835\,i)$, in exact agreement with the frequency-domain BHPT calculation performed by Ma \emph{et al.}~\cite{ma2024excitation}, who considered only a single regular mode excitation at first order, which corresponds in our case to setting $ A_{\lOne; -\mOne; \QOne} = 0$.

\begin{figure*}
    \includegraphics[scale=0.66]{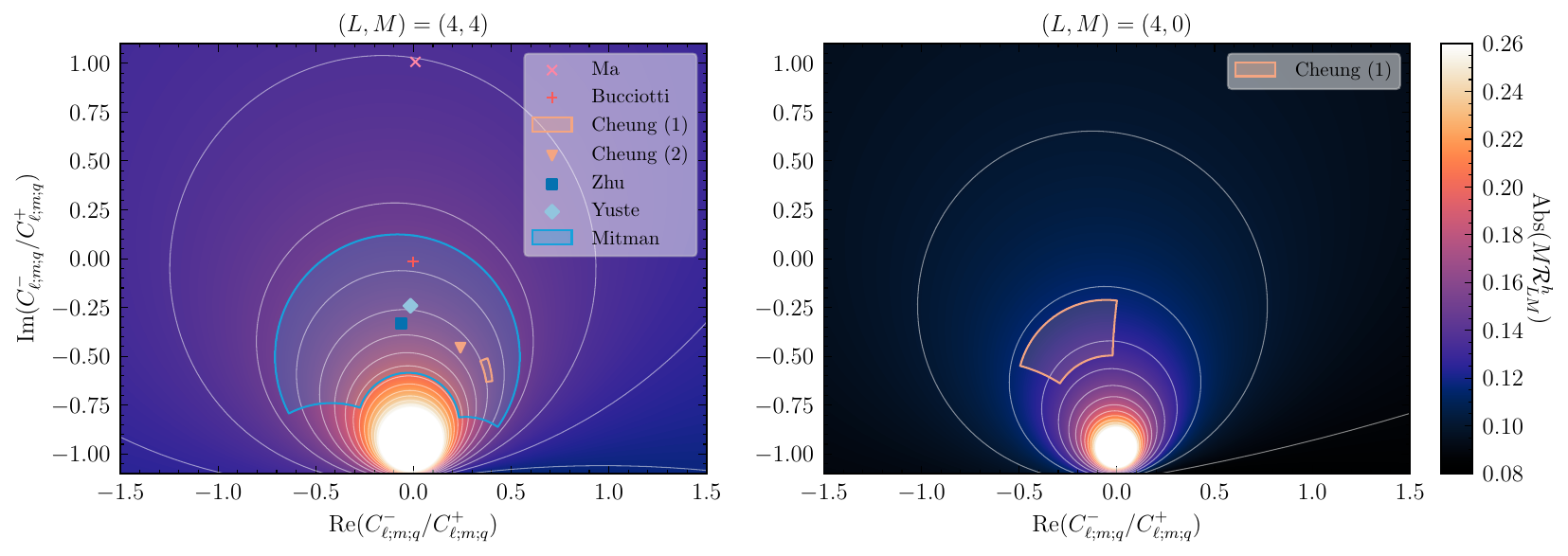}
    \caption{Contour plot of the quadratic-to-linear strain ratio $ \mathcal{R}^h_{\L, 2 \mOne}$ as a function of the odd-to-even complex amplitude ratio $C^-_{\lOne; \mOne; \qOne} / C^+_{\lOne; \mOne; \qOne}$. Solid lines indicate values of constant amplitude of $ \mathcal{R}^h_{\L, 2 \mOne}$. {\em Left Panel:} Quadratic mode $(\L,\M) = (4, 4)$ from parent linear mode $(\lOne, \mOne, \qOne) = (2,2,0)$. Also shown are prior results employing second-order BHPT:
    Redondo-Yuste \emph{et al.}~\cite{Redondo-Yuste:2023seq};
    Ma \emph{et al.}~\cite{ma2024excitation};
    Bucciotti \emph{et al.}~\cite{Bucciotti:2024zyp};
    Zhu \emph{et al.}~\cite{Zhu:2024rej}, as well as
    results from NR:
    Cheung \emph{et al.}~\cite{Cheung:2022rbm};
    Cheung \emph{et al.}~\cite{Cheung:2023vki};
    Mitman \emph{et al} \cite{Mitman:2022qdl}.
    Note that the NR results are for spinning BHs, while our method of placing them on this plot assumes a non-spinning BH. 
    {\em Right Panel:} Quadratic mode 
    $(\L,\M) = (4, 0)$ from parent linear mode $(\lOne, \mOne, \qOne) = (2,0,0)$. We added prior NR results for head-on collisions yielding a final non-spinning BH. Figure reproduced from our letter~\cite{Bourg:2024jme}.}
    \label{fig:QQNM_44}
\end{figure*}

In Fig.~\ref{fig:QQNM_44}, we show contour plots of $ \mathcal{R}^h_{\L, 2\mOne} $ as a function of the ratio $C^-_{\lOne ;\mOne; \qOne} / C^+_{\lOne; \mOne; \qOne}$ for the case $(\lOne;\mOne;\qOne) = (2,2,0)$ (left plot) and $(\lOne,\mOne,\qOne) = (2,0,0)$ (right plot).
For comparison, in these plots we include values from Refs.~\cite{Perrone:2023jzq,Nakano:2007cj,ma2024excitation,Bucciotti:2024zyp,Cheung:2022rbm,Cheung:2023vki,Zhu:2024rej,Mitman:2022qdl,Redondo-Yuste:2023seq}, in which the QQNM ratio was computed either using BHPT or by fitting NR data. We place these data on the plots in Fig.~\ref{fig:QQNM_44} using our formulae~\eqref{eqn:RPsi4_1} and \eqref{eqn:Rh_1}. For each reported value of the complex ratio $ {\cal R}^h_{\L, 2\mOne}$, Eqs.~\eqref{eqn:RPsi4_1} and \eqref{eqn:Rh_1} allow us to compute the (unique) corresponding value of $ z_{\lOne ;\mOne; \qOne}$, and thus from Eqs.~\eqref{eqn:AtoC1} and \eqref{eqn:AtoC2}, the corresponding ratio $C^-_{\lOne; \mOne; \qOne} / C^+_{\lOne; \mOne; \qOne}$, which we then plot in the figure~\cite{Bourg:2024jme}.

We first highlight the frequency-domain BHPT calculations by Ma \emph{et al}.~\cite{ma2024excitation} (pink cross) and Bucciotti et al.~\cite{Bucciotti:2024zyp} (light red plus), which should be perfectly consistent with our results (up to numerical error). As mentioned above, Ma \emph{et al.} report a ratio $\approx 0.137e^{-0.083i}$. Their computation of this result assumed that $\Psi_4^{(1)}$ is composed of a single regular frequency, $\lp_{2;2;0}^{(1)}$, so that $ A_{2;-2;1}^{(1)} = 0$. Equivalently, they assume $C^-_{2;2;0} = i C^+_{2;2;0}$. Starting from their result for $ \mathcal{R}^h_{44}$ and inverting Eqs.~\eqref{eqn:RPsi4_1} and \eqref{eqn:Rh_1} to obtain $C^-_{2;2;0}/C^+_{2;2;0}$, we exactly recover their input $C^-_{2;2;0}/C^+_{2;2;0}=i$.
Similarly, Bucciotti \emph{et al.} report a ratio $\approx 0.154e^{-0.068\,i}$. Their computation instead enforced even-parity modes only, $C^-_{2;2;0} = 0$, which is equivalent to setting $ A_{2;2;0}^{(1)} =  A_{2;-2;1}^{(1)}$. Again, we consistently recover $C^-_{2;2;0} = 0$ from their reported ratio.

Now turning our attention to the NR data, we note the reported values of $|| {\cal R}^h_{\L, 2\mOne}||$ are typically in the range $\approx 0.15$--$0.20$.
For the NR data we show in the figure, the simulated binary systems possess up-down symmetry; hence, the odd-parity modes should identically vanish for even values of $\lOne+\mOne$. Therefore, data for these systems should lie precisely at $C^-_{2;2;\qOne} = C^-_{2;0;\qOne} = 0$. For the case $(2;2;0)\times (2;2;0)\rightarrow (4,4)$, the reported NR values might appear relatively far away from this point. However, it is essential to realise
that the NR simulations of these binary inspirals yield spinning remnant BHs, while we use a Schwarzchild BH in computing the coefficient appearing in the relationship~\eqref{eqn:Psi42ndOrder++}.
One may therefore expect that this explains at least part of the deviation of the NR data away from the origin in the left panel of the figure.

To eliminate this uncertainty due to spin, in the right panel of Fig.~\ref{fig:QQNM_44}, we show results for the mode coupling $(2;0;0)\times (2;0;0)\rightarrow (4,0)$. Unlike the NR data in the left panel, the NR data we display in the right panel comes from simulations of head-on collisions~\cite{Cheung:2022rbm}, for which the remnant BHs are non-spinning. Nonetheless, the systems remain up-down symmetric, implying the NR data should precisely lie at $C^-_{2;0;\qOne} = 0$ if we neglect any numerical errors or systematic biases. This prediction is nearly within the lower bound for the error bars reported by the authors. Since these error bars mainly estimate contributions from the fitting, we expect the theoretical prediction $C^-_{2;0;\qOne} = 0$ to lie within the true error, which would also account for any other systematic errors. Importantly, the mean deviation from the origin is comparable to the deviation of the NR data shown in the left panel, suggesting that the spin of the remnant BH might not be the dominant cause of the deviation there.

\subsection{Two first-order mode excitations}
\label{sec:two modes}

We now generalize the results of the previous section by considering first-order perturbations generated by two regular modes, $(\lOne, \mOne, \qOne)$ and $(\lTwo, \mTwo, \qTwo)$, and their associated mirror modes, $(\lOne, -\mOne, \QOne)$ and $(\lTwo, -\mTwo, \QTwo)$. In particular, the first-order Weyl scalar now takes the form
\begin{align} \label{eqn:Psi4_Ansatz2}
    \tilde{\Psi}^{(1)}&(\tau,\sigma,\theta,\varphi) \nonumber\\
    &= A_{\lOne ;\mOne; \qOne}^{(1)} \tilde{\psi}_{\lOne ;\mOne; \qOne}^{(1)}(\sigma) e^{\lp_{\lOne; \mOne; \qOne}^{(1)} \tau} {}_{-2} Y_{\lOne \mOne}\nonumber \\
    & + A_{\lOne; -\mOne; \QOne}^{(1)}  \tilde{\psi}_{\lOne; -\mOne;\QOne}^{(1)}(\sigma)e^{\lp_{\lOne; -\mOne; \QOne}^{(1)} \tau} {}_{-2} Y_{\lOne -\mOne}, \nonumber \\
    &+ A_{\lTwo; \mTwo; \qTwo}^{(1)}  \tilde{\psi}_{\lTwo; \mTwo; \qTwo}^{(1)}(\sigma) e^{\lp_{\lTwo; \mTwo; \qTwo}^{(1)} \tau} {}_{-2} Y_{\lTwo \mTwo}  \nonumber \\
    & + A_{\lTwo; -\mTwo; \QTwo}^{(1)}  \tilde{\psi}_{\lTwo; -\mTwo;\QTwo}^{(1)}(\sigma)e^{\lp_{\lTwo; -\mTwo; \QTwo}^{(1)} \tau} {}_{-2} Y_{\lTwo -\mTwo},
\end{align}
where $ A_{\lOne; \mOne; \qOne}^{(1)}$, $ A_{\lOne; -\mOne; \QOne}^{(1)}$, $ A_{\lTwo; \mTwo; \qTwo}^{(1)}$, and $ A_{\lTwo; -\mTwo; \QTwo}^{(1)}$ are arbitrary complex amplitudes.

The second-order source again takes the generic form~\eqref{eq:SecOrdSource_FullDecomposition}.
As before, this sum over all the combinations $I, I'$ that contribute to a given $(\L, \M)$ mode at second order can be made more explicit. 
Specifically, since the second-order source depends on the first-order fields quadratically, a first class of contributions only involves products of one of the regular modes and its associated mirror mode.
For the $(\lOne, \mOne, \qOne)$ for example, there are altogether four such products, which can be schematically written as
\begin{align} \label{eqn:QQNM-pure_contribution-1}
    &(\lOne, \mOne, \qOne) \times (\lOne, \mOne, \qOne), \\
    &(\lOne, \mOne, \qOne) \times (\lOne, -\mOne, \QOne), \\
    &(\lOne, -\mOne, \QOne) \times (\lOne, \mOne, \qOne),\\
    &(\lOne, -\mOne, \QOne) \times (\lOne, -\mOne, \QOne), \label{eqn:QQNM-pure_contribution-3}
\end{align}
and similarly for terms only involving $(\lTwo, \mTwo, \qTwo)$ and its mirror mode. Since no mixing between the two mode excitations is involved, they will contribute to the second-order Weyl scalar as described in the previous section~\ref{sec:Single_mode_excitation}. Specifically, this part of the second-order source, and thus the corresponding QQNM solution, takes the same form as in Eqs. \eqref{eqn:2ndOrderSourceDecomposition_1} and \eqref{eqn:2ndOrderPsiDecomposition_1}. In particular, the amplitudes of the QQNM solution are precisely described by Eqs.~\eqref{eqn:Psi42ndOrder++}--\eqref{eqn:Psi42ndOrder--}. These no-mixing contributions will, therefore, not be discussed here any further. More interestingly, the second-order source is also composed of terms involving the mixing of the $(\lOne, \mOne, \qOne)$ and $(\lTwo, \mTwo, \qTwo)$ mode. Altogether, there are four different such contributions:
\begin{align} \label{eqn:QQNM-mixing_contribution-1}
    &(\lOne, \mOne, \qOne) \times (\lTwo, \mTwo, \qTwo), \\
    &(\lOne, \mOne, \qOne) \times (\lTwo, -\mTwo, \QTwo), \\ 
    &(\lOne, -\mOne, \QOne) \times (\lTwo, \mTwo, \qTwo), \\ 
    &(\lOne, -\mOne, \QOne) \times (\lTwo, -\mTwo, \QTwo). \label{eqn:QQNM-mixing_contribution-4}
\end{align}
In particular, $M$ takes four possible values, namely, $M= \mOne+\mTwo$, $\mOne-\mTwo$, $-\mOne+\mTwo$, and $-\mOne-\mTwo$.
The second-order source~\eqref{eq:SecOrdSource_FullDecomposition} now takes the form
\begin{align} \label{eqn:2ndOrderSourceDecomposition_2}
    & \tilde{\cal S}^{(2)}(\tau,\sigma,\theta, \varphi) = \sum_{\L} \nonumber \\
    \Bigl\{& \tilde{S}_{\L, \mOne+\mTwo}^{(2)}(\sigma) e^{(\lp_{\lOne; \mOne; \qOne}^{(1)}+\lp_{\lTwo; \mTwo; \qTwo}^{(1)}) \tau} {}_{-2} Y_{\L, \mOne+\mTwo} \nonumber\\
    &+ \tilde{S}_{\L, \mOne-\mTwo}^{(2)}(\sigma) e^{(\lp_{\lOne; \mOne; \qOne}^{(1)}+\lp_{\lTwo; -\mTwo; \QTwo}^{(1)}) \tau} {}_{-2} Y_{\L, \mOne-\mTwo} \nonumber \\
    &+ \tilde{S}_{\L, -\mOne+\mTwo}^{(2)}(\sigma) e^{(\lp_{\lOne; -\mOne; \QOne}^{(1)}+\lp_{\lTwo; \mTwo; \qTwo}^{(1)}) \tau} {}_{-2} Y_{\L, -\mOne+\mTwo} \nonumber \\
    &+ \tilde{S}_{\L, -\mOne-\mTwo}^{(2)}(\sigma) e^{(\lp_{\lOne; -\mOne; \QOne}^{(1)}+\lp_{\lTwo; -\mTwo; \QTwo}^{(1)}) \tau} {}_{-2} Y_{\L, -\mOne-\mTwo} \Bigr\},
\end{align}
where in the above, we ignored the single-mode-contributions of the form \eqref{eqn:QQNM-pure_contribution-1}--\eqref{eqn:QQNM-pure_contribution-3}.
Note that the above reduces to the decomposition~\eqref{eqn:2ndOrderSourceDecomposition_1} by taking $(\lOne,\mOne,\qOne) = (\lTwo,\mTwo,\qTwo)$.
The corresponding QQNM solution then inherits the same structure,
\begin{align} \label{eqn:2ndOrderPsiDecomposition_2}
    & \tilde{\Psi}^{(2)}(\tau,\sigma,\theta, \varphi) = \sum_{\L} \nonumber \\
    \Bigl\{& \tilde{\cal A}_{\L, \mOne+\mTwo}^{(2)}(\sigma) e^{(\lp_{\lOne; \mOne; \qOne}^{(1)} +\lp_{\lTwo; \mTwo; \qTwo}^{(1)}) \tau} {}_{-2} Y_{\L, \mOne+\mTwo} \nonumber\\
    &+ \tilde{\cal A}_{\L, \mOne-\mTwo}^{(2)}(\sigma) e^{(\lp_{\lOne; \mOne; \qOne}^{(1)} +\lp_{\lTwo; -\mTwo; \QTwo}^{(1)}) \tau} {}_{-2} Y_{\L, \mOne-\mTwo} \nonumber \\
    &+ \tilde{\cal A}_{\L, -\mOne+\mTwo}^{(2)}(\sigma) e^{(\lp_{\lOne; -\mOne; \QOne}^{(1)} +\lp_{\lTwo; \mTwo; \qTwo}^{(1)}) \tau} {}_{-2} Y_{\L, -\mOne+\mTwo} \nonumber \\
    &+ \tilde{\cal A}_{\L, -\mOne-\mTwo}^{(2)}(\sigma) e^{(\lp_{\lOne; -\mOne; \QOne}^{(1)} +\lp_{\lTwo; -\mTwo; \QTwo}^{(1)}) \tau} {}_{-2} Y_{\L, -\mOne-\mTwo} \Bigr\},
\end{align}
and the QQNM amplitudes depend on the first-order amplitude according to%
\begingroup%
\allowdisplaybreaks
\begin{align}
    \tilde{\cal A}_{\L, \mOne+\mTwo}^{(2)}(\sigma) &=  \tilde{a}_{\L, \mOne+\mTwo}  (\sigma)  A_{\lOne; \mOne; \qOne}^{(1)}  A_{\lTwo; \mTwo; \qTwo}^{(1)} \nonumber \\*
    &+  \tilde{b}_{\L, \mOne+\mTwo} (\sigma)  A_{\lOne; \mOne; \qOne}^{(1)} \bracket{ A_{\lTwo; -\mTwo; \QTwo}^{(1)}}^\star, \nonumber \\*
    &+  \tilde{c}_{\L, \mOne+\mTwo} (\sigma) \bracket{ A_{\lOne; -\mOne; \QOne}^{(1)}}^\star  A_{\lTwo; \mTwo; \qTwo}^{(1)}, \label{eqn:A_1} \\
     \tilde{\cal A}_{\L, \mOne-\mTwo}^{(2)}(\sigma) &=  \tilde{d}_{\L, \mOne-\mTwo}  (\sigma)  A_{\lOne; \mOne; \qOne}^{(1)}  A_{\lTwo; -\mTwo; \QTwo}^{(1)} \nonumber \\*
    &+  \tilde{e}_{\L, \mOne-\mTwo} (\sigma)  A_{\lOne; \mOne; \qOne}^{(1)} \bracket{ A_{\lTwo; \mTwo; \qTwo}^{(1)}}^\star \nonumber \\*
    &+  \tilde{f}_{\L, \mOne-\mTwo} (\sigma) \bracket{ A_{\lOne; -\mOne; \QOne}^{(1)}}^\star  A_{\lTwo; -\mTwo; \QTwo}^{(1)}, \\
     \tilde{\cal A}_{\L, -\mOne+\mTwo}^{(2)}(\sigma) &= (-1)^{\lOne+\lTwo+\L}  \Big[ \nonumber \\* 
    & \tilde{d}_{\L, -\mOne+\mTwo}^\star (\sigma)  A_{\lOne; -\mOne; \QOne}^{(1)}  A_{\lTwo; \mTwo; \qTwo}^{(1)} \nonumber \\*
    &+  \tilde{e}_{\L, -\mOne+\mTwo}^\star (\sigma)  A_{\lOne; -\mOne; \QOne}^{(1)} \bracket{ A_{\lTwo; -\mTwo; \QTwo}^{(1)}}^\star \nonumber \\*
    &+  \tilde{f}_{\L, -\mOne+\mTwo}^\star (\sigma) \bracket{ A_{\lOne; \mOne; \qOne}^{(1)}}^\star  A_{\lTwo; \mTwo; \qTwo}^{(1)} \Big], \\
     \tilde{\cal A}_{\L, -\mOne-\mTwo}^{(2)}(\sigma) &= (-1)^{\lOne+\lTwo+\L}  \Big[ \nonumber \\*
    & \tilde{a}_{\L, -\mOne-\mTwo}^\star (\sigma)  A_{\lOne; -\mOne; \QOne}^{(1)}  A_{\lTwo; -\mTwo; \QTwo}^{(1)} \nonumber \\*
    &+  \tilde{b}_{\L, -\mOne-\mTwo}^\star (\sigma)  A_{\lOne; -\mOne; \QOne}^{(1)} \bracket{ A_{\lTwo; \mTwo; \qTwo}^{(1)}}^\star \nonumber \\*
    &+  \tilde{c}_{\L, -\mOne-\mTwo}^\star (\sigma) \bracket{ A_{\lOne; \mOne; \qOne}^{(1)}}^\star  A_{\lTwo; -\mTwo; \QTwo}^{(1)} \Big] \label{eqn:A_4}.
\end{align}
\endgroup
The factor $(-1)^{\lOne+\lTwo+\L}$ generalises the factor $(-1)^\L$ appearing in Eq.~\eqref{eqn:Psi42ndOrder--}.
Note that the above formulae reduce to those appearing in Eqs.~\eqref{eqn:Psi42ndOrder++}--\eqref{eqn:Psi42ndOrder--} when setting $(\lOne,\mOne,\qOne) = (\lTwo,\mTwo,\qTwo)$. In particular, $ a_{\L, 2\mOne} =  \tilde{a}_{\L, 2\mOne}$, $ b_{\L, 2\mOne} =  \tilde{b}_{\L, 2\mOne} +  \tilde{c}_{\L, 2\mOne}$, $ d_{\L 0} =  \tilde{f}_{\L 0}^\star =  \tilde{e}_{\L 0}$, and $ c_{\L 0} =  \tilde{d}_{\L 0}$.
We provide the above coefficients $\tilde{a},\tilde{b},\tilde{c},\tilde{d},\tilde{e},\tilde{f}$ evaluated at $\scri^+$ in Table~\ref{table:abcdef_2}, for a selection of first-order mode excitations $(\lOne,\mOne,\qOne)$ and $(\lTwo,\mTwo,\qTwo)$.

\begin{table}[tb] \centering
\begin{tabular}{ c c| c c } 
\toprule
$(\lOne,\mOne,\qOne)$ & & \multicolumn{2}{c}{} \\
$\times (\lTwo,\mTwo,\qTwo)$ & $\L$ & \multicolumn{2}{c}{at $\scri^+ \, (\sigma=0)$} \\
\midrule
\multirow{5}{4em}{$(2,0,0)$ $\times (2,0,2)$} & \multirow{5}{2.5em}{\ \ \ $2$}
& $\tilde{a}_{20}$ & $ 2.38635 - 4.44036\,i $ \\
&& $\tilde{b}_{20}$ & $ -0.0129848 - 0.0958113\,i $ \\
&& $\tilde{c}_{20}$ & $ -0.294480 - 0.182355\,i $ \\
&& $\tilde{d}_{20}$ & $ (-3.39753-0.715685\,i)*10^5 $ \\
&& $\tilde{e}_{20}$ & $ (-0.755970+10.9458\,i)*10^3 $ \\
&& $\tilde{f}_{20}$ & $ (-3.23480-0.967544\,i)*10^5 $ \\
\midrule
\multirow{5}{4em}{$(2,0,0)$ $\times (4,0,0)$} & \multirow{5}{2.5em}{\ \ \ $4$}
& $\tilde{a}_{40}$ & $ 0.634554 - 1.51403\,i $ \\
&& $\tilde{b}_{40}$ & $ -0.0121117 - 0.0205308\,i $ \\
&& $\tilde{c}_{40}$ & $ -0.0137579 + 0.499035\,i $ \\
&& $\tilde{d}_{40}$ & $ 0.0464563 - 0.000954264\,i $ \\
&& $\tilde{e}_{40}$ & $ 0.00595266 + 0.0143701\,i $ \\
&& $\tilde{f}_{40}$ & $ -0.309025 + 0.201908\,i $ \\
\midrule
\multirow{5}{4em}{$(2,2,0)$ $\times (2,2,2)$} & \multirow{5}{2.5em}{\ \ \ $4$}
& $\tilde{a}_{44}$ & $ 17.6382 + 15.8957\,i $ \\
&& $\tilde{b}_{44}$ & $ 1.0879 + 1.19354\,i $ \\
&& $\tilde{c}_{44}$ & $ 0.723966 + 0.99525\,i $ \\
&& $\tilde{d}_{40}$ & $ 455.669 + 638.857\,i $ \\
&& $\tilde{e}_{40}$ & $ 17.7772 - 18.2189\,i $ \\
&& $\tilde{f}_{40}$ & $ 427.197 + 657.056\,i $ \\
\midrule
\multirow{5}{4em}{$(2,2,0)$ $\times (3,2,0)$} & \multirow{5}{2.5em}{\ \ \ $4$}
& $\tilde{a}_{44}$ & $ 0.516926 - 1.68240\,i $ \\
&& $\tilde{b}_{44}$ & $ 0.177091 + 0.19585\,i $ \\
&& $\tilde{c}_{44}$ & $ -0.731825 - 0.458078\,i $ \\
&& $\tilde{d}_{40}$ & $ 0.00139546 - 0.000780128\,i $ \\
&& $\tilde{e}_{40}$ & $ -0.00204365 - 0.00678924\,i $ \\
&& $\tilde{f}_{40}$ & $ -0.0909477 - 0.0168087\,i $ \\
\midrule
\multirow{5}{4em}{$(2,2,0)$ $\times (3,3,0)$} & \multirow{5}{2.5em}{\ \ \ $5$}
& $\tilde{a}_{55}$ & $ 24.9740 + 7.42180\,i $ \\
&& $\tilde{b}_{55}$ & $ -0.569561 - 0.342118\,i $ \\
&& $\tilde{c}_{55}$ & $ 3.11586 + 1.86907\,i $ \\
&& $\tilde{d}_{5,-1}$ & $ -0.000164949 + 0.000973865\,i $ \\
&& $\tilde{e}_{5,-1}$ & $ 0.0000891774 + 0.000267646\,i $ \\
&& $\tilde{f}_{5,-1}$ & $ 0.00375721 - 0.00126218\,i $ \\
\bottomrule
\end{tabular}
\caption{Values of the coefficients at $\scri^+$ appearing in Eqs.~\eqref{eqn:A_1}--\eqref{eqn:A_4} for a selection of mode numbers $((\lOne,\mOne,\qOne) \times (\lTwo,\mTwo,\qTwo))_{LM}$. The values for $\tilde{a}_{55},\tilde{b}_{55}, \tilde{c}_{55}$ are consistent with those in \cite[Table I and II]{Khera:2023oyf}.}
\label{table:abcdef_2}
\end{table}

\begin{figure*}
    \includegraphics[scale=0.66]{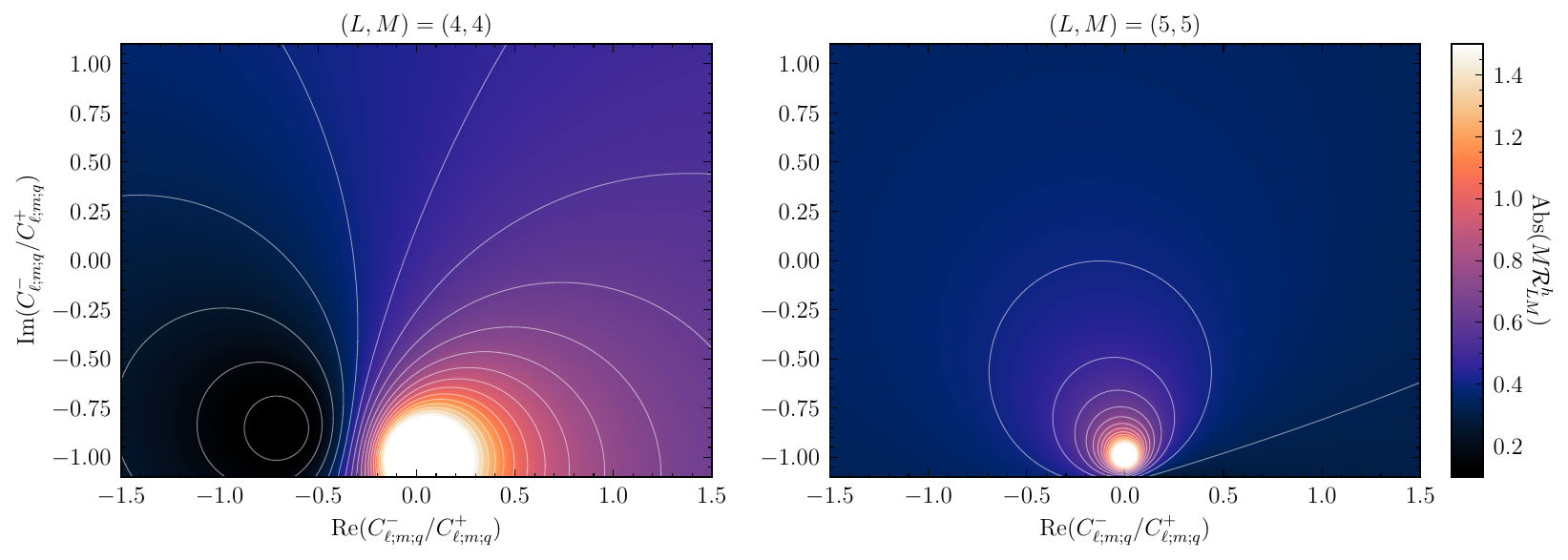}  
    \caption{Contour plot of the ratio $ \mathcal{R}^h_{\lOne+\lTwo, \mOne+\mTwo}$ as a function of the odd-to-even complex amplitude ratio $C^-_{\l ;\m; \q} / C^+_{\l; \m; \q}$, with the solid lines indicating values of constant amplitude of the QQNM ratio.
    Since the ratio $ \mathcal{R}^h_{\L, \mOne+\mTwo}$ depends on two complex numbers, $ z_{\lOne; \mOne; \qOne}$ and $ z_{\lTwo; \mTwo; \qTwo}$, we have chosen to set $ z_{\l; \m; \q} :=  z_{\lOne; \mOne; \qOne} =  z_{\lTwo; \mTwo; \qTwo}$.
    {\em Left Panel:} Quadratic mode $(\L,\M) = (4, 4)$ from linear modes $(\lOne, \mOne, \qOne) = (2,2,0)$ and $(\lTwo, \mTwo, \qTwo) = (3,2,0)$.
    {\em Right Panel:} Quadratic mode 
    $(\L,\M) = (5, 5)$ from linear modes $(\lOne, \mOne, \qOne) = (2,2,0)$ and $(\lTwo, \mTwo, \qTwo) = (3,3,0)$.}
    \label{fig:QQNM_Mixing_44_55}
\end{figure*}

We can then define the ratio
\begin{align} \label{eqn:QQNMRatio_Psi4Reg_TwoMode}
     \mathcal{R}^{ \tilde{\Psi}}_{\L, \mOne+\mTwo}(\sigma) &:= \frac{\left( \tilde{\Psi}^{(2)}\right)_{\L, \mOne+\mTwo}}{\left( \tilde{\Psi}^{(1)}\right)_{\lOne; \mOne; \qOne} \times \left( \tilde{\Psi}^{(1)}\right)_{\lTwo; \mTwo;\qTwo}} \nonumber \\
    &= \frac{ \tilde{\cal A}_{\L, \mOne+\mTwo}^{(2)}(\sigma)}{  \tilde{\cal A}_{\lOne; \mOne; \qOne}^{(1)}(\sigma) \times  \tilde{\cal A}_{\lTwo; \mTwo; \qTwo}^{(1)}(\sigma)}.
\end{align}
This ratio depends on the progenitor system's up-down (a)symmetry via two complex amplitudes, constructed from the ratio of even-to-odd mode parity of the two-mode excitation at first order,
\begin{align} \label{eqn:RPsi4_Mixing_1}
     \mathcal{R}^{ \tilde{\Psi}}_{\L, \mOne+\mTwo}(\sigma) &=  \tilde{a}_{\L, \mOne+\mTwo} (\sigma) \nonumber \\
    &+  \tilde{b}_{\L, \mOne+\mTwo} (\sigma)  z_{\lOne; \mOne; \qOne} \nonumber \\
    &+  \tilde{c}_{\L, \mOne+\mTwo} (\sigma)  z_{\lTwo; \mTwo; \qTwo},
\end{align}
where we define
\begin{align} \label{eqn:z12_Mixing}
     z_{\lOne; \mOne; \qOne} &:= \frac{\bracket{ A_{\lOne ;-\mOne; \QOne}^{(1)}}^\star}{ A_{\lOne; \mOne; \qOne}^{(1)}}, \\
     z_{\lTwo; \mTwo; \qTwo} &:= \frac{\bracket{ A_{\lTwo; -\mTwo; \QTwo}^{(1)}}^\star}{ A_{\lTwo; \mTwo; \qTwo}^{(1)}}.
\end{align}

In Fig.~\ref{fig:QQNM_Mixing_44_55}, we show a contour plot of the ratio $ \mathcal{R}^h_{\lOne+\lTwo, \mOne+\mTwo}$ as a function of $C^-_{\l;\m; \q} / C^+_{\l ;\m; \q}$ for different values of $(\lOne,\mOne, \qOne) \times (\lTwo,\mTwo, \qTwo)$. As the ratio $ \mathcal{R}^h_{\L, \mOne+\mTwo}$ depends in general on two complex numbers~\eqref{eqn:RPsi4_Mixing_1}, we have chosen to set $ z_{\lOne; \mOne; \qOne} =  z_{\lTwo; \mTwo; \qTwo}$, or equivalently, $C^-_{\lOne; \mOne; \qOne}/C^+_{\lOne; \mOne; \qOne}=C^-_{\lTwo; \mTwo; \qTwo}/C^+_{\lTwo; \mTwo; \qTwo}$.

\subsection{Frequency versus time-domain codes comparison} 
\label{sec:comparison}

\begin{figure*}[t] \centering
\includegraphics[width= \linewidth]
{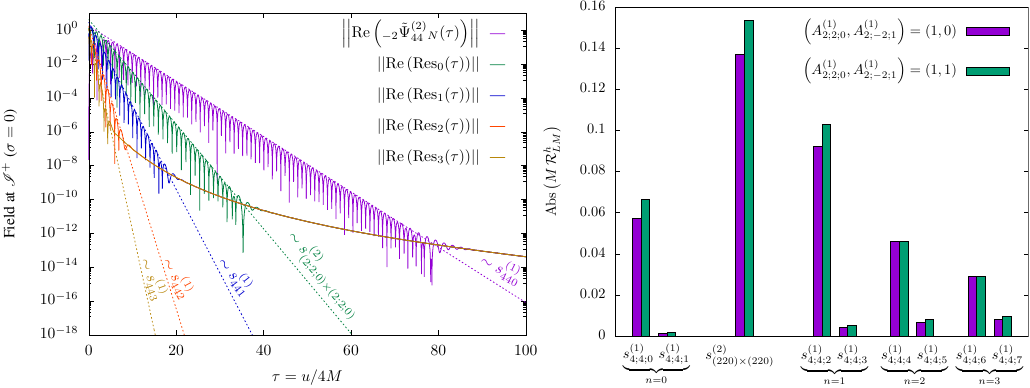}

  \caption{{\em Left Panel:} Dynamics at second order in perturbation theory driven by a source quadratic in the singly excited QNMs. The signal is dominated by the slowest decaying mode $\lp^{(1)}_{4;4;0}$, and $\lp^{(1)}_{4;4;1}$ ($\lp^{(1)}_{440}$ -purple line) until late times, when the power-law tail dominates. Subdominant residual dynamics require filtering the signal via Eq.~\eqref{eq:res}. With second-order QNM amplitudes and QQNM amplitudes predicted by the frequency-domain infrastructure, the residuals unveil the QQNM decay with frequency $\lp^{(2)}_{(2;2;0)\times(2;2;0)}$ (green line), as well as higher overtones $ \lp^{(1)}_{4;4;2}$, $\lp^{(1)}_{4;4;3}$ ($\lp^{(1)}_{441}$-blue line), $\lp^{(1)}_{4;4;4}$, $\lp^{(1)}_{4;4;5}$ ($\lp^{(1)}_{442}$-orange line) and $\lp^{(1)}_{4;4;6}$, $\lp^{(1)}_{4;4;7}$ ($\lp^{(1)}_{443}$-dark yellow). The filtered signal also unveils the contribution from the tail decay at earlier times.
  {\em Right Panel:} Spectral analysis of the second-order field at future null infinity arising from a first-order dynamics containing only the regular QQM (purple), or regular and mirror modes (green). The QQNM is the most excited mode at second order, though its time-domain contribution diminishes rapidly due to faster decay compared to the fundamental QNM, cf.~left panel. The contribution from the regular QNM ($\q=0,2,4,6$) consistently dominates across overtone orders ($\n=0,1,2,3$), but accounting for mirror modes ($\q=1,3,5,7$) is critical for accurate filtering of the time-domain signal via Eq.~\eqref{eq:res}.
  }    \label{fig:Amplitudes_histogram}
\end{figure*}

Having isolated and analyzed individual pieces of the second-order waveform using our frequency-domain infrastructure, we now examine the complete second-order time-domain signal using our time-domain code. This serves to establish confidence in both methods. More importantly, it allows us to directly explore the relative contributions of various QQNMs and linear second-order QNMs in a complete time-domain signal. 

We consider a second-order time evolution driven by a source term resulting from the quadratic coupling of the regular linear QNM $\lp_{2;2;0}^{(1)}$, i.e. Eq.~\eqref{eqn:Psi4_Ansatz1} with $(\lOne,\mOne,\qOne) = (2,2,0)$ and $\left( A_{2; 2; 0}^{(1)}, A_{2; -2; 1}^{(1)} \right) = (1,0)$. 
In this particular case, the first-order data yields a second-order dynamics with $L=4$ and we concentrate on the dynamics in the sector $M=+4$.

The right panel of Fig.~\ref{fig:Amplitudes_histogram} shows the numerical time-domain evolution $ \tilde{\Psi}_{\l\m}^{(2)}{}_{N}(\tau,0)$ at $\scri^+$ driven purely by the quadratic second-order source resulting from this pure QNM solution at first order. As discussed in Sec.~\ref{sec:second order dynamics}, the full second-order signal consists of contributions from both the QNMs and tail decay, as well as QQNMs, resulting, respectively, from the homogenous and the particular solution to the wave equation. Therefore, in this particular case, the time evolution contains oscillations with QNM frequencies $\lp^{(1)}_{4;4;\q}$ and $\lp^{(1)}_{4;4;\Q}$, as well as the QQNM frequency $\lp^{(2)}_{(2;2;0)\times(2;2;0)}$, and an eventual power law decay. 

The purple line displays the full signal with all the aforementioned contributions, as obtained from our time-domain code. Because the fundamental modes $(\n=0)$
\begin{equation}
\lp^{(1)}_{4;4;0} \approx -0.377 + 3.237\,i, \quad
\lp^{(1)}_{4;4;1}\approx -0.377 - 3.237\,i,
\end{equation}
decay more slowly than the QQNM
\begin{equation}
\lp^{(2)}_{(2; 2; 0)\times(2; 2; 0)} \approx -0.712 + 2.989\,i,
\end{equation}
the second-order contribution is not directly identified in the time evolution, without a careful control of the subdominant signatures.

To unveil the contribution of faster-decaying modes, we exploit the frequency-domain framework that {\em predicts} the corresponding second-order QNM and QQNM excitation amplitudes, cf.~Eqs.~\eqref{eq:2ndOrderQNM} and \eqref{eqn:QQNM_equation} respectively. This approach allows us to introduce the residual
\begin{equation}
\label{eq:res}
{\rm Res}_{\upsilon}(\tau) =  \tilde{\Psi}_{\l\m}^{(2)}{}_{N}(\tau,0) - \sum_{\upsilon'=0}^\upsilon   A_\upsilon \, e^{\tau \lp_\upsilon},
\end{equation}
with $\upsilon$ an overall label associated with frequencies ordered according to their exponential decaying rate; i.e.,
\begin{align}
\upsilon' &= 0 \longleftrightarrow \lp^{(1)}_{4;4;0} \text{  and } \lp^{(1)}_{4;4;1},\\
\upsilon' &= 1 \longleftrightarrow \lp^{(2)}_{(2; 2; 0)\times(2; 2; 0)},\\
\upsilon' &= 2 \longleftrightarrow \lp^{(1)}_{4;4;2} \text{  and } \lp^{(1)}_{4;4;3},
\end{align}
etc. In this way, we can now filter the contribution from the fundamental modes $\upsilon = 0$, and observe the signal decaying with the expected QQNM frequency (green line) up to $\tau \sim 45$, when the tail phase takes over. Consistently, the QQNM contribution  $\upsilon = 1$ can be also filtered, with the blue line unveiling the contribution from the overtone $\n=1$,
\begin{equation}
\lp^{(1)}_{4;4;2}\approx -1.137 + 3.186\,i, \quad \lp^{(1)}_{4;4;3}\approx -1.137 - 3.186\,i.
\end{equation}
Systematically, the left panel of Fig.~\ref{fig:Amplitudes_histogram} shows the results up to $\upsilon  = 3$, where we further observe the underlying contribution from the linear overtone $\n = 2$, 
\begin{equation}
\lp^{(1)}_{4; 4; 4}\approx -1.920 + 3.091\,i, \quad \lp^{(1)}_{4; 4; 5}\approx -1.920 - 3.091\,i,   
\end{equation}
and the linear overtone $\n = 3$, 
\begin{equation}
\lp^{(1)}_{4, 4, 6}\approx -2.736 + 2.959\,i, \quad \lp^{(1)}_{4, 4, 7}\approx -2.736 - 2.959\,i,    
\end{equation}
for $\upsilon = 2$ (orange line) and $\upsilon = 3$ (dark yellow line), respectively.

We recall that the results in the left panel of Fig.~\ref{fig:Amplitudes_histogram} correspond to a configuration $( A_{2;2;0}, A_{2;-2;1})=(1,0)$, i.e. a dynamic containing only the regular mode excited at first order, and its effect on the second-order dynamics. A similar analysis could be performed for the configuration $( A_{2;2;0},  A_{2;-2;1})=(1,1)$, with results being qualitatively identical to the ones discussed so far. 

To better quantify the difference between both configurations, Fig.~\ref{fig:Amplitudes_histogram}'s right panel presents a histogram for the spectral analysis of the second-order field at $\scri^{+}$. The $x$~axis is ordered according to the modes' exponential decaying rate; i.e., with $\upsilon = 0, \dots, 3$. The $y$~axis shows the relative ratio between the second-order QNM or QQNM amplitudes to the first-order QNM amplitude, with a normalisation according to Eq.~\eqref{eqn:Rh_1}.

One observes that the QQNM $\lp^{(2)}_{(2; 2; 0)\times(2; 2; 0)}$ is indeed the most excited mode. As explained, however, its contribution is not straightforwardly observed in the time-domain signal, as the QQNM contribution quickly becomes subdominant because its decay rate is faster than the fundamental modes $\lp^{(1)}_{4;4;0}$ and $\lp^{(1)}_{4;4;1}$. In fact, the contribution from the first overtones $\lp^{(1)}_{4;4;2}$, $\lp^{(1)}_{4;4;3}$ is also bigger than the fundamental mode. Higher overtones then show a smaller contribution to the signal. 

Interestingly, the second-order source does not excite the regular and mirror modes evenly. Even though we observe a predominance of the regular mode $\q=0,2,4$ and $6$ at all orders in the overtones $n=0,1,2$ and $3$, correctly accounting for the contribution from the mirror modes $\q=1,3,5$ and $7$ is crucial for the correct and complete filtering in the time-domain signal via Eq.~\eqref{eq:res}.

\section{Conclusion}\label{sec:Conclusion}

In this paper, we have provided a complete description of quadratic QNMs in Schwarzschild spacetime, fleshing out the details of our letter~\cite{Bourg:2024jme} and complementing broadly similar recent studies in Refs.~\cite{ma2024excitation,Bucciotti:2024jrv,Khera:2024yrk}.

Although BHPT calculations have only recently begun to focus on second-order effects during ringdown, there are now several established methods of performing such calculations. This is particularly the case for QQNMs in the frequency domain. We now take the opportunity to highlight the differences between these various methods and the novel aspects of our treatment. 

References~\cite{ma2024excitation,Bucciotti:2024jrv} present two methods in the frequency domain, using ordinary $(t,r,\theta,\phi)$ coordinates rather than hyperboloidal slicing. Both begin from scalar field equations that we may write generically as $\hat{\mathcal{O}}\Psi^{(2)} = S^{(2)}$ -- the Regge-Wheeler and Zerilli equations in Ref.~\cite{Bucciotti:2024jrv}, and the Teukolsky equation in Ref.~\cite{ma2024excitation}. For two coupled first-order QNMs, this becomes 
\begin{equation}\label{eq:OPsi=Swewt}
 \hat{\mathcal{O}}\Psi^{(2)} = S^{(2)}_{\omega}e^{-i\omega t},
\end{equation}
where $\omega=\omega_{\l_1;\m_1;\q_1}+\omega_{\l_2;\m_2;\q_2}$ is the sum of the two QNM frequencies. By adopting an ansatz $\Psi^{(2)}=\Psi^{(2)}_\omega e^{-i\omega t}$ and factoring the exponential out of the equation, Refs.~\cite{ma2024excitation,Bucciotti:2024jrv} reduce Eq.~\eqref{eq:OPsi=Swewt} to a frequency-domain equation of the form %
\begin{equation}\label{eq:OwPsiw=Sw}%
\hat{\mathcal{O}}_{\omega}\Psi^{(2)}_{\omega}=S^{(2)}_{\omega}. 
\end{equation}

In the case of the Teukolsky equation, the source $S^{(2)}_{\omega}$ is singular at the boundaries due to the fact that every constant-$t$ slice runs between the bifurcation sphere and spatial infinity, where the first-order QNM solutions are singular. Reference~\cite{ma2024excitation} overcomes this problem by extending the radial coordinate into the complex plane and integrating Eq.~\eqref{eq:OwPsiw=Sw} over a complex radial contour that picks out the QQNM amplitude, taking inspiration from Refs.~\cite{Zimmerman:2014aha,Green:2022htq,London:2023aeo}.

In the case of the Regge-Wheeler and Zerilli equations, the blow-up of the QNM solutions at the boundaries is subsidiary to the fact that the Regge-Wheeler and Zerilli master variables are innately ill-behaved at second order~\cite{Brizuela:2009qd}. Reference~\cite{Bucciotti:2024jrv} resolves these singularities, following earlier work~\cite{Nakano:2007cj}, using a subtraction method (akin to a puncture method in self-force theory~\cite{Cunningham:2024dog}). They find a singular quadratic combination of first-order fields, call it $\Psi^{S}_\omega$, that cancels the divergences in the source. Defining the residual field $\Psi^{R}_\omega:=\Psi^{(2)}_\omega-\Psi^{S}_\omega$ and moving $\Psi^{S}_\omega$ to the right-hand side of Eq.~\eqref{eq:OwPsiw=Sw}, they then solve for the well-behaved field $\Psi^{R}_\omega$. When the metric perturbation $h^{(2)}_{ab}$ is reconstructed from the master scalars, the singularities cancel, and the physical GW is obtained from the residual field.

Reference~\cite{Khera:2024yrk} follows an approach most similar to our own, using hyperboloidal slicing and spectral methods. In this instance, the coefficient $S^{(2)}_\omega$ in $S^{(2)}_\omega e^{-i\omega\tau}$ is well behaved at the boundaries. The ansatz $\Psi^{(2)}=\Psi^{(2)}_\omega e^{-i\omega\tau}$ then leads to a well-behaved version of Eq.~\eqref{eq:OwPsiw=Sw} that can be solved with regular boundary conditions.

Our approach goes some distance beyond the above by considering an initial-value problem, restricting our analysis to the future domain of dependence of some slice $\tau=0$. This is the physically relevant problem following the formation of the ringing BH. By taking a Laplace transform of $\hat{\mathcal{O}}\Psi^{(2)}=S^{(2)}_\omega e^{-i\omega\tau}$, we are able, in principle, to calculate the complete solution $\Psi^{(2)}$, not only the QQNM contribution. This includes the contribution from initial data and the second-order linear QNMs, for example. In this setting, in Sec.~\ref{sec:QQNMs}, we have shown how to isolate and compute the linear QNMs and QQNMs within the full solution.

Our use of a Laplace transform also allows us to directly link our frequency-domain results to simulations in the time domain that solve the same initial-value problem. In Sec.~\ref{sec:Results}, we have illustrated how access to frequency-domain results allows us to cleanly deconstruct the time-domain signal obtained from our independent time-domain code. This synergy has significant advantages, both in accuracy and in physical insight, over either time-domain evolutions or frequency-domain calculations on their own. Given only a time-domain evolution, the various QNM, QQNM, and other contributions to the GW can only be extracted through fitting~\cite{Ripley:2020xby,Redondo-Yuste:2023seq,Zhu:2024rej}; given only frequency-domain calculations, one cannot easily be sure of whether one has included all significant contributions (e.g., tail effects, as we return to below, or the contributions from the high-frequency arc).

Given the number of perturbative second-order ringdown calculations now available in the literature, our paper's main new contribution is its overarching formalism. Consequently, our presentation has focused on the technical details of the formalism and its implementation. However, we have also further fleshed out the key result in our earlier letter~\cite{Bourg:2024jme}: the dependence of the QQNM/QNM ratio on the ratio between even- and odd-parity content in the first-order QNM.

There are several natural followups. One is the extension of our calculations to a Kerr background. Our methods apply almost without change in Kerr: the separability of the Teukolsky equation, the method of metric reconstruction to calculate the second-order source, the hyperboloidal and Laplace-transform infrastructure, and the pole structure of the Green's function and of the second-order source all carry over directly. The only new aspect is the more complicated mode coupling in the second-order source, which arises from the lack of spherical symmetry and causes any single (spheroidal) $\l\m$ first-order mode to generate a formally infinite number of second-order source modes. This can be dealt with by (i) avoiding a full separation of variables at second order, instead working with partial differential equations in $r$ and $\theta$~\cite{Assaad2025}, or (ii) numerically (or semi-analytically~\cite{Spiers:2024src}) projecting the source into individual spheroidal-harmonic modes, relying on the decay of high-$\l$ modes to neglect them in the sum. Such calculations in Kerr spacetime would provide a valuable independent verification of Ref.~\cite{Khera:2024yrk}'s results.

Another natural follow-up would be to include the contributions of branch cuts in the Green's function. These lead to late-time power-law tails in the solution and can have nontrivial behaviour at intermediate times~\cite{Casals:2015nja}. Their relevance in nonlinear ringdowns has been studied in Refs.~\cite{Cardoso:2024jme,DeAmicis:2024eoy} in the time domain, and our approach would provide an important complement in the frequency domain.

A third avenue of further study would be to explore quadratic effects at the BH horizon. These are known to exhibit the same behaviour as at future null infinity~\cite{Khera:2023oyf,RibesMetidieri:2025lxr}. Our hyperboloidal framework, which cleanly links ${\cal H}^+$ to $\scri^+$, would be a natural setting to examine these relationships.

\begin{acknowledgments}
This work makes use of the Black Hole Perturbation Toolkit. 
PB acknowledges support from the Dutch Research Council (NWO) with file number OCENW.M.21.119. 
RPM acknowledges support from the Villum Investigator program supported by the VILLUM Foundation (grant no. VIL37766) and the DNRF Chair program (grant no. DNRF162) by the Danish National Research Foundation. 
The Center of Gravity is a Center of Excellence funded by the Danish National Research Foundation under grant No. 184.
AS acknowledges support from the STFC Consolidated Grant no. ST/V005596/1.
AP acknowledges the support of a Royal Society University Research Fellowship and a UKRI Frontier Research Grant (as selected by the ERC) under the Horizon Europe Guarantee scheme (Grant No. EP/Y008251/1). 

\end{acknowledgments}

\appendix

\section{The Newman-Penrose and Geroch-Held-Penrose formalisms}
\label{app:np-ghp}

The Teukolsky equations in Sec.~\ref{sec:BHPT_Review} are derived within the NP or GHP formalism, as is the metric reconstruction method reviewed in \cref{app:MetricReconstruction}. In this appendix, we briefly review the relevant elements of these formalisms.

\subsection{The Newman--Penrose formalism}\label{sec:NP}
The NP formalism~\cite{Newman:1961qr} is a tetrad formalism for studying general relativistic spacetimes. A tetrad consists of four orthonormal null vector fields, which form a basis throughout a spacetime.
The NP basis vectors are labelled 
\begin{align}
    e_{[a]}^a:=\{e_{[1]}^a,e_{[2]}^a,e_{[3]}^a,e_{[4]}^a\} := \{l^a,n^a,m^a,\bar{m}^a\},
\end{align}
where indices in square brackets denote tetrad indices. The vectors $l^a$ and $n^a$ are real, whereas $m^a$ is complex, and overbars denote complex conjugation. Conventionally, for a positive metric signature, the orthonormal relationship of the tetrad takes the form
\begin{equation}\label{eq:orthonormal}
l^an_a=-1, \quad m^a\bar{m}_a=1;
\end{equation}
all other contractions of tetrad vectors and co-vectors are zero. Following Eq.~\eqref{eq:orthonormal}, the metric can be expressed as~\cite{Newman:1961qr},
\begin{equation}
  g_{ab} = -2 l_{(a} n_{b)} +2 m_{(a} \bar{m}_{b)} .
\end{equation}

The NP formalism uses Ricci rotation coefficients to express the connection~\cite{Newman:1961qr, waldbook},
\begin{equation}
\gamma_{[c][a][b]}=e_{[c]}^ke_{[a]k;i}e_{[b]}^i.\label{eq:RRC}
\end{equation}
There are 24 independent Ricci rotation coefficients. In the NP formalism, these are expressed as 12 independent complex scalars, known as spin coefficients:
\begin{align}
\kappa&=-\gamma_{[3][1][1]},\ \tau=-\gamma_{[3][1][2]},  \ \sigma=-\gamma_{[3][1][3]}, \notag \\
\rho&=-\gamma_{[3][1][4]},\ \varpi=-\gamma_{[2][4][1]},\ \nu=-\gamma_{[2][4][2]}, \notag
\\  \mu&=-\gamma_{[2][4][3]},\  
\lambda=-\gamma_{[2][4][4]}, \notag
\\  \ \epsilon&=-\frac{\gamma_{[2][1][1]}+\gamma_{[3][4][1]}}{2},  \ \gamma=-\frac{\gamma_{[2][1][2]}+\gamma_{[3][4][2]}}{2}, \notag
\\ 
\beta&=-\frac{\gamma_{[2][1][3]}+\gamma_{[3][4][3]}}{2},  \ \alpha=-\frac{\gamma_{[2][1][4]}+\gamma_{[3][4][4]}}{2}. \label{eq:spincoeffs}
\end{align}

Curvature quantities are also represented using scalars in the NP formalism. The vacuum curvature, contained in the ten degrees of freedom of the Weyl tensor, is expressed using five complex scalars, known as Weyl scalars:
\begin{align}
\Psi_0 = C_{[1][3][1][3]}&, \
\Psi_1 = C_{[1][3][1][2]},  \
\Psi_2 = C_{[1][3][4][2]},  \ \notag \\
\Psi_3 &= C_{[1][2][4][2]},  \
\Psi_4 = C_{[2][4][2][4]}  .
\end{align}
%
The covariant derivative is also expressed using tetrad components, as
\begin{align}
&D
:=l^a \nabla_a,
\ \ \ \ 
\Delta
:=n^a \nabla_a
\notag \\ 
&\delta
:= m^a \nabla_a,
\ \ \ \ \overline{\delta} 
:=\overline{m}^a \nabla_a.
\label{paradervs}
\end{align}

One convenience of the NP formalism is that many of the curvature scalars and spin coefficients can be set to zero in Kerr spacetime. This is because the Kerr spacetime admits two principal null vectors (two pairs of principal null vectors which coincide)~\cite{Petrov:2000bs, chandrabook}; that is, Kerr is Petrov type D. A tetrad can be chosen such that $l^a$ and $n^a$ are tangent to the principal null directions. In such tetrads, four of the Weyl scalars and four spin coefficients vanish~\cite{chandrabook}:
\begin{align}\label{petrovconds1}
\Psi_0=0,\;\Psi_1=0&,\;\Psi_3=0,\;\Psi_4=0, \\
\kappa=0,\; \lambda=0&,\; \nu=0, \;\sigma=0.\label{petrovconds2}
\end{align}

In this paper, we choose to work in the Kinnersley tetrad~\cite{Kinnersley:1969zza}. Expressed in Schwarzschild coordinates $(t,r,\theta,\varphi)$, the tetrad is
\begin{align} \label{eq:tetrad,Kinn,t}
    l^\alpha &= \frac{1}{f} \biggl(1,f,0,0  \biggr),\\
    n^\alpha &= \frac{1}{2} \biggl(1,-f,0,0  \biggr),\\
    m^\alpha &= \frac{1}{r\sqrt{2}} \left(0,0,1,\frac{i}{\sin \theta}  \right),
\end{align}
where $f=1-\frac{2M}{r}$. The Kinnersley tetrad is principal-null-direction aligned, obeying \cref{petrovconds1,petrovconds2}, and also satisfies $\epsilon=0$.

\subsection{Perturbative expansion}

Similarly to \cref{eq:metricexpansion}, we expand $\Psi_4$ as follows:
\begin{align}
    \Psi_4=\Psi_4^{(0)}+\varepsilon\Psi_4^{(1)}+\varepsilon^2\Psi_4^{(2)}+\ldots+ \varepsilon^n \Psi_4^{(n)}+\ldots
\end{align}
This is in contrast to other works on QQNMs~\cite{Campanelli:1998jv, Loutrel:2020wbw, Ripley:2020xby, ma2024excitation, Khera:2024yrk}, which define their perturbative expansion similarly to Campanelli and Lousto~\cite{Campanelli:1998jv}, as
\begin{multline}
    \Psi_4=\Psi_4^{(0)CL}+\varepsilon\Psi_4^{(1)CL}+\frac{\varepsilon^2}{2}\Psi_4^{(2)CL}\\
    +\ldots+ \frac{\varepsilon^n}{n!} \Psi_4^{(n)CL}+\ldots
\end{multline}
The background and first-order perturbations are identical in both expansions. However, at second order and beyond, they differ as $\Psi_4^{(n)}=\frac{1}{n!} \Psi_4^{(n)CL}$.

Conventionally, the ${(0)}$ superscript on background quantities is dropped for legibility. In the remainder of this paper, we adopt such a notation; that is, $\Psi_4:=\Psi_4^{(0)}$.

\subsection{The Geroch-Held-Penrose formalism}\label{subsec:GHP}

The GHP formalism refines the NP formalism by specialising to tetrads that are aligned with principal null directions and then working with expressions that are covariant under the transformations within that class.
In a Petrov type D spacetime, $l^a$ and $n^a$ are chosen to point along the two principal null directions. The remaining freedoms are associated with spin and boost transformations, which are isomorphic to the group of multiplications by a complex number, $\vartheta$~\cite{Geroch:1973am}. To represent the conserved quantities corresponding to the spin and boost transformations, GHP weights $p$ and $q$ are introduced. A quantity $f$ with weights $f \GHPwt \{p,q\}$, transforms under a spin and boost transformation as
  \begin{equation}
      f \xrightarrow{} \vartheta^p \bar{\vartheta}^q f .
  \end{equation}
GHP weights $p$ and $q$ can be equated to the spin weight $s=\frac{1}{2}(p-q)$ and boost weight $b=\frac{1}{2}(p+q)$. The $[b,s]$ weights of the tetrad vectors are, $[1,0]$, $[-1,0]$, $[0,1]$, and $[0,-1]$ for $l^a$, $n^a$, $m^a$, and $\bar{m}^a$ respectively. Similarly, the $\{p,q\}$ weights of the tetrad vectors are $\{1,1\}$, $\{-1,-1\}$, $\{1,-1\}$, and $\{-1,1\}$ respectively. 

Additionally, there is a further freedom to interchange $l^a$ and $n^a$, denoted by a prime operation. The prime operation affects the GHP weights according to $f' \circeq \{-p,-q\}$. Complex conjugation affects the weights as $\bar{f}\circeq \{q,p\}$. Half of the NP spin coefficients are relabelled using the prime notation:
\begin{align}
\kappa^\prime:=-\nu, \ \sigma^\prime:=-\lambda, \ \rho^\prime:=-\mu, \notag \\ \tau^\prime:=-\pi, \ \beta^\prime:=-\alpha, \ \epsilon^\prime:=-\gamma.
\end{align}
The GHP weights of the spin coefficients (and their primes) follow directly from the weights of the tetrad vectors (using Eqs.~\eqref{eq:spincoeffs}); e.g.,
\begin{align}
\kappa\circeq\{3,1\},  \ \sigma\circeq\{3,-1\}, \ \rho\circeq\{1,1\},\  \tau\circeq\{1,-1\}.
\end{align}
The spin coefficients $\epsilon$, $\epsilon^\prime$, $\beta$, and $\beta^\prime$ do not have well-defined weights. Similarly, the NP derivative operators do not have well-defined weights. GHP found by combining these poorly defined weighted quantities, one can produce derivative operators with well-defined weights,
\begin{align}
\thorn &= (D-p\epsilon-q\bar{\epsilon}), \\
\thorn^\prime &= (\Delta+p\epsilon^\prime+q\bar{\epsilon}^\prime), \\
\edth &= (\delta-p\beta+q\bar{\beta}^\prime), \\
\edth^\prime &= (\bar{\delta}+p\beta^\prime-q\bar{\beta}). \label{eq:GHPderivatives} 
\end{align}

Two clear advantages of the GHP formalism are that the equations are more condensed than in NP form, and they offer a straightforward consistency check by checking that the weights of an equation are consistent.

\section{Metric reconstruction and quadratic source}
\label{app:MetricReconstruction}

The bulk of the paper deals with solving the second-order Teukolsky equation~\eqref{eq:masterTeuk} (with $i=2$ and $s=-2$), with the quadratic source term~\eqref{eq:masterreducedTeuk s=-2}. Constructing that source term requires the complete first-order metric perturbation $h^{(1)}_{ab}$. In this appendix, we summarize how $h^{(1)}_{ab}$ is reconstructed from a first-order Weyl scalar and the resulting structure of the quadratic source. 

In vacuum, $h^{(1)}_{ab}$ can be obtained from $\Psi_0^{(1)}$ or $\Psi_4^{(1)}$ using the Chrzanowski-Cohen-Kegeles (CCK) metric reconstruction procedure~\cite{cohen1975space, Chrzanowski:1975wv, Kegeles:1979an,Wald:1973wwa}.
There are two forms of CCK metric reconstruction associated with different gauge choices for $h^{(1)}_{ab}$: the \textit{ingoing radiation gauge} (IRG), for which a CCK metric perturbation is regular at the BH future horizon, and the \textit{outgoing radiation gauge} (ORG), for which a CCK metric perturbation is regular at future null infinity. We are motivated to use the ORG CCK metric reconstruction as, unlike the IRG CCK metric reconstruction\footnote{The metric perturbations in the IRG can produce a regular second-order source if it is produced using an alternative method to CCK metric reconstruction~\cite{Loutrel:2020wbw, Ripley:2020xby, hollands2024metric}.}, it ensures the second-order source~\eqref{eq:masterreducedTeuk s=-2} is regular in hyperboloidal coordinates; see Fig.~\ref{fig:2ndOrderSource}.

\subsection{Weyl scalar to Hertz potential}

In CCK metric reconstruction, the first-order metric perturbation is computed from the Hertz potential $\Phi^{(1)}$. The Hertz potential in ORG is a solution to the vacuum Teukolsky equation~\eqref{eq:teuk},
\begin{align}\label{eq:teukolskyPhi}
    \hat \O'[\Phi^{(1)}]=0 \, .
\end{align}
As a result, it is related to $\Psi_0^{(1)}$ or $\Psi_4^{(1)}$ via so-called radial and angular \textit{inversion} relations, respectively. The angular inversion is the simplest of the two as it only involves angular and time derivatives, which become algebraic at the level of spherical-harmonic modes in the frequency domain. In Schwarzschild, it is given by
\begin{align}\label{eq:angularInversion}
\Psi_0^{(1)} &= \frac{1}{4} \eth^4 (\Phi^{(1)})^\star + \frac{3}{4} M r^{-4} \mathcal{L}_\xi \Phi^{(1)},
\end{align}
where the $\mathcal{L}_\xi$ is the differential operator
\begin{equation}
\mathcal{L}_\xi = -r (-\rho' \pth + \rho \pth') - \frac{p+q}{2} r \Psi_2^{(1)}.
\end{equation}
In hyperboloidal coordinates, this simply reduces to a time derivative,
\begin{equation}
\mathcal{L}_\xi = \frac{1}{\lambda} \partial_\tau.
\end{equation}

Following the discussion in Sec.~\ref{section:Regular_and_mirror_modes}, we decompose the Weyl scalar ${}_2 \Psi^{(1)} = \Psi_0^{(1)}$ as
\begin{align}
    {}_{2} \Psi^{(1)} &= {}_{2} \mathcal{Z}\sum_{\l,\m,\q} {}_2 A_{\l; \m;\q}^{(1)} \, {}_{2} \tilde{\psi}_{\l ;\m;\q}^{(1)} e^{\lp_{\l ;\m;\q}^{(1)} \tau}.
\end{align}
From the angular inversion~\eqref{eq:angularInversion}, the Hertz potential $\Phi^{(1)}$ will have a similar decomposition, namely,
\begin{align}
    \Phi^{(1)} &= r^4 {}_{2} \mathcal{Z}\sum_{\l,\m,\q} {}_{2} \tilde{\phi}_{\l ;\m;\q}^{(1)} e^{\lp_{\l ;\m;\q}^{(1)} \tau}.
\end{align}

Plugging these decompositions into the angular inversion~\eqref{eq:angularInversion}, we obtain algebraic equations,
\begin{align}
        16 \, {}_2 A_{\l; \m;\q}^{(1)} {}_2 \tilde{\psi}_{\l ;\m;\q}^{(1)} &= (-1)^\m (\l-1)_4 \bracket{{}_2 \tilde{\phi}_{\l; -\m;\Q}^{(1)}}^\star \nonumber\\
        &\quad + 12 M \lambda^{-1} \lp_{\l; \m;\q}^{(1)} \; {}_2 \tilde{\phi}_{\l; \m;\q}^{(1)},
\end{align}
where $(\l-1)_4 = (\l-1) \l (\l+1)(\l+2)$ denotes the Pochhammer symbol.
Solving for $\tilde{\phi}_{\l; \m;\q}^{(1)}$, we find
\begin{align} \label{eq:Weyl_to_Hertz}
    {}_2 \tilde{\phi}_{\l ;\m;\q}^{(1)} &= \frac{16 \, {}_2 \tilde{\psi}_{\l ;\m ;\q}^{(1)}} {(\l-1)_4^2 - \bracket{12 M \lambda^{-1} \lp_{\l ;\m;\q}^{(1)}}^2} \nonumber\\
    &\quad\times\Bigl[(-1)^\m (\l-1)_4 \bracket{{}_2 A_{\l; -\m ;\Q}^{(1)}}^\star \nonumber\\
    &\qquad\qquad\qquad - 12 M \lambda^{-1} \lp_{\l; \m ;\q}^{(1)}\; {}_2 A_{\l; \m ;\q}^{(1)}\Bigr],
\end{align}
where we made use of relations~\eqref{QNM_frequency_new_relation} and \eqref{Q_relation}.

The above relation relates the modes of the Hertz potential to the modes of $\Psi_0^{(1)}$, which depend in particular on the amplitudes ${}_2 A_{\l;\m;\q}^{(1)}$. They are themselves related to the amplitudes for $\Psi_4^{(1)}$.
Specifically, consider the decomposition of the Weyl scalar $ \Psi^{(1)} = r^4 \Psi_4^{(1)}$,
\begin{align}
    {}_{-2} \Psi^{(1)} &= {}_{-2} \mathcal{Z}\sum_{\l,\m,\q} {}_{-2} A_{\l; \m;\q}^{(1)} \, {}_{-2} \tilde{\psi}_{\l; \m;\q}^{(1)} e^{\lp_{\l; \m;\q}^{(1)} \tau}.
\end{align}
Via the Teukolsky-Starobinsky relation~\cite{Pound:2021qin}, and making use of our choice of normalisation of the QNM eigenfunctions (cf. Eq.~\eqref{eq:QNM_func_norm}), we find the following algebraic relation between the excitation amplitudes ${}_{\pm 2} A_{\l; \m;\q}^{(1)}$:
\begin{align} \label{eq:QNM_ampltidue_relation_between_opposite_spin}
    1024 \frac{M^8}{\lambda^8} \bracket{\lp_{\l;\m;\q}^{(1)}}^4 & {}_2 A_{\l; \m;\q}^{(1)} = (\l-1)_4 {}_{-2} A_{\l; \m;\q}^{(1)} \nonumber \\
    &+ 12 (-1)^\m \lp_{\l;\m;\q}^{(1)} \frac{M}{\lambda}  \bracket{{}_{-2} A_{\l ;-\m;\Q}^{(1)}}^\star.
\end{align}
This relation implies that solving for ${}_{-2}\Psi^{(1)}$ or ${}_2\Psi^{(1)}$ is effectively equivalent: given the amplitudes of one, we can obtain the amplitudes of the other.

\subsection{Hertz potential to metric perturbation}

The first-order metric perturbation in the ORG is expressed in terms of the Hertz potential via the formula
\begin{align} \label{eqn:hToHertz}
h_{ab}^{(1)} &= 2 {\rm Re}(\mathcal{S}_4^\dagger \Phi^{(1)})_{ab},
\end{align}
where
\begin{align}
\label{eqn:hhatORG}
(\mathcal{S}_4^\dagger \Phi^{(1)})_{ab} &= -\frac{1}{2} n_a n_b {\eth'}^2 \Phi^{(1)} \nonumber \\
&- \frac{1}{2} \bar{m}_a \bar{m}_b (\pth'-\rho')(\pth'+3\rho') \Phi^{(1)} \nonumber \\
&+ \frac{1}{4} (n_a \bar{m}_b + n_b \bar{m}_a) [\pth' \eth' + \eth' (\pth' + 3 \rho')] \Phi^{(1)}.
\end{align}
Hence, there are three independent, non-zero components in the metric perturbation:
\begin{align}
h_{ll}^{(1)} &= (\mathcal{S}_4^\dagger \Phi^{(1)})_{ll} + (\mathcal{S}_4^\dagger \Phi^{(1)})_{ll}^\star, \\
h_{lm}^{(1)} &= (\mathcal{S}_4^\dagger \Phi^{(1)})_{lm}, \\
h_{mm}^{(1)} &= (\mathcal{S}_4^\dagger \Phi^{(1)})_{mm},
\end{align}
and their complex conjugations. This perturbation satisfies the traceless ORG gauge conditions
\begin{equation}\label{eq:ORG}
    h^{(1)}_{ab}n^b = 0 = h^{(1)}_{ab}g^{ab}.
\end{equation}

In Schwarzschild, one can easily relate the modes of the metric perturbation to the modes of the Hertz potential.
We decompose the tetrad components of the metric perturbation as
\begin{equation}
    h_{[a][b]}^{(1)} = \sum_{\l,\m,\q} (h_{[a][b]}^{(1)})_{\l; \m;\q} e^{\lp_{\l ;\m;\q}^{(1)} \tau} {}_{\NPs} Y_{\l\m}(\theta, \varphi),
\end{equation}
where $\NPs$ is the spin weight of the tetrad component $h^{(1)}_{[a][b]}$.

Similarly, as was done for the Weyl scalars and source term in Eqs.~\eqref{eq:ConfMasterFunc} and \eqref{eq:ConfTeukMast_Gen}, we define the regularized version of the tetrad components $h^{(1)}_{[a][b]}$,
\begin{align}
(h_{ll}^{(1)})_{\l ;\m;\q} &= \frac{M^2 {}_2 \mathcal{Z}}{\sigma^2} (\tilde{h}_{ll}^{(1)})_{\l ;\m;\q}, \\
(h_{lm}^{(1)})_{\l ;\m;\q} &= \frac{M^2 {}_2 \mathcal{Z} (1-\sigma)}{\sigma^3} (\tilde{h}_{lm}^{(1)})_{\l ;\m;\q} ,\\
(h_{mm}^{(1)})_{\l ;\m;\q} &= \frac{M^2 {}_2 \mathcal{Z} (1-\sigma)^2}{\sigma^4} (\tilde{h}_{mm}^{(1)})_{\l; \m;\q},
\end{align}
where we recall that the tetrad components $h^{(1)}_{[a][b]}$ are in Kinnersley.

By construction, the dimensionless quantities $(\tilde{h}_{[a][b]}^{(1)})_{\l; \m;\q}$ are regular quantities everywhere (for our asymptotically flat, horizon-regular metric perturbations in the ORG), including at the endpoints $\sigma=0$ and $\sigma=1$. They are related to $\tilde{\phi}_{\l ;\m;\q}^{(1)}$ by
\begingroup%
\allowdisplaybreaks%
\begin{subequations}\label{eq:Hertz_to_MP}
\begin{align}
    (\tilde{h}_{ll}^{(1)})_{\l; \m;\q} &= -\mu_\l^1 \mu^2_\l \bracket{ \tilde{\phi}_{\l; \m;\q}^{(1)} + (-1)^\m \bracket{\tilde{\phi}_{\l; -\m;\Q}^{(1)}}^\star}, \\
    (\tilde{h}_{lm}^{(1)})_{\l ;\m;\q} &= -\frac{\mu^2_\l}{\sqrt{2}} \Biggl[\bracket{3 \sigma + 4 \frac{M}{\lambda} (1+\sigma) \lp_{\l; \m;\q}^{(1)}} \nonumber\\*
    &\qquad\qquad + \sigma^2 \partial_\sigma \Biggr] \tilde{\phi}_{\l; \m;\q}^{(1)}, \\
    (\tilde{h}_{mm}^{(1)})_{\l; \m;\q} &= -\frac{1}{2} \Biggl[ 4 \Bigl(\sigma^2 + \frac{M}{\lambda} \sigma (4+5 \sigma) \lp_{\l ;\m;\q}^{(1)} \nonumber\\*
    &\qquad\quad + 4 \frac{M^2}{\lambda^2} (1+\sigma)^2 \bracket{\lp_{\l ;\m;\q}^{(1)}}^2 \Bigr) \nonumber \\*
    &\qquad\quad + \sigma^2 \bracket{6 \sigma+8\frac{M}{\lambda}(1+\sigma) \lp_{\l ;\m;\q}^{(1)}} \partial_\sigma \nonumber \\*
    &\qquad\quad + \sigma^4 \partial^2_\sigma \Biggr] \tilde{\phi}_{\l; \m;\q}^{(1)}, \\
    (\tilde{h}_{l \bar{m}}^{(1)})_{\l; \m;\q} &= (-1)^{\m+1} \Bigl[(\tilde{h}_{lm}^{(1)})_{\l; -\m;\Q}\Bigr]^\star, \\
    (\tilde{h}_{\bar{m} \bar{m}}^{(1)})_{\l ;\m;\q} &= (-1)^\m \Bigl[(\tilde{h}_{mm}^{(1)})_{\l; -\m;\Q}\Bigr]^\star,
\end{align}
\end{subequations}
\endgroup
where $\mu_{\l}^{\NPs} := \sqrt{(\l+\NPs)(\l-\NPs+1)}$.

We can express the metric perturbation modes~\eqref{eq:Hertz_to_MP} directly in terms of the modes of $\Psi^{(1)}_0$ using the relationship~\eqref{eq:Weyl_to_Hertz}.

\subsection{Quadratic source}
\label{app:source calculation}

The quadratic source for ${}_{-2}\Psi^{(2)}$~\eqref{eq:masterreducedTeuk s=-2}, has the schematic form
\begin{equation}
    {}_{-2}S^{(2)} \sim \nabla\nabla\left(h^{(1)}\nabla\nabla h^{(1)}+\nabla h^{(1)}\nabla h^{(1)}\right).
\end{equation}
Each $\l\m$ mode of the source, as appearing in Eq.~\eqref{eq:ConfTeukMast_Gen_lm}, then takes the form
\begin{equation}\label{eq:Slm coupling}
    {}_{-2}\tilde S^{(2)}_{\l\m} = \sum_{\substack{\NPs_1\l_1\m_1\\ \NPs_2\l_2\m_2}} {}_{-2}\tilde S^{\NPs_1\l_1\m_1 \NPs_2\l_2\m_2}_{\l\m}\left[{}_{\NPs_1}h^{(1)}_{\l_1\m_1},\;{}_{\NPs_2}h^{(1)}_{\l_2\m_2}\right],
\end{equation}
where we have restricted to the ORG and defined 
\begin{align}
{}_{0}h^{(1)}_{\l\m}&=(h^{(1)}_{ll})_{\l\m}, \nonumber\\ {}_{1}h^{(1)}_{\l\m}&=(h^{(1)}_{lm})_{\l\m}, \quad\ 
{}_{-1}h^{(1)}_{\l\m}=(h^{(1)}_{l\bar m})_{\l\m},  \nonumber\\ 
{}_{2}h^{(1)}_{\l\m}&=(h^{(1)}_{mm})_{\l\m}, \quad 
{}_{-2}h^{(1)}_{\l\m}=(h^{(1)}_{\bar m\bar m})_{\l\m}.
\end{align}
${}_{-2}\tilde S^{\NPs_1\l_1\m_1 \NPs_2\l_2\m_2}_{\l\m}$ is a bilinear differential operator involving $\tau$ and $\sigma$ derivatives, satisfying the standard properties for coupling of angular momenta: ${}_{-2}\tilde S^{\NPs_1\l_1\m_1 \NPs_2\l_2\m_2}_{\l\m}$ vanishes unless
\begin{align}
 |\l_1-\l_2| \leq \l\leq \l_1+\l_2 \quad\text{and}\quad \m_1+\m_2 = \m. 
\end{align}
Note we do not have $\NPs_1+\NPs_2=\NPs$ because the operator $\mathcal{S}_4$ in Eq.~\eqref{eq:masterreducedTeuk s=-2} involves spin-raising and -lowering operations that ensure each term has net spin weight $-2$.

Equation~\eqref{eq:Slm coupling} is given explicitly in the \emph{PerturbationEquations} package~\cite{warburton2023perturbationequations}, up to the rescaling in Eq.~\eqref{eq:ConfTeukMast_Gen}. In our case, with $ h^{(1)}_{\l, \m}=\sum_{\l, \m, \q} h^{(1)}_{\l; \m;\q} e^{\lp_{\l ;\m;\q}^{(1)} \tau}$, each term in Eq.~\eqref{eq:Slm coupling} becomes proportional to $\exp\left[ \tau\left(\lp_{\l_1;\m_1;\q_1}^{(1)}+\lp_{\l_2;\m_2;\q_2}^{(1)}\right)\right]$. This, then, is the source as it appears in Eq.~\eqref{eq:SecOrdSource_TimeDep}. Each term in the source is expressed directly in terms of QNM modes of $\Psi^{(1)}_0$ using Eqs.~\eqref{eq:Hertz_to_MP} and ~\eqref{eq:Weyl_to_Hertz}.

\section{Choice of second-order field variable: $\psi^{(2)}_{4}$ versus $\psi^{(2)}_{4L}$} 
\label{sec:psi4 vs psi4L}

At second order, there are two common forms of the Teukolsky equation for each spin weight $\pm2$. As explained in Sec.~\ref{sec:BHPT_Review}, we follow Refs.~\cite{Green:2019nam,Spiers:2023cip} in using the \emph{reduced} second-order Teukolsky variable $\Psi^{(2)}_{4L}$, satisfying the reduced second-order Teukolsky equation~\eqref{eq:teuk}. A common alternative~\cite{Campanelli:1998jv, ma2024excitation, Zhu:2024rej, Loutrel:2020wbw, Ripley:2020xby} is to use the complete second-order Weyl scalar, 
\begin{align}
\label{eq:psi4_2ndorder}
\Psi_4^{(2)}&=\T_4[h^{(2)}_{ab}]+\delta^2\Psi_4[h^{(1)}_{ab},e^{(1)a}_{[a]}] = \Psi_{4L}^{(2)} +\Psi_{4Q}^{(2)},
\end{align}
where $\Psi_{4Q}^{(2)}:=\delta^2\Psi_4[h^{(1)}_{ab},e^{(1)a}_{[a]}]$ is quadratic in $h^{(1)}_{ab}$ and involves products of $h^{(1)}_{ab}$ and the tetrad perturbations $e^{(1)a}_{[a]}$. Reference~\cite{Spiers:2023cip} compares the two choices of field variable, $\Psi^{(2)}_{4L}$ and $\Psi^{(2)}_{4}$. Here, we provide additional details.

Unlike $\Psi^{(2)}_{4L}$, the field $\Psi^{(2)}_{4}$ depends on the choice of perturbed tetrad $e^a_{[a]}+\varepsilon e^{(1)a}_{[a]}$ through the operator $\delta^2\Psi_4$. 
The second-order Weyl scalar, $\Psi^{(2)}_{4}$, satisfies the Campanelli-Lousto form of the second-order Teukolsky equation~\cite{Campanelli:1998jv}, which, in vacuum, is \begin{align}\label{eq:LCT}
\hat{\mathcal{O}}'[\Psi^{(2)}_4] &= S^{(2)}_{CL}\Bigl[h^{(1)}_{ab},e^{(1)a}_{[a]}\Bigr],
\end{align}
where
\begin{align}\label{eq:LCT-Source}
S^{(2)}_{CL}&\Bigl[h^{(1)}_{ab},e^{(1)a}_{[a]}\Bigr]\nonumber\\
 &= \Bigl[\bar{d}_3^{(0)} (\edth -\tau)^{(1)} - \bar{d}_4^{(0)}(\pth-\rho)^{(1)}\Bigr]\Psi_4^{(1)} \nonumber\\
 &\quad-\Bigl[\bar{d}_3^{(0)} (\pth' +4\mu )^{(1)} - \bar{d}_4^{(0)}(\edth'+4\pi)^{(1)}\Bigr]\Psi_3^{(1)}\nonumber\\
&\quad +3\Big[ \bar{d}_3^{(0)} \nu^{(1)} - \bar{d}_4^{(0)}\lambda^{(1)}\Big]\Psi_2^{(1)} \nonumber \\
&\quad+ 3\Big[(\bar{d}_3-3\pi)^{(1)}\nu^{(1)}-(\bar{d}_4 - 3\mu)^{(1)}\lambda^{(1)}\Big] \Psi_2^{(0)},
\end{align}
with $\bar{d}_3 :=\edth'-\bar{\tau}+4\pi$ and $\bar{d}_4 :=\pth'+4\mu+\bar{\mu}$.
The source here depends on the tetrad perturbations through the dependence on $\Psi^{(1)}_2$ and $\Psi^{(1)}_3$, perturbations to the spin coefficients, and perturbations to the GHP derivatives $\pth$, $\pth'$, $\edth$, and $\edth'$.

\subsection{Specialization to the Chrzanowski tetrad and outgoing radiation gauge}

Reference~\cite{ma2024excitation} solves the Campanelli-Lousto equation~\eqref{eq:LCT} with two specializations: the Chrzanowski perturbed tetrad~\cite{Chrzanowski:1975wv} and the ORG.

If we restrict to the Chrzanowski tetrad, then the tetrad perturbations become explicit functions of the metric perturbations:
\begingroup%
\allowdisplaybreaks%
\begin{align}
  l^{(1)a} &= \frac{1}{2} h_{ll} n^a, \\
  n^{(1)a} &= \frac{1}{2} h_{nn} l^{a} + h_{nl} n^a, \\
  m^{(1)a} &= -\frac{1}{2} h_{mm} \bar{m}^{a} - \frac{1}{2} h_{m \bar{m}} m^a  + h_{m l} n^a +  h_{m n} l^a, \\
  \bar{m}^{(1)a} &= -\frac{1}{2} h_{\bar{m}\bar{m}} m^{a} - \frac{1}{2} h_{m \bar{m}} \bar{m}^a  + h_{\bar{m} l} n^a + h_{\bar{m} n} l^a.
\end{align}%
\endgroup%
The operator $\delta^2\Psi_4$ is given in Ref.~\cite{2nd-order-notebook} for this choice of tetrad. 

If we further restrict to the ORG (satisfying $h^{(1)}_{ab}n^b=0$), then we find $\Psi_{4Q}^{(2)}=0$. This can be shown by enforcing \cref{eq:ORG} in the expression for $\delta^2\Psi_4$ in the Mathematica notebook~\cite{2nd-order-notebook} associated with Ref.~\cite{Spiers:2023cip}. That is, $\Psi_{4}^{(2)}=\Psi_{4L}^{(2)}$ when in the Chrzanowski perturbed tetrad and ORG. 

It follows that with these choices, the two variants of the second-order Teukolsky equation~\eqref{eq:teuk,eq:LCT}, should be identical, though the equivalence of \cref{eq:teuk,eq:LCT} in the ORG and Chrzanowski perturbed tetrad has not been explicitly shown algebraically.

\subsection{General case}

Even if we do not specialize the perturbed tetrad or gauge, $\Psi^{(2)}_{4L}$ and $\Psi^{(2)}_{4}$ contain identical GW content~\cite{Campanelli:1998jv,Spiers:2023cip}. 
We infer this from the fact that, in a good gauge, $\Psi^{(2)}_{4Q}$ decays at least as $1/r^2$ because it is quadratic in first-order perturbations. It should then follow that the coefficients of $1/r$ are the same in $\Psi^{(2)}_{4L}$ as in $\Psi^{(2)}_{4}$. 

However, there is some subtlety in this conclusion, related to the choice of initial data. Suppose we solve Eq.~\eqref{eq:LCT} for $\Psi^{(2)}_4$, with some choice of initial data, and Eq.~\eqref{eq:teuk} for $\Psi^{(2)}_{4L}$, also with some choice of initial data. The difference between the two fields, $\Delta\Psi^{(2)}_4:=\Psi^{(2)}_{4}-\Psi^{(2)}_{4L}$, is then a solution to 
\begin{equation}
     \hat{\cal O}'[\Delta\Psi^{(2)}_{4}] = S_{CL} - {\cal S}_4[\delta^2 G_{ab}].
\end{equation}
A generic solution to this equation will contain nonzero GW content, suggesting that the GWs in $\Psi^{(2)}_{4}$ differ from those in $\Psi^{(2)}_{4L}$. However, as alluded to above,
the quadratic field 
$\Psi^{(2)}_{4Q}$, which decays as $1/r^2$ and therefore contains no GWs, is \emph{also} a particular solution to the same equation:
\begin{equation}
     \hat{\cal O}'[\Psi^{(2)}_{4Q}] = S_{CL} - {\cal S}_4[\delta^2 G_{ab}],
\end{equation}
as can be confirmed by direct computation. Therefore, we are guaranteed that $\Delta\Psi^{(2)}_4 = \Psi^{(2)}_{4Q}$, and that $\Delta\Psi^{(2)}_4$ contains no GW content, if $\Delta\Psi^{(2)}_4$ has the same initial data as $\Psi^{(2)}_{4Q}$. 

Hence, we can say that $\Psi^{(2)}_{4}$ and $\Psi^{(2)}_{4L}$ will contain the same GW content \emph{if} their initial data are related by 
\begin{multline}
\left\{\Psi^{(2)}_{4},\dot\Psi^{(2)}_{4}\right\}\Bigr|_{\Sigma_0}=\left\{\Psi^{(2)}_{4L},\dot\Psi^{(2)}_{4L}\right\}\Bigr|_{\Sigma_0}\\
+\left\{\Psi^{(2)}_{4Q},\dot\Psi^{(2)}_{4Q}\right\}\Bigr|_{\Sigma_0}, 
\end{multline}
where $\Sigma_0$ is the initial-data surface, and an overdot denotes a derivative orthogonal to this surface. Since $\Psi^{(2)}_{4Q}$ is a specified function of first-order quantities, one can always enforce this relationship in principle when comparing between two calculations.

\section{Relation between the QQNM ratio of $\Psi_4$ and gravitational wave strain}
\label{Appendix:SecondOrderQNMAmplitude}

Our results in Sec.~\ref{sec:Results} are for the (conformally rescaled) Weyl scalar perturbations $\Psi^{(1)}_{4}$ and $\Psi^{(2)}_{4L}$, yielding a QQNM ratio in terms of the amplitudes of these perturbations at $\scri^+$. For comparison with calculations in the literature, we convert to the QQNM ratio defined from the GW strain using Eq.~\eqref{eqn:Rh_1}. Here, we derive that relationship.

As given in Eq.~\eqref{eqn:QQNMRatio_Psi4Reg_TwoMode},
for arbitrary QNM modes $I_1 = (\l_1, \m_1, \q_1)$ and $I_2 = (\l_2, \m_2, \q_2)$ at first order, and an arbitrary QQNM mode excitation $(\L,\M)$ at second order, the quadratic coupling coefficient ${\cal R}^v_{(I_1 \times I_2)_{\L\M}}$\footnote{again, we suppress the spin-$-2$ index} relates the QQNM mode amplitude $ \mathcal{A}_{(I_1 \times I_2)_{\L\M}}$ to its parent first-order amplitudes $ \mathcal{A}_{I_1}$ and $ \mathcal{A}_{I_2}$ via the formula
\begin{equation}
 \mathcal{A}^{(2),v}_{(I_1 \times I_2)_{\L\M}} =  {\cal R}^{v}_{(I_1 \times I_2)_{\L\M}}  \mathcal{A}^{(1),v}_{I_1}  \mathcal{A}^{(1),v}_{I_2}.
\end{equation}
In the above, the superscript $v$ can refer to several different quantities evaluated at null infinity, such as $r \Psi_4$, the regularised Weyl scalar~\eqref{eq:ConfMasterFunc} with spin $\NPs=-2$, ${}_{-2} \tilde{\Psi}$ as is the case in Eqs.~\eqref{eqn:QQNMRatio_Psi4Reg_OneMode} and \eqref{eqn:QQNMRatio_Psi4Reg_TwoMode}, or the gravitational wave strain $h$.

In our work, we use the Kinnersley tetrad, which has the following asymptotic form in the usual Boyer-Lindquist coordinates:
\begin{align}
l^\mu_{K} &\stackrel{r \to \infty}{\longrightarrow} \partial_t + \partial_r, \\
n^\mu_{K} &\stackrel{r \to \infty}{\longrightarrow} \frac{1}{2} \bracket{\partial_t - \partial_r}, \\
m^\mu_{K} &\stackrel{r \to \infty}{\longrightarrow} \frac{1}{\sqrt{2} r} \bracket{\partial_\theta + \frac{i}{\sin \theta}\partial_\varphi}.
\end{align}
In most NR simulations (such as in the SpEC code~\cite{Boyle:2019kee}), a rescaled tetrad is used instead,
\begin{align}
l^\mu_{\rm SpEC} &\stackrel{r \to \infty}{\longrightarrow} \frac{1}{\sqrt{2}} \bracket{\partial_t + \partial_r}, \\
n^\mu_{\rm SpEC} &\stackrel{r \to \infty}{\longrightarrow} \frac{1}{\sqrt{2}} \bracket{\partial_t - \partial_r}, \\
m^\mu_{\rm SpEC} &\stackrel{r \to \infty}{\longrightarrow} m^\mu_K.
\end{align}
As a result, the Weyl scalar in the Kinnersley and SpEC tetrads differ by an overall factor,
\begin{equation}
    \Psi_4^{\rm SpEC} = 
    2 \Psi_4^K.
\end{equation}
It then follows that the corresponding QQNM ratios are related by
\begin{equation} \label{eq:RPsiSpeC_To_PsiK}
     {\cal R}^{r\Psi_4^{\rm SpEC}}_{(I_1 \times I_2)_{\L\M}} = \frac{1}{2}  {\cal R}^{r \Psi_4^K}_{(I_1 \times I_2)_{\L\M}}.
\end{equation}

On the other hand, from Eq.~\eqref{eq:ConfMasterFunc}, the modes of $r \Psi_4^K$ are related to the modes of $ \tilde{\Psi}$ by
\begin{equation}
(r \Psi_4^K)_{\l;\m;\q} = \frac{\lambda^2}{8 M^3} \bracket{1-\frac{2M}{r}}^2  \tilde{\psi}_{\l;\m;\q}.
\end{equation}
Therefore, at null infinity, we have
\begin{equation}
(r \Psi_4^K)_{\l;\m;\q} \stackrel{r \to \infty}{\longrightarrow} \frac{\lambda^2}{8 M^3}  \tilde{\psi}_{\l;\m;\q}.
\end{equation}
So,
\begin{align}
 {\cal R}^{r \Psi_4^K}_{(I_1 \times I_2)_{\L\M}} &= \frac{ \mathcal{A}^{(2), r \Psi_4^K}_{(I_1 \times I_2)_{\L\M}}}{ \mathcal{A}^{(1), r \Psi_4^K}_{I_1}  \mathcal{A}^{(1), r \Psi_4^K}_{I_2}} \\
&= \frac{8 M^3}{\lambda^2} \frac{ \mathcal{A}^{(2),  \tilde{\Psi}}_{(I_1 \times I_2)_{\L\M}}}{ \mathcal{A}^{(1),  \tilde{\Psi}}_{I_1}  \mathcal{A}^{(1),  \tilde{\Psi}}_{I_2}} \\
&= \frac{8 M^3}{\lambda^2}  {\cal R}^{ \tilde{\Psi}}_{(I_1 \times I_2)_{\L\M}}. \label{eq:RPsiK_To_RPsiReg}
\end{align}

Finally, we can relate the QQNM ratio of the amplitudes of the Weyl scalar against those for the gravitational strain. From the SpEC conventions, $\Psi_4^{\rm SpEC}$ and the strain $h=h_+-i h_\times$ are related by
\begin{equation}
    \Psi_4^{SpEC} = - \partial_u^2 h = \omega^2 h = -\frac{\lp^2}{\lambda^2} h.
\end{equation}
It then follows that
\begin{align}
 {\cal R}^{h}_{(I_1 \times I_2)_{\L\M}} &= \frac{\frac{1}{\bracket{\omega^{(2)}_{(I_1 \times I_2)_{\L\M}}}^2}}{\frac{1}{\bracket{\omega^{(1)}_{I_1}}^2} \frac{1}{\bracket{\omega^{(1)}_{I_2}}^2}}  {\cal R}^{r\Psi_4^{SpEC}}_{(I_1 \times I_2)_{\L\M}}.
\end{align}
This implies, from Eq.~\eqref{eq:RPsiSpeC_To_PsiK},
\begin{equation}
 {\cal R}^{h}_{(I_1 \times I_2)_{\L\M}} = \frac{1}{2}
\frac{\bracket{\omega^{(2)}_{I_1}}^2 \bracket{\omega^{(2)}_{I_2}}^2}{\bracket{\omega^{(2)}_{(I_1 \times I_2)_{\L\M}}}^2}  {\cal R}^{r \Psi_4^K}_{(I_1 \times I_2)_{\L\M}},
\end{equation}
and so, from Eq.~\eqref{eq:RPsiK_To_RPsiReg}, we have the following relation for the QQNM ratio of the regularised Weyl scalar $ \tilde{\Psi}$ and strain $h$:
\begin{equation}
 {\cal R}^{h}_{(I_1 \times I_2)_{\L\M}} = -\frac{4 M^3}{\lambda^4}
\bracket{\frac{\lp^{(1)}_{I_1} \lp^{(1)}_{I_2}}{\lp^{(2)}_{(I_1 \times I_2)_{\L\M}}}}^2  {\cal R}^{ \tilde{\Psi}}_{(I_1 \times I_2)_{\L\M}}.
\end{equation}
For the case where $I_1 = I_2$, this immediately simplifies to Eq.~\eqref{eqn:Rh_1}.

\section{Relation between $\Psi_4$ amplitudes and even- and odd-parity contributions to the waveform}
\label{app:even and odd}

In the body of the paper, we perform expansions in spin-weighted spherical harmonics, leading to Teukolsky amplitudes and GW strain decomposed in that form. However, some properties of the GW-emitting system are more easily understood by decomposing the GW in even- and odd-parity tensor harmonics. In this appendix we relate the Teukolsky QNM amplitudes in terms of the corresponding even- and odd-parity GW mode amplitudes. 

As given in Eq.~\eqref{eq:QNM_sum_q} with Eqs.~\eqref{eq:masterTeuk} and~\eqref{eq:ConfMasterFunc}, the most general time-domain solution to the first-order Teukolsky equation, containing only QNMs, is given by
\begin{align}
  \Psi_4^{(1)}&= \frac{{}_{-2}\mathcal{Z}(\sigma)}{r^4} \sum_{\l,\m,\q} {}_{-2} A^{(1)}_{\l;\m;\q} \, {}_{-2} \tilde \psi_{\l;\m;\q}^{(1)}(\sigma) \, e^{ \lp^{(1)}_{\l;\m;\q} \tau} {}_{-2} Y_{\l\m},
\end{align}
where ${}_{-2} A^{(1)}_{\l;\m;\q}$ are arbitrary coefficients, and we note that Eq.~\eqref{eq:ConfMasterFunc} implies
\begin{equation}
\frac{{}_{-2}\mathcal{Z}(\sigma)}{r^4} = \frac{\lambda^2}{8M^3r} + \mathcal{O}\!\left(\frac{1}{r^{2}}\right).
\end{equation}
This mode expansion can be immediately translated to the GW strain using the fact that ${\displaystyle\lim_{r\rightarrow \infty} (r\Psi_4^{(1)})} = -\frac{1}{2}{\displaystyle\lim_{r\rightarrow \infty} (r\partial_u^2{h}^{(1)}_{\bar{m} \bar{m}})}$, where $\bar m^a=\frac{1}{\sqrt{2}r}(0,0,1,i\csc\theta)$:
\begin{equation}\label{eq:spin-weighted GW}
    h^{(1)}_{\bar m\bar m} \stackrel{r\to\infty}{=} -\frac{\lambda^4}{4M^3r} \sum_{\l,\m,\q} \frac{{}_{-2} A^{(1)}_{\l;\m;\q}}{(\lp^{(1)}_{\l;\m;\q})^2}\, e^{ \lp^{(1)}_{\l;\m;\q} \tau} {}_{-2} Y_{\l\m}.
\end{equation}
For simplicity, we have assumed the normalization~\eqref{eq:QNM_func_norm} $\, {}_{-2} \tilde \psi_{\l;\m;\q}^{(1)}(0)=1$.

Alternatively, at $\scri^+$, the GW (or more strictly, the shear~\cite{Madler:2016xju}) can be naturally decomposed into even-parity ($Y_{AB}$) and odd-parity ($X_{AB}$) tensor harmonics~\cite{Martel:2005ir,Spiers:2023mor},
\begin{equation}\label{eq:tensor-harmonic GW}
	h_{AB}^{(1)} = \frac{\lambda^4 r}{8M^3}\sum_{\l,\m,\q} \left( C^+_{\l;\m;\q} Y_{AB}^{\l\m} + C^-_{\l;\m;\q} X_{AB}^{\l\m} \right) e^{ \lp_{\l;\m;\q}^{(1)} \tau},
\end{equation}
where it is understood that this applies over the sphere with coordinates $\theta^A=(\theta,\varphi)$, in the limit $r\to\infty$ ($\sigma\to0$). The factor of $r$, rather than $1/r$, is due to the scaling of angular components, and $C^\pm_{\l;\m;\q}$ are constant (complex) amplitudes. We have introduced the factor $\lambda^4/(8M^3)$ for ease of comparison with Eq.~\eqref{eq:spin-weighted GW}.

The balance of even- and odd-parity content in the waveform is directly related to the system's degree of up-down symmetry. Under reflection through the equatorial plane, $\theta\to \pi-\theta$, the harmonics transform as%
\begin{align}%
    Y^{\l\m}_{AB} &\to (-1)^{\l+\m}Y^{\l\m}_{AB},\\
    X^{\l\m}_{AB} &\to -(-1)^{\l+\m}X^{\l\m}_{AB}.
\end{align}
It follows that for an up-down symmetric system, $C^+_{\l;\m;\q}$ must vanish for odd values of $\l+\m$, and $C^-_{\l;\m;\q}$ must vanish for even values of $\l+\m$. In contrast, for an up-down \emph{anti}symmetric system, $C^+_{\l;\m;\q}$ must vanish for \emph{even} values of $\l+\m$, and $C^-_{\l;\m;\q}$ must vanish for \emph{odd} values of $\l+\m$.

In order to relate the amplitudes ${}_{-2} A_{\l;\m;\q}^{(1)}$ to $C_{\l;\m;\q}^\pm$, we contract Eq.~\eqref{eq:tensor-harmonic GW} with $\bar m^A \bar m^B$, use the relations between harmonics in Ref.~\cite{Spiers:2023mor}, and compare to Eq.~\eqref{eq:spin-weighted GW}. A short calculation reveals
\begin{align} \label{eqn:AtoC1}
	{}_{-2} A_{\l;\m;\q}^{(1)} &= -\frac{\bracket{\lp_{\l;\m;\q}^{(1)}}^2}{4} \lambda_{\l,2} \left( C^+_{\l;\m;\q} - i C^-_{\l;\m;\q} \right), \\
	\bracket{{}_{-2} A_{\l; -\m;\Q}^{(1)}}^\star &= -(-1)^\m \frac{\bracket{\lp_{\l;\m;\q}^{(1)}}^2}{4} \nonumber\\
    &\quad \times\lambda_{\l,2} \left( C^+_{\l;\m;\q} + i C^-_{\l;\m;\q} \right), \label{eqn:AtoC2}
\end{align}
where $\lambda_{\l,2} = \sqrt{(\l+2)! / (\l-2)!}$.
To derive Eq.~\eqref{eqn:AtoC2}, we have also made use of the mirror relation~\eqref{QNM_frequency_new_relation} and the fact that $h_{AB}^{(1)}$ is real, which together imply $C^\pm_{\l; -\m;\Q} = (-1)^m (C^\pm_{\l;\m;\q})^\star$.

\bibliography{references}

\end{document}